# Visualizing nanodomain superlattices in halide perovskites giving picosecond quantum transients

Dengyang Guo[1,2]*, Thomas A. Selby[1]*, Simon Kahmann[1]†, Sebastian Gorgon[2], Linjie Dai[1,2], Milos Dubajic[1], Terry Chien-Jen Yang[1], Simon M. Fairclough[3], Thomas Marsh[2], Ian E. Jacobs[2], Baohu Wu[4], Renjun Guo[5], Satyawan Nagane[1], Tiarnan A. S. Doherty[1,3], Kangyu Ji[2], Cheng Liu[2], Yang Lu [1,2], Taeheon Kang[1], Capucine Mamak[1], Jian Mao[1], Peter Müller-Buschbaum[4,5], Henning Sirringhaus[2], Paul A. Midgley[3], Samuel D. Stranks[1,2]🖂

**Affiliations:**

[1]Department of Chemical Engineering and Biotechnology, University of Cambridge; Cambridge, UK.

[2]Department of Physics, Cavendish Laboratory, University of Cambridge; Cambridge, UK.

[3]Department of Materials Science and Metallurgy, University of Cambridge; Cambridge, UK.

[4]Forschungszentrum Jülich, JCNS at MLZ; Garching, Germany.

[5]Chair for Functional Materials, Department of Physics, TUM School of Natural Sciences, Technical University of Munich; Garching, Germany.

†Present address: Institute of Physics, Chemnitz University of Technology, Chemnitz, Germany

*These authors contributed equally to this work.

🖂 Corresponding author. Email: sds65@cam.ac.uk

**Abstract:**
The high optoelectronic quality of halide perovskites lends them to be utilized in optoelectronic devices and recently in emerging quantum emission applications[1–4]. Advancements in perovskite nanomaterials have led to the discovery of processes in which luminescence decay times are sub-100 picoseconds[5–7], stimulating the exploration of even faster radiative rates for advanced quantum applications, which have only been prominently realised in III-V materials grown through costly epitaxial growth methods[8]. Here, we discovered ultrafast quantum transients of time scales ~2 picoseconds at low temperature in bulk formamidinium lead iodide films grown through scalable solution or vapour approaches. Using a multimodal strategy, combining ultrafast spectroscopy, optical and electron microscopy, we show that these transients originate from quantum tunnelling in nanodomain superlattices. The outcome of the transient decays, photoluminescence, mirrors the photoabsorption of the states, with an ultra-narrow linewidth at low temperature as low as <2 nm (~4 meV). Localized correlation of the emission and structure reveals that the nanodomain superlattices are formed by alternating ordered layers of corner sharing and face sharing octahedra. This discovery opens new applications leveraging intrinsic quantum properties and demonstrates powerful multimodal approaches for quantum investigations.

## Main Text:

Halide perovskites have emerged as a promising class of materials for optoelectronic devices, including as solar cells, light-emitting diodes (LEDs) and X-Ray detectors[2,4]. Specifically, the unique combination of excellent light absorption, long charge carrier diffusion lengths, facile fabrication and bandgap tunability has led to the fastest rise in power conversion efficiencies recorded in solar cell research history[3]. When considering tuning at the nanoscale, quantum confinement, combined with unique optoelectronic properties, extends their application to a diverse array of fields including quantum emitters, which are critical components in the development of ultrafast communication, computing, and sensing devices[1,5,6]. Ideal quantum emitters should possess both coherent photon production and ultrafast dynamics, both of which require fine determination over the quantum confinement in the material. Single-photon emission has been reported in colloidal perovskite quantum dots, with sub-100-picosecond superradiance in superlattice structures[9], propelling perovskites to the forefront of ultrafast quantum applications.

Beyond nanostructured perovskites, surprising peaked features above the bandgap have been recently reported in the absorption and emission spectra of bulk $FAPbI_3$ films at cryogenic temperatures[9,10]. These exotic features are proposed to originate from discrete optical transitions of electrons from quantum confined states, with confinement length scales estimated to be within ~10 – 20 nm. However, due to the steady-state nature of absorption spectroscopy[9], and the temporal resolution limit of photoluminescence spectroscopy[10], the precise photophysical evolution and nature of the quantum confined states has yet to be elucidated. The origin of the quantum confinement is speculated to be from the formation of nanodomains that are either ferroelastic and/or ferroelectric, or δ-phase inclusions, yet direct evidence and structural understanding of these features, and connection with photophysical behaviour, is lacking[9]. Herein, we combine an ultrafast spectroscopic strategy with correlative optical and electron microscopy to report ultrafast quantum processes in bulk $FAPbI_3$. We directly reveal that the origin of the quantum effects are layered, nanotwinned α- and δ-phases, which, when periodic, can be regarded as a high-order hexagonal polytype. This multimodal microscopy study establishes nanoscale control of bulk perovskite materials as a key lever to enable new quantum technologies by leveraging the intrinsic and deterministic relationship between nanoscale structure and ultrafast photophysics.

## Picosecond quantum transients

We first monitored the ultrafast evolution of the quantized features (quantum transients) in bulk $FAPbI_3$ on a picosecond time scale using temperature-dependent transient absorption (TA) spectroscopy. $FAPbI_3$ bulk films were deposited on quartz substrates first using thermal co-evaporation (see Materials and Methods), showing high optoelectronic quality as demonstrated by solar cell performance (Supplementary Fig. 1). TA is a pump-probe technique that probes the wavelength-resolved and time-dependent difference in photo-absorption of the measured material, before and after pulsed laser-pump photoexcitation. This difference in wavelength pinpoints which energy levels the excited states occupy, and how fast the excited states leave the levels. We first used a pump energy of 3.1 eV (400 nm), well above the absorption onset of the $FAPbI_3$ films (1.52 eV; 817 nm, see Supplementary Fig. 1), to excite the thin film at a series of temperatures. With the sample at 5 K, we observed particularly distinct wavelength- and time-dependent oscillatory features (red stripes in Fig. 1a). These stripes demonstrate that, above the bandgap, there exists quantized energy levels and the signal from these excited states is attenuated over time on the order of pico-seconds.

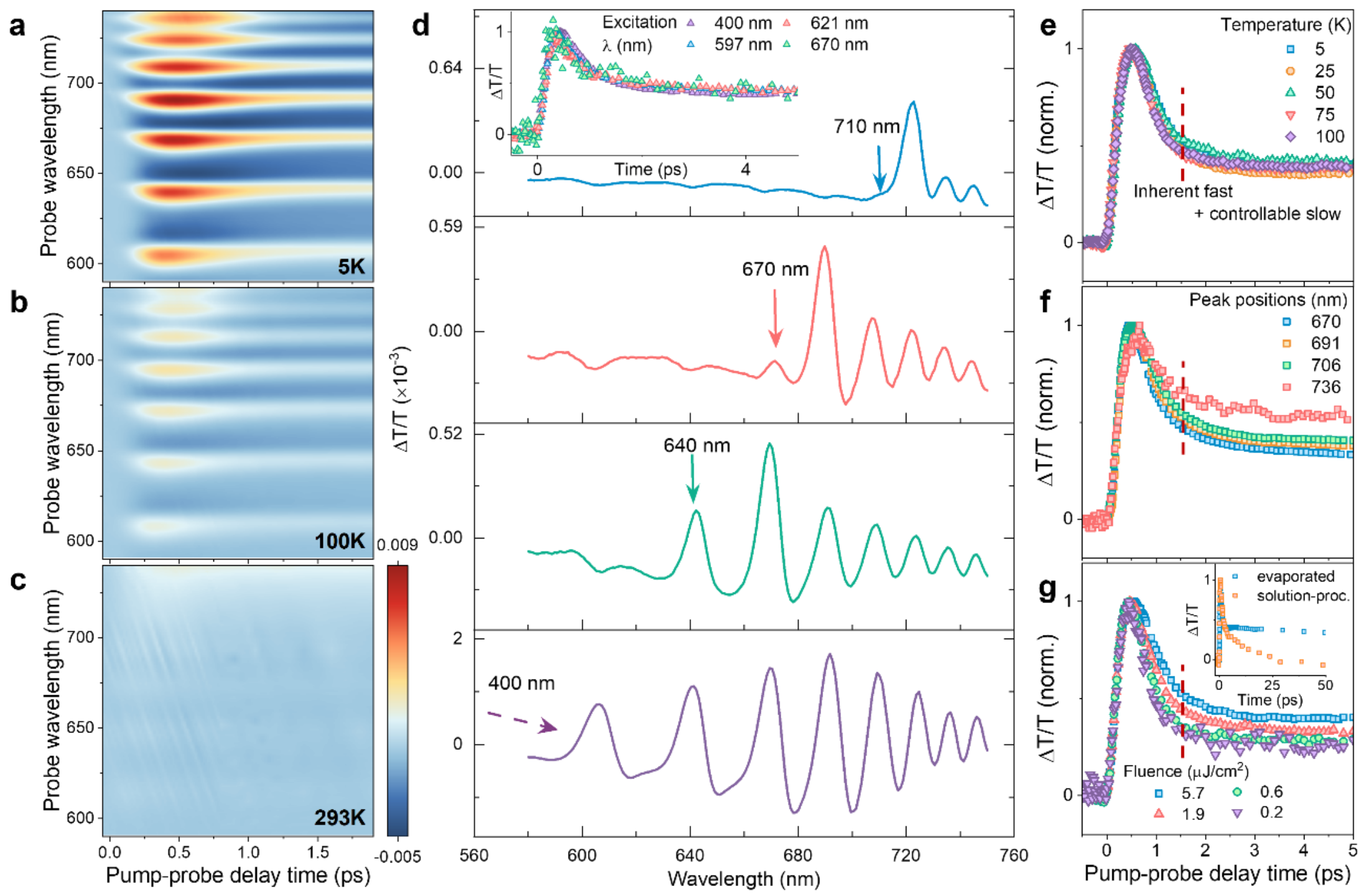

**Fig. 1 | Bulk $FAPbI_3$ thin films exhibit ultrafast quantum transients.** (**a-c**) Temperature-dependent TA maps of vapour-processed $FAPbI_3$ films show that discrete absorption peaks are attenuated as the temperature increases. The colour bar is shared for comparison, ranging from 5 to $9\times10^{-3}$, far above the $\Delta T/T$ sensitivity of $\sim10^{-6}$ of the home-built TA setup (**d**) Excitation-wavelength-dependent absorption spectra within a consistent time window. Only peaks at lower energy are clearly observed. Temperature: 5 K. Excitation wavelength (inset of **d**) and temperature-dependent (**e**) transients at peak position 723 nm. (**f**) Transients at different quantum peaks, excited by a fixed 400 nm laser beam with pulse fluence of 5.7 $uJ/cm^2$. (**g**) Excitation-fluence-dependent transients following the peak at 723 nm, excited by 400-nm laser excitation. Inset: comparison of the transients between the solution-processed and evaporated samples over a time window of 50 ps. Dashed line in (**e-g**) marks the turning point at around 2 ps.

These quantum features emerged gradually as we decreased the temperature from ambient conditions to 5 K (Fig. 1a-c). Consequently, all spectra at low temperatures, extracted vertically from time zero to a point near the centre of the stripe, displayed an oscillatory pattern that gradually emerged, transitioning from a smooth curve at room temperature (Supplementary Fig. 2). The same emergence of the oscillatory pattern was also observed in stabilised solution-processed $FAPbI_3$ films (Supplementary Fig. 2)[11]. Importantly, each peak of the pattern corresponds to the centre of a stripe, indicating the position of a quantum energy level. We also observed a slight blue shift in these levels with decreasing temperature (Supplementary Fig. 3), which is independent of a simultaneous red shift in bandgap, and thus was attributed to the temperature-induced changes in the lattice parameter relating to the quantum levels[9].

To explore the temporal characteristics, we extracted the time evolution of the quantum peaks using the linked minima of the stripes as a baseline (Supplementary Fig. 4). After excitation with the same excitation density at different wavelengths, but at the same temperature (Fig. 1d, inset), or after excitation at the same wavelength but at different temperatures (Fig. 1e), we find overlapping transients in the same quantum level (723 nm). Such an overlap is further observed in the transients from other quantum levels (Supplementary Fig. 5). Therefore, the temperature- and excitation-independent transients of the excited quantum states are inherently determined, indicating intrinsic transients via either transition to a corresponding ground state, or transfer to another energy level[12].

To distinguish between these two possibilities, we set the excitation wavelength to the individual peak or valley positions along the oscillatory profile (Fig. 1d, Supplementary Fig. 6). Only quantum signatures with a longer wavelength (lower energy) than the excitation wavelength are present in each spectrum, indicating the absence of shared ground states among the individual quantum levels. Physically, this result suggests that these quantum levels do not stem from different integer quantum numbers within the same confinement, but from distinct confinement regions within the material[12]. Structurally, this segregation implies that the quantum confinement is not a bulk property of the material, but rather emerges from separate domains with dimensions sufficiently small to induce confinement, in line with the proposal of a confinement length scale of ~10-20 nm[9].

To further explore the independence of the quantum levels, we extracted the full decay characteristics at each peak under identical excitation wavelength and temperature (Fig. 1f, Supplementary Fig. 7). Interestingly, at the quantum peaks of longer wavelengths (lower energy), we observed a correspondingly longer decay time, which may mean excited states transfer from higher to lower energy levels. However, such an assumption implies a close attachment between these regions, which could enable the direct transfer. Such a close proximity is not guaranteed, as further demonstrated by the distinct emission peaks observed in Fig. 2, as well as the necessity of physical barriers separating the domains from the bulk. Therefore, the distinct decays at different peaks indicate that the time-evolution of the excited states can be an intrinsic character of the quantum confinement, in analogy to the size-dependent photocarrier lifetime of short-period ErSb/GaSb nanoparticle superlattices[13].

To further substantiate that the quantum levels arise from confinement within individual domains, we examined the decay behaviour at the same quantum peak (Fig. 1g). Importantly, the decays for transients occurred within the initial 2 ps. Interestingly, similar 2 ps transients were also observed in solution-processed films (Supplementary Fig. 8). Moreover, in both evaporated (Fig. 1g) and solution-processed (Supplementary Fig. 6) films, the excitation density does not influence the timescale of these fast transients, with the turnover point from a fast to slow component remaining approximately at 2 ps. The constancy of the initial ultrafast process agrees with the quantum transients being intrinsic to each level. In contrast, the slow part of the full decay after the initial 2 ps transients depends on the sample processing method (Fig. 1g inset), showing a timescale of ~1 ns in evaporated films and ~50 ps in solution-processed films (Supplementary Fig. 7). Such variability of the long-tail slow process suggests that this second decay component is controllable and may relate to different trap densities. Such a combination of fast and slow components on picosecond timescales is in line with the photocarrier lifetime in GaAs superlattice grown by molecular beam epitaxy, of which the slow part is controlled by extrinsic effects (in that case, ErAs doping)[13,14].

## Isolated emissions of the quantum transients

To understand the isolated transient processes originating from individual domains, we recorded cryogenic hyperspectral photoluminescence (PL) images. Surprisingly, we see individual PL spots at distinct spatial locations, with the distributions particularly stark in the solution-processed samples in which the emission points are spread randomly several microns away from each other within the PL map (Fig. 2a-e and Supplementary Movie 1; see

Supplementary Fig. 9 for similar qualitative behaviour on evaporated samples). The PL signals exhibit distinct and sharp emission peaks with ultra-narrow full width at half maximum (FWHM) of 1 to 5 nm (~ 4 - 10 meV), as depicted in Fig. 2f-j. These PL peaks align precisely with the corresponding peak segments of the photoabsorption spectrum (cf. Fig. 1), shown as dashed lines in each region. The precise alignment indicates the absence of a Stokes shift, which is typically observed in perovskite films due to carrier relaxation in a continuous band. Hence, this finding further highlights the quantum nature of the observed peaks, underscoring their individual and isolated characteristics.

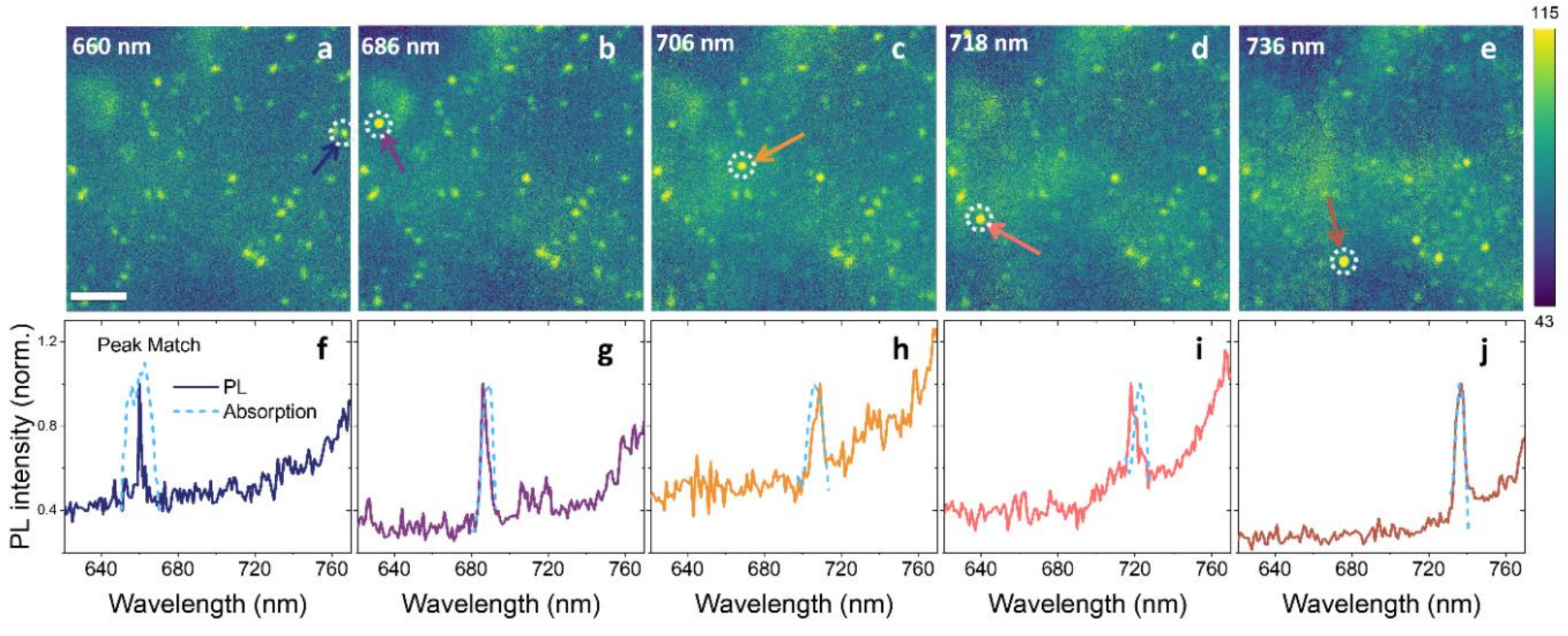

**Fig. 2. Bulk $FAPbI_3$ thin films exhibit isolated sites of quantum emission with extremely narrow linewidths**. (**a-e**) At 5K, the hyperspectral wavelength-selective PL map of solution-processed $FAPbI_3$ films exhibits individual emission spots at various wavelengths, shown here with excitation at 405 nm with power ~1.8 mW/cm$^2$. Scale bar: 10 µm. The map is recorded from a fixed region of the film, and the wavelength of the individual map is denoted on each panel. The individual spots appear at different positions when the map is displayed at changing wavelengths (see also Supplementary Movie 1). The intensity bar denotes PL intensity. (**e-h**): Photoluminescence spectra showing emission peaks with extremely narrow FWHMs at individual spots denoted with white dashed circles in the map. Dashed cyan lines are region-selected TA peaks, of which the original data is normalized to one. The TA and PL peaks show a precise match.

We note that these isolated characteristics cannot be resolved by bulk structural tools, including cryogenic X-ray diffraction (cryo-XRD), cryogenic wide-angle X-ray scattering (cryo-WAXS), and cryogenic small-angle X-ray scattering (cryo-SAXS), as shown in Supplementary Fig. 10-12. This suggests that the confinement requires a local and nanoscale structural resolution beyond these bulk structure characteristics, especially considering that quantum tunnelling typically occurs when the barrier thickness is below ~1-3 nanometres[15,16].

## Ubiquitous spread of nanotwinning domains

To gain a structural understanding of the isolated emission spots, spatially correlated hyperspectral PL and scanning electron diffraction (SED) measurements were performed at both ambient and cryogenic temperatures on $FAPbI_3$ thin films thermally evaporated onto $SiN_x$ membrane grids compatible with both measurement modalities. SED is a variant of 4D-scanning transmission electron microscopy (4D-STEM) where an electron probe with low convergence angle (typically < 1 mrad) is rastered across a thin sample and a transmitted electron diffraction pattern obtained at each pixel position. This provides spatially resolved information on crystallographic phase, orientation and extended defects at low electron dose[11,17,18]. When performing measurements of hybrid perovskites using the TEM, electron beam induced damage was constantly considered and the beam current was maintained at ~3 pA, therefore providing a sufficiently low dose (~11e/Å$^2$) to capture the innate state of the film considering previously reported dose limits (Supplementary Text 1)[19–21].

At both cryogenic and ambient temperatures, when virtual bright field (VBF) images are formed by taking intensity only from the direct beam in the diffraction pattern, intragrain planar features are observed (zoomed example shown in Fig. 3a & b and Supplementary Fig. 13 & 14). These features are explained by the presence of stacking faults along a common crystallographic direction. Indeed, diffuse scattering along a $<111>_c$ direction is also present in the corresponding diffraction patterns, explained by the occurrence of nanoscale planar features leading to the elongation of reciprocal space lattice points into rods (Fig. 3c). In addition to this, archetypal diffraction patterns consisting of two overlaid $<110>_c$ zone axis patterns sharing spots along a $<111>_c$ direction are also observed, suggesting the presence of twinning along the said $<111>_c$ direction (Supplementary Fig. 15). It is important to note that a perfect $\{111\}_c$ type twin boundary is equivalent to a planar layer of face-sharing octahedra (δ phase) separating regions of corner-sharing octahedra (α phase), as illustrated in Fig. 3f. Furthermore, depending on the number of face-sharing layers present, the planar defect can be regarded as a twin or stacking fault[22] and henceforth we shall use the all-encompassing term 'nanotwinning' to broadly describe the different types of these planar defects.

To link the streaking observed in the diffraction patterns to the real space planar features, a rotation calibration was applied to the SED data. As expected after this calibration, the diffuse scattering along a $<111>_c$ direction is perpendicular to the planes observed in the VBF images (Fig. 3b & c). Additionally, if virtual dark field (VDF) images are created by placing two apertures over complementary reflections, planar stripes that approximately anti-correlate with respect to each other are observed (Fig. 3d & e and Supplementary Fig. 16). To understand the contrast in these VDF images, masks were applied to the data to extract the averaged diffraction pattern from each complementary reflection (Supplementary Fig. 16). These patterns display features still indicative of nanoscale twins, allowing us to conclude that the SED probe is likely encompassing multiple twin boundaries at each pixel position and the contrast observed in the VDFs is from $\{111\}_c$ type intra-grain nanotwinning which is largely aperiodic throughout the grain (Fig. 3d & e, and Supplementary Fig. 16). Linking the structural and photophysical observations presented thus far, we propose that many of these coplanar $\{111\}_c$ type nanotwins form an array of wells and barriers analogous to a Krönig-Penney (KP) type quantum superlattice within grains of alternating corner and face sharing octahedra (Fig. 3f), as previously theorized[9]. This is further consistent with prior literature with the occurrence of $\{111\}_c$ twinning being shown to be present in $FAPbI_3$[22,23]. Interestingly, reports have shown that when quantum confinement effects are observed this is detrimental to solar cell performance[37]. Intuitively this is consistent with our observation as when the nanotwins are oriented in the plane of the film they will act as barriers to the vertical movement of charge carriers, leading to worse device efficiency[34]. Furthermore, it has been shown that A-site alloying with Cs, which is known to reduce the prevalence of nanotwin domains,[38] attenuates the above bandgap oscillations in the absorption and emission spectra.[10]

Using the capability of the local mapping, we now explore the spatial distribution of the nanotwinning across the film. Given the enormity and richness of the SED datasets acquired (totalling 428 Gb after compression from 122 individual SED scans) manually indexing and interpreting each diffraction pattern is intractable. We therefore developed automated analysis pipelines which are computationally inexpensive, general and robust (Supplementary Text 2). Inspired by simple linear iterative clustering (SLIC) used in the remote sensing field[24,25], we first cluster the diffraction patterns before taking the average pattern from each cluster. This takes the 262,144 (512 × 512) individual patterns recorded in a single SED measurement to approximately 500 averaged patterns. Strikingly, despite the quantity of data, this process can be optimized to only take minutes for each SED scan (Supplementary Text 2). Once the diffraction patterns are clustered, we observe the extensive presence of $\{111\}_c$ type nanotwinning through the observation of many diffraction patterns with diffuse scattering consistent with the proposed structural model (Fig. 3g and Supplementary Fig. 13c & 14c). Similarly, planar features in reconstructed VBF images are also observed across the entirety of the large area imaged (27 µm × 14 µm; Fig. 3g and Supplementary Fig. 13a & 14a). Furthermore, these observations are consistent across multiple samples prepared in different batches (Fig. 3g & 4a) allowing us to conclude $\{111\}_c$ intragrain nanoscale twins are a highly ubiquitous structural feature in $FAPbI_3$ films.

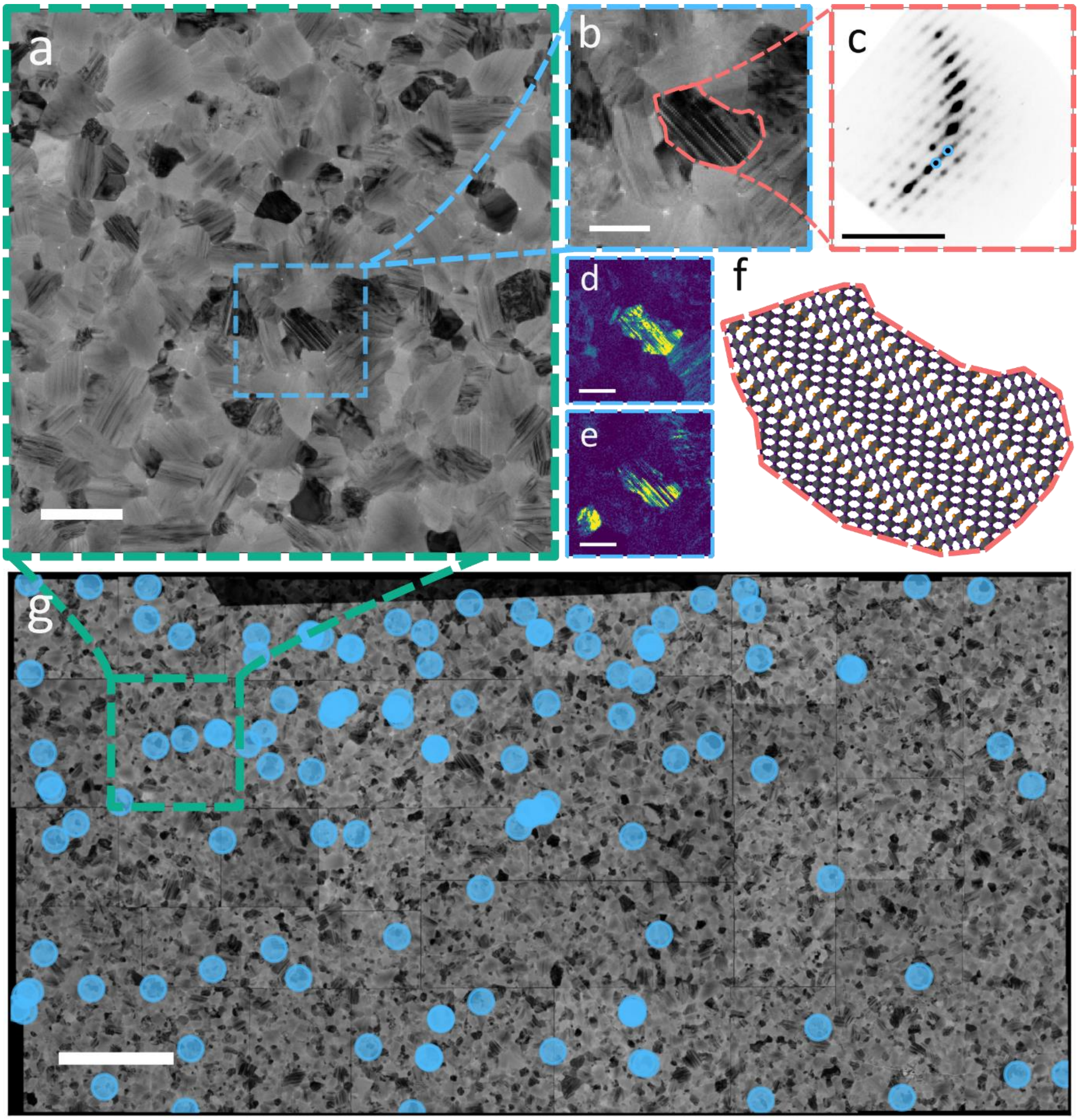

**Fig. 3. $\{111\}_c$ type nanotwinning is ubiquitous in bulk $FAPbI_3$ thin films as observed through spatial nanoscale structural maps**. (**a**) A virtual bright field (VBF) image formed from a typical area of a vapour-deposited $FAPbI_3$ film as displayed in (**g**), taken at ambient temperature. (**b**) Zoom-in on the region marked in (**a**). (**c**) The average diffraction pattern from the grain marked in (**b**). (**d&e**) Virtual dark field images (VDF) formed when 'apertures' are placed over the spots corresponding to opposing twins, marked by the circles in (**c**) (blue). (**f**) A structural cartoon of the nanotwinned structure within the grain marked in (**b**), where corner sharing octahedra (purple) are separated by $\{111\}_c$ type twins or equivalently layers of face sharing octahedra (orange). (**g**) Virtual bright field (VBF) images from the SED dataset taken at ambient temperature are stitched together to form a large area map with the blue points exhaustively showing where grains are oriented close to a $<110>_c$ zone axis such that the observation of nanotwinning is most pronounced. Scale bars are: 500 nm (**a**); 200 nm (**b**, **d & e**) 3μm (**g**); and 1$Å^{-1}$ (**c**).

SED was performed at both ambient temperature and when cooled to a stable temperature of ~90 K such that any changes induced upon cooling could be observed. Indeed, even though $\{111\}_c$ nanotwinning remains alike between temperatures (Supplementary Fig. 17), the appearance of additional superstructure reflections can be observed upon cooling, marking a phase transition to a lower symmetry phase (Supplementary Fig. 14). If octahedral tilting is considered to be the only structural distortion occurring upon cooling, we assign the low-temperature phase of $FAPbI_3$ to be an in-phase tilt system about more than one pseudocubic axis; if we consider a perfectly corner-sharing Pb-I sublattice this would result in a space group of either *Immm* ($a^+b^+c^+$); *I*4/*mmm* ($a^0b^+b^+$); or $Im\overline{3}$ ($a^+a^+a^+$) (Supplementary Text 3). As the appearance of superstructure reflections in these space groups is dictated by the body-centring operation, extra reflections appear at identical positions[26]. Linking this observation to the PL spectra at cryogenic temperatures, where a red-shift of the band edge is observed compared to room temperature[9], we can now provide a nuanced structural understanding of how this PL shift is related to octahedral tilting at low temperature (Supplementary Text 3). This result is consistent with variable temperature powder XRD results where a shift in peak position is observed upon cooling attributed to structural contraction (Supplementary Fig. 10).

## Correlation of the localised emission and nanotwinning superlattice

To connect the structural observations with the localized photophysical properties, SED measurements were performed on the same area to that imaged by hyperspectral PL, both at cryogenic temperatures. We show the hyperspectral PL maps of the same region in Fig. 4a & b, representing the emission at 706 nm and 734 nm, respectively. We observe local above-bandgap oscillations in the PL spectra throughout the films examined (Supplementary Fig. 9) as well as particularly bright isolated emitters from each wavelength that vary spatially. We use uniquely shaped gold fiducial markers to spatially correlate between datasets from the two techniques (Fig. 4d, Supplementary Text 4)[27]. Due to the spatial resolution difference between the two imaging modalities, many SED scans were acquired before being stitched together such that a large area of the hyperspectral data could be covered (Fig. 4a & b and Supplementary Fig. 13 & 14). To do this accurately, as shown in Fig. 3g and 4a & b, a method combining computer vision techniques was used, namely the scale invariant feature transform, random sample consensus algorithm, iterative closest point algorithm and modern pretrained machine learning models (HardNet and AdaLAM; Supplementary Text 4)[28,29]. Having stitched the multiple contiguous SED scans together we then co-register the large area image recovered with the hyperspectral PL data. To achieve this the two datasets were cropped to only include the Au fiducial marker and an affine transform was defined where the normalized cross correlation was maximal (Supplementary Text 4).

Crucially, when the cryo-hyperspectral PL and cryo-SED datasets are overlaid (illustrated in Fig. 4a and Supplementary Fig. 18), the structural properties of the isolated emitters can be understood. Firstly, we note that grains oriented close to a $<111>_c$ (equivalent to a $<001>_h$) zone axis of $FAPbI_3$ are prevalent in areas of pronounced isolated emission and that this is true for multiple separate sample batches (Fig. 4a & b, and Supplementary Fig. 22). We propose that the isolated emission is predominately dictated by the ordering and periodicity of the superlattice. Considering the fact that superstructure reflections appear in $<111>_c$ patterns consistent with the octahedral tilting framework discussed above upon cooling (Fig. 4a and inset of Supplementary Fig. 18e), and that no more additional reflections are observed, we assign the grains oriented close to $<111>_c$ to be either a 3C, 6H, 9R, 12H, 12R or 18H polytype (or combination thereof, Supplementary Fig. 20)[33,34]. It is important to note that if we consider grains to be a high-order hexagonal polytypes, where nanotwinned layers are periodic, we cannot assign the space group to be formally $Immm$ ($a^+b^+c^+$); $I4/mmm$ ($a^0b^+b^+$); or $Im\bar{3}$ ($a^+a^+a^+$) which relate to the cubic 3C aristotype (Supplementary Text 3). We therefore simulate the diffraction pattern of a 12H polytype possessing tilted corner sharing octahedra oriented down the $<001>_h$ zone axis to confirm the assignment of in-phase tilting about more than one pseudocubic axis (Supplementary Text 3). This assignment is in agreement with prior literature where high order polytypes (12H) of $CsPbI_3$ were found to possess a direct bandgap whereas the lower order equivalents have indirect bandgaps, and are thus would not be expected to be emissive[34]. We note definitive assignment of the emission to a structural feature is challenging due to the resolution differences between the two techniques.

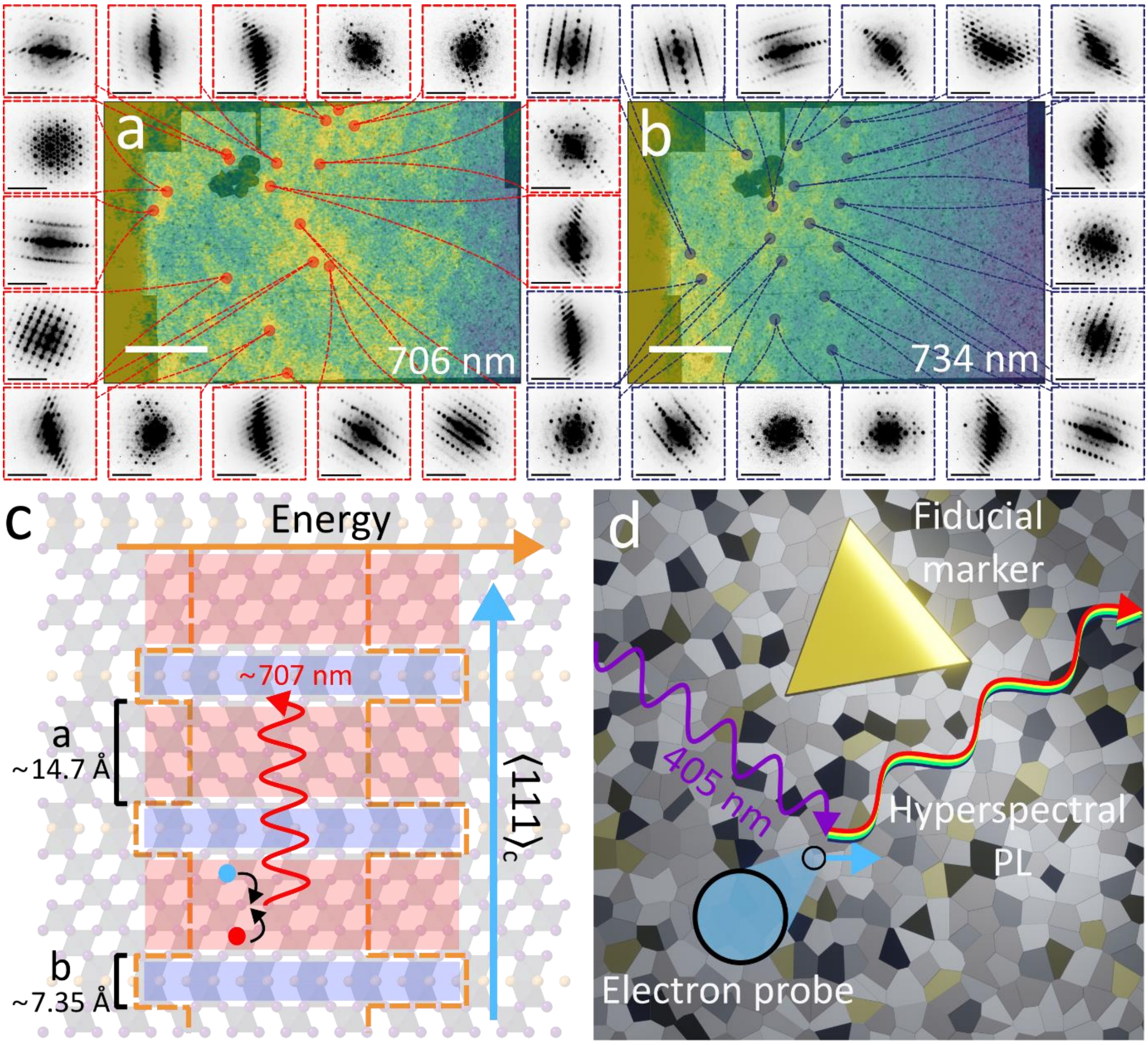

**Fig. 4. Spatial correlations between local photophysics (PL) and structure (SED) reveal bright isolated emission arising from grains which show ordered nanoscale twinning.** (**a & b**) Stitched SED (cryogenic temperature) correlated with the hyperspectral PL (4.5 K) on a vapour-processed $FAPbI_3$ film. Maps are shown from the same scan area of the PL at 706 and 734 nm. For each wavelength the PL intensity shown is at a 65% threshold of the maximum value. (**c**) The structural model proposed (12H polytype illustrated) where isolated emitters of ~707 nm are observed if highly ordered twin domains are present. (**d**) A schematic of how a common region is found between the electron and optical microscopes, with the use of a gold fiducial marker. Scale bars are: 5 μm and 1Å$^{-1}$.

Considering the proposal that the isolated emission originates from a superlattice, the wavelength of emission (706 or 734 nm presented) should be modulated by the spacing between corner and face sharing layers. Thus, we attempt to understand the size, ordering and periodicity of the nanotwinned array collectively from the ensemble of diffraction patterns collected and the reconstructed VBF and VDF images for isolated emission at both 706 nm and 734 nm. If the diffraction pattern of a perfectly regular superlattice viewed along a $[100]_h$ zone axis is examined, in principle the Bragg peaks along the twinning direction converge into an unresolvable streak as the number of corner sharing layers between face sharing nanotwins increases (Supplementary Fig. 19). It is therefore difficult to discern between a high-order hexagonal polytype and the purely diffuse scattering which can result from a disordered arrangement of nanotwins. Furthermore, as each SED pixel (6.5 nm) is likely to encompass multiple twin boundaries we can only ascertain an estimate of the upper limit (~5-10 nm) for the thickness of corner sharing octahedral layers upon which the superlattice forms, which is then used to inform subsequent calculations. In Fig. 4c, we show an illustrative structure of the $\{111\}_c$ type nanotwins viewed in cross-section into the film; if this is considered to be perfectly ordered, it can be viewed as a high-order hexagonal polytype (12H as presented consistent with the assignment of the observed $<111>_c$ patterns). We propose that the alternating α- and δ-phase serve as the quantum well and barrier in the KP model. By assuming the width of the well and barrier occurs at integer numbers of the octahedral layers along a $<111>_c$ direction (intervals of $u$ = 0.37 nm at room temperature), we find the simulated quantum well and barrier parameters to model a 12H polytype are $a$ = 1.48 nm (4$u$) and $b$ = 0.74 nm (2$u$). By implementing this configuration in the KP model, along with an effective mass of $0.21m_0$ (where $m_0$ is the mass of an electron)[36], we calculate a first discrete energy level from confinement to be 707 nm, which is consistent with the experimentally observed peak at 706 nm. Similarly, the 734 nm emission could originate from a combination of $a$ = 1.85 nm (5$u$) and $b$ = 1.11 nm (3$u$). We note that an 18H polytype is another example that would yield consistent results, and that the calculation depends on the effective mass value used, making it difficult to unambiguously assign a specific polytype structure to a specific quantum peak (Supplementary Text 5, Table S1).

Finally, we seek insight into the transients with these structural and energy parameters. Given the observed ultrafast transients are ~2 ps, regardless of the material conditions, and barrier thicknesses are below 1 nm,[15] a reasonable mechanism facilitating the ultrafast transients of the excited states from the quantum levels involves quantum tunnelling. We thereby apply a semi-classical approach using the Wentzel–Kramers–Brillouin (WKB) approximation. This approach takes the states as classical particles for calculating the frequency of bouncing at the barriers but treats the states as wavefunctions for estimating the chance of quantum tunnelling through the barriers. The lifetimes of the states are inherently determined by quantum tunnelling and modulated by the width of the wall and the thickness of the barrier. By applying the barrier and well parameters within the WKB approximation, we estimate the mean lifetime of the excited carriers in the quantum confined levels to be approximately ~2.0 ps for a peak at 706 nm and ~2.8 ps for a peak at 756 nm. This estimation aligns well with the experimental values presented in Fig. 1 (Supplementary Text 6). Such consistency of the experimental data and calculated values based on the KP model, along with the timescales (accounting for the picosecond transients and the WKB approximation), supports the concept that the observed quantum transients relate directly to the nanotwinned structures forming quantum wells.

## Discussion

Both the KP model and WKB approximation are fundamental concepts in quantum physics. Our work demonstrates that these concepts, i.e., ultrafast dynamics of photoexcited carriers in quantum levels within a KP superlattice, can be intrinsically obtained in an inexpensive and easily fabricated metal halide perovskite bulk film. The ultrafast lifetime of excited states in a confined KP superlattice is ultimately regulated by the WKB approximation, determined by the length of the well and thickness of the barrier. As the inherent lifetime is defined by these structural parameters, the quest for achieving ultrafast lifetimes has in the past relied on constructing ideal KP superlattices such as through expensively grown molecular beam epitaxy, or introduction of extra doping as a modification step[14]. The discovery of ~2 picosecond quantum transients originating from the intrinsic nanotwinning highlights opportunities for facile strategies to achieve nanoscale control of ultrafast quantum emitters. Future work would focus on developing strategies to controllably grow twinned nanostructures into uniform, vertical arrays, maintaining a well-ordered and common alternating number so that ultrafast dynamics can be achieved at a fixed emission wavelength. Such a uniform arrangement could be further applied for circuit integration for ultrafast device applications. Alternatively, an isolated and horizontal growth of a chosen alternating number, surrounded by a thicker barrier to control the quantum tunnelling probability, could provide further insights into controlling the inherent lifetime of excited carriers. These identical but isolated ultrafast emitters could also be utilized for quantum-related studies. With the direction of growth under control, the transition dipole moment of emission could also be further analysed, for example by applying a polarized excitation.[32,39]

Furthermore, our multimodal strategy, which combines ultrafast spectroscopy and correlative optical and electron microscopy, can serve as a generalized platform for other material systems. Such a platform offers a top-down routine, capable of resolving ultrafast processes while resolving the correlated nanoscale structure. By leveraging these computer vision-based analysis pipelines and clustering methodologies, one can investigate and understand the nano-structural photophysics of a wide array of advanced materials, leveraging analyses on extremely rich but large data sets. Furthermore, the correlation of structural attributes and optoelectronic properties offers a unique tool for precision engineering at the nanoscale. This could lead to the discovery of novel materials with tailored properties for specific applications in quantum computing, solar energy harvesting, and light-emitting devices, among others.

## References

1. Kaplan, A. E. K. *et al.* Hong–Ou–Mandel interference in colloidal CsPbBr3 perovskite nanocrystals. *Nat. Photonics* **17**, 775–780 (2023).
2. Stranks, S. D. & Snaith, H. J. Metal-halide perovskites for photovoltaic and light-emitting devices. *Nat. Nanotechnol.* **10**, 391–402 (2015).
3. Snaith, H. J. Present status and future prospects of perovskite photovoltaics. *Nat. Mater.* **17**, 372–376 (2018).
4. Yi, L., Hou, B., Zhao, H. & Liu, X. X-ray-to-visible light-field detection through pixelated colour conversion. *Nature* **618**, 281–286 (2023).
5. Utzat, H. *et al.* Coherent single-photon emission from colloidal lead halide perovskite quantum dots. *Science* **363**, 1068–1072 (2019).
6. Zhu, C. *et al.* Single-photon superradiance in individual caesium lead halide quantum dots. *Nature* **626**, 535–541 (2024).
7. Rainò, G. *et al.* Superfluorescence from lead halide perovskite quantum dot superlattices. *Nature* **563**, 671–675 (2018).
8. García De Arquer, F. P. *et al.* Semiconductor quantum dots: Technological progress and future challenges. *Science* **373**, eaaz8541 (2021).
9. Wright, A. D. *et al.* Intrinsic quantum confinement in formamidinium lead triiodide perovskite. *Nat. Mater.* **19**, 1201–1206 (2020).
10. Elmestekawy, K. A. *et al.* Controlling Intrinsic Quantum Confinement in Formamidinium Lead Triiodide Perovskite through Cs Substitution. *ACS Nano* **16**, 9640–9650 (2022).
11. Doherty, T. A. S. *et al.* Stabilized tilted-octahedra halide perovskites inhibit local formation of performance-limiting phases. *Science* **374**, 1598–1605 (2021).
12. Dai, L. *et al.* Slow carrier relaxation in tin-based perovskite nanocrystals. *Nat. Photonics* **15**, 696–702 (2021).

13. Hanson, M. P., Driscoll, D. C., Zimmerman, J. D., Gossard, A. C. & Brown, E. R. Subpicosecond photocarrier lifetimes in GaSb∕ErSb nanoparticle superlattices at 1.55μm. *Appl. Phys. Lett.* **85**, 3110–3112 (2004).
14. Kadow, C. *et al.* Self-assembled ErAs islands in GaAs for THz applications. *Phys E Low-Dimens Syst Nanostructures* **7**, 97–100 (2000).
15. Shankar, R. *Principles of Quantum Mechanics*. (Plenum Press, 1994).
16. Lerner, R. G. & Trigg, G. L. *Encyclopedia of Physics*. (New York : VCH, 1991).
17. Doherty, T. A. S. *et al.* Performance-limiting nanoscale trap clusters at grain junctions in halide perovskites. *Nature* **580**, 360–366 (2020).
18. Macpherson, S. *et al.* Local nanoscale phase impurities are degradation sites in halide perovskites. *Nature* **607**, 294–300 (2022).
19. Ferrer Orri, J. *et al.* Unveiling the Interaction Mechanisms of Electron and X-ray Radiation with Halide Perovskite Semiconductors using Scanning Nanoprobe Diffraction. *Adv. Mater.* **34**, 2200383 (2022).
20. Rothmann, M. U. *et al.* Structural and Chemical Changes to $CH_3NH_3PbI_3$ Induced by Electron and Gallium Ion Beams. *Adv. Mater.* **30**, 1800629 (2018).
21. Li, Y. *et al.* Unravelling Degradation Mechanisms and Atomic Structure of Organic-Inorganic Halide Perovskites by Cryo-EM. *Joule* **3**, 2854–2866 (2019).
22. Li, W. *et al.* The critical role of composition-dependent intragrain planar defects in the performance of MA1–xFAxPbI3 perovskite solar cells. *Nat. Energy* **6**, 624–632 (2021).
23. Pham, H. T. *et al.* Unraveling the influence of CsCl/MACl on the formation of nanotwins, stacking faults and cubic supercell structure in FA-based perovskite solar cells. *Nano Energy* **87**, 106226 (2021).
24. Shi, Y., Wang, W., Gong, Q. & Li, D. Superpixel segmentation and machine learning classification algorithm for cloud detection in remote-sensing images. *J. Eng.* **2019**, 6675–6679 (2019).
25. Achanta, R. *et al.* SLIC Superpixels Compared to State-of-the-Art Superpixel Methods. *IEEE Trans. Pattern Anal. Mach. Intell.* **34**, 2274–2282 (2012).
26. Woodward, D. I. & Reaney, I. M. Electron diffraction of tilted perovskites. *Acta Crystallogr. B* **61**, 387–399 (2005).
27. Guo, Z. *et al.* Facile synthesis of micrometer-sized gold nanoplates through an aniline-assisted route in ethylene glycol solution. *Colloids Surf. Physicochem. Eng. Asp.* **278**, 33–38 (2006).
28. Riba, E., Mishkin, D., Ponsa, D., Rublee, E. & Bradski, G. Kornia: an Open Source Differentiable Computer Vision Library for PyTorch. (2019) doi:10.48550/ARXIV.1910.02190.
29. Liu, J. & Bu, F. Improved RANSAC features image-matching method based on SURF. *J. Eng.* **2019**, 9118–9122 (2019).
30. Kim, K. & Kim, J. Origin and Control of Orientation of Phosphorescent and TADF Dyes for High-Efficiency OLEDs. *Adv. Mater.* **30**, 1705600 (2018).
31. Kumar, S. *et al.* Anisotropic nanocrystal superlattices overcoming intrinsic light outcoupling efficiency limit in perovskite quantum dot light-emitting diodes. *Nat. Commun.* **13**, 2106 (2022).
32. Ye, J. *et al.* Direct linearly polarized electroluminescence from perovskite nanoplatelet superlattices. *Nat. Photonics* (2024) doi:10.1038/s41566-024-01398-y.
33. Li, Z., Park, J.-S. & Walsh, A. Evolutionary exploration of polytypism in lead halide perovskites. *Chem. Sci.* **12**, 12165–12173 (2021).
34. Li, Z., Park, J.-S., Ganose, A. M. & Walsh, A. From Cubic to Hexagonal: Electronic Trends across Metal Halide Perovskite Polytypes. *J. Phys. Chem. C* **127**, 12695–12701 (2023).

35. Weller, M. T., Weber, O. J., Frost, J. M. & Walsh, A. Cubic Perovskite Structure of Black Formamidinium Lead Iodide, α-[$HC(NH_2)_2$]$PbI_3$, at 298 K. *J. Phys. Chem. Lett.* **6**, 3209–3212 (2015).
36. Wang, S., Xiao, W. & Wang, F. Structural, electronic, and optical properties of cubic formamidinium lead iodide perovskite: a first-principles investigation. *RSC Adv.* **10**, 32364–32369 (2020).
37. Elmestekawy, K. A. *et al.* Photovoltaic Performance of $FAPbI_3$ Perovskite Is Hampered by Intrinsic Quantum Confinement. *ACS Energy Lett.* **8**, 2543–2551 (2023).
38. Iqbal, A. N. *et al.* Composition Dictates Octahedral Tilt and Photostability in Halide Perovskites. *Adv. Mater.* 2307508 (2024) doi:10.1002/adma.202307508.
39. Gudiksen, M. S., Lauhon, L. J., Wang, J., Smith, D. C. & Lieber, C. M. Growth of nanowire superlattice structures for nanoscale photonics and electronics. *Nature* **415**, 617–620 (2002).

# Methods

**(Cryo-) Transient absorption spectroscopy**

Transient absorption spectroscopy measurements were performed via a home-built platform. The output from a titanium:sapphire amplifier system (Spectra Physics Solstice Ace) was divided into pump and probe beam paths. This system operates at 1 kHz and generates pulses of approximately 100 fs. The 400 nm pump pulses were produced by directing the 800 nm fundamental beam of the Solstice Ace through a 1 mm thick SHG BBO crystal (Eksma Optics). Pump pulses with tuneable wavelengths (the ones used in Fig.1) were generated by directing the 800 nm fundamental beam through a TOPAS optical parametric amplifier and subsequent suitable spectral filtering. A chopper wheel blocked every other pump pulse to provide pump-on and pump-off referencing, while a computer-controlled mechanical delay stage (Thorlabs DDS300-E/M) adjusted the temporal delay up to 2 ns between the pump and the probe. A visible broadband beam (525–775 nm) was generated in a custom-built noncollinear optical parametric amplifier (NOPA), and the white light was divided into two probe and reference beams using a 50/50 beamsplitter. The reference beam, which did not interact with the pump, passed through the sample. With this setup, it was possible to measure small signals with a $\Delta T/T$ of approximately $10^{-5}$. The transmitted probe and reference pulses were collected with an InGaAs dual-line array detector (Hamamatsu G11608-512DA; spectrograph, Andor Shamrock SR-303i-B), which was driven and read out by a custom-built board (Stresing Entwicklungsbüro).

The temporal resolution is limited by the pump pulse duration only. For the 400 nm pump, the pulse duration is ~100 fs and for visible NOPA pumps (all the others that are not 400 nm), the pulse duration is ~120 fs, overall, they are definitely all less than 150fs. We stretched the probe beam in time using fused silica (24 mm) and an AR coated quartz (24 mm) to get a broadband amplification from 530 nm to 780 nm. The probe beam do not affect the temporal resolution, because we collect wavelength-dependent broadband probe data and we have chirp correction for each probe wavelength, so that the temporal stretch due to group velocity dispersion of the probe is corrected for each wavelength in the software. As per the measurements, the camera is time-independent and collects all the light from the probe, at each position that the probe has a certain delay with respect to the pump due to different optical paths in real space. Additionally, we can clearly see that the rise of the TA signal is sharper (steeper) than the decay, this also indicates that our decay timescale is not limited by the IRF/temporal resolution.

Temperature-dependent TA measurements were performed by mounting the sample in a closed-circuit pressurized helium cryostat (Optistat Dry BL4, Oxford Instruments) placed at the focal point of the probe and reference beams. The cryostat was driven by a compressor (HC-4E2, Sumitomo) and a temperature controller (Mercury iTC, Oxford Instruments). The vacuum level inside the cryostat was below $10^{-5}$ mbar.

**(Cryo-) Scanning Electron Diffraction (SED)**

Scanning electron diffraction (SED) was performed using the ThermoFisher Spectra 300 STEM operated at 200kV accelerating voltage and 300 μrad convergence angle giving a probe size of ~5 nm. A Quantum Detectors Merlin/Medipix3 single chip direct electron detector was used to record diffraction patterns with a dwell time of 1ms equating to a dose of ~11e/Å$^2$ which was maintained throughout the measurement. The ThermoFisher Maps software was used to record annular dark field (ADF) images concurrently to SED data with an inner and outer detector collection semiangle of 62 and 200 mrad respectively. Reciprocal space and rotation calibrations were performed via the use of a MAG*I*CAL® calibration sample. For measurements at cryogenic temperatures the Gatan 613 cryogenic holder was used with liquid $N_2$ used to achieve low temperatures which once cooled stabilized at a ~90K. Post-acquisition analysis of SED data was performed using pyXem 0.16 and py4DSTEM 0.14 (Supplementary Note 3).[40,41]

**Au Marker Synthesis**

Au markers were synthesized[27,42] by dissolving $HAuCl_4$•$3H_2O$ (0.024g) in ethylene glycol (25 mL) before leaving the solution to be stirred at 350 rpm at 70°C for 30 minutes. Aniline (1.12mL, 0.1M) in ethylene glycol was then added and a brown suspension formed instantaneously. Stirring was then stopped but heating continued for 3 hours. After this time the resulting suspension was cooled to room temperature and the supernatant removed. Ethanol (20 mL) was then added, and the resulting suspension bath sonicated for 1 minute before being centrifuged at 7500 rpm for 3 minutes and the supernatant discarded. A further two washing steps were carried out whereby 10 mL of ethanol was added, and the previous centrifuging step repeated. Finally, the resultant fiducial markers were dispersed in 1 mL of cholobenzene and 10 µL was spun coat at 1000 rpm for 30 seconds onto a glass substrate to assess size and coverage. These conditions also proved optimal to obtain adequate coverage of markers on $SiN_x$ TEM grids (Norcada NT025X).

**(Cryo-) XRD**

PheniX cryostat was used for X-ray diffraction (XRD) analysis. This closed cycle helium system, designed for low-temperature powder diffraction, allows for measurement of a flat, static sample between 12 - 310 K with automated temperature control (+/- 1 K) through the XRD control software. It houses a two-stage Gifford McMahon cooler, operating with a sealed helium gas circuit, ensuring no helium consumption. Sample cooling is achieved through heat conduction, with temperature measured at the sample stage. The system minimizes heat leaks with a radiation shield and a sturdy lid with X-ray transparent windows. The PheniX Front Loader variant allows for cold-loading of samples, offering a quick turnaround. The XRD was obtained by a Bruker D8 Advance powder X-ray diffractometer. The XRD facility is equipped with Bruker EVA software, which is used for phase identification and qualitative analysis. Additionally, it utilizes Bruker Topas software for conducting quantitative analysis.

**(Cryo-) Hyperspectral Microscopy**

Wide-field, hyperspectral microscopy measurements were performed on a Photon Etc. IMA system. Measurements were carried out with an Olympus LMPlanFL N 100× (NA = 0.8) or a Nikon TU Plan Fluor 20× (NA = 0.45) objective lens. The sample was kept under vacuum for the temperature-dependent experiments. Excitation occurred with a 405 nm continuous wave laser (unpolarised), which was filtered out by a dichroic longpass for the detection. The emitted light was directed towards a volume Bragg grating, which dispersed the light spectrally onto a CCD camera. The detector was a 1040 × 1392 resolution silicon CCD camera kept at 0 °C with a thermoelectric cooler and has an operational wavelength range of 400–1000 nm. By scanning the angle of the grating relative to the incident light, the spectrum of light coming from each point on the sample could be obtained.

For the temperature-dependent experiment, the sample was fixed with silver paste to the cold finger of an Oxford HiRes cryostat cooled with liquid helium. The cryostat was attached to the microscope with a self-made holder allowing for focus correction. The sample was held at the set temperature for at least 15 minutes prior to every measurement.

**(Cryo-) UV-VIS**

Variable temperature UV-vis measurements were collected on a Varian Cary 6000i dual beam spectrometer, using an Oxford Instruments Optistat CF-V optical cryostat equipped with quartz windows. Measurements were collected in transmission geometry under high vacuum (~$10^{-6}$ mbar).

**(Cryo-) WAX and SAXS**

SAXS-WAXS experiments were performed by a laboratory based SAXS-WAXS beamline KWS-X (XENOCS XUESS 3.0 XL) at JCNS MLZ. The MetalJet X-ray source (Excillum $D^{2+}$) with a liquid metal anode was operated at 70 kV and 3.57 mA with Ga-Kα radiation (wavelength $\lambda$ = 1.314 Å). Thin film samples were measured in a Linkam temperature-controlled stage (HFS350) with a liquid nitrogen pump which allows for an ultra-low temperature up to -150 $^{o}$C. The sample to detector distances from 0.1 m to 1.70 m which cover the scattering vector q range from 0.003 to 4.5 $Å^{-1}$ (Q is the scattering vector, $Q=(4\pi/\lambda)\sin(\theta)$, $2\theta$ is the scattering angle). The SAXS patterns were normalized to an absolute scale and azimuthally averaged to obtain the intensity profiles, and the glass background was subtracted.

**Solution deposition of $FAPbI_3$ thin films**

A mixture of lead iodide ($PbI_2$, 0.346 g, 1.5 mmol), formamidinium iodide (FAI, 0.155 g, 1.8 mmol), and EDTA (5 mol% relative to $PbI_2$) was dissolved in 0.5 mL of dimethyl sulfoxide (DMSO). This mixture was continuously stirred and heated at 75$^{o}$C until a clear dark yellow solution was obtained. This solution was then applied to a UV-Ozone-cleaned quartz substrate using a spin coating process (4000 rpm for 40 seconds). To ensure the uniformity of the film, nitrogen gas was blown onto the surface of the spinning films for 20 seconds, starting 5 seconds after the beginning of the spinning process. The nitrogen gun was initially positioned ~7 cm away from the film for the first 10 seconds, and then the distance was reduced to ~5 cm for the remaining 10 seconds. The spin coated films were then annealed at 150$^{o}$C for 1 hour in a nitrogen-filled glovebox.

**Vapor deposition of $FAPbI_3$ thin films**

*Materials*

N, N-Dimethylformamide (DMF, anhydrous, 99.8%), dimethyl sulfoxide (DMSO, anhydrous, 99.9%), ethanol (anhydrous, 99.9%), and ethylenediaminetetraacetic acid (EDTA, anhydrous, 99.99%) were purchased from Sigma-Aldrich and used without further purification. Formamidinium iodide (FAI, 99.9%) was purchased from Greatcell Solar Materials and used without further purification. Lead iodide ($PbI_2$, 98%), and MeO-2PACz (C16H18NO5P, >98.0%) was purchased from TCI and used without further purification.

**Perovskite Film Co-evaporation**

The quartz substrates were cleaned with detergent (Decon 90), deionized water, acetone, and isopropanol in an ultra-sonication bath for 15 minutes at each. The clean substrates were treated by UV-ozone for 15 minutes. TEM grids were placed on the quartz substrates via carbon tape (TEM grids were *Single Window Silicon Nitride TEM Grids*, NT025X from Norcada). The deposition was done in a CreaPhys PEROvap evaporator inside an MBraun $N_2$ glovebox ($O_2$ and $H_2O$ levels <0.5 ppm) to avoid exposure of precursors and deposited films to oxygen and water during sample fabrication and handling. The evaporator chamber was pumped down to a pressure below $2 \times 10^{-6}$ mbar for all depositions. The evaporator system was specifically designed with a cooling system that maintains the evaporator walls, source shutters and shields at -20 °C throughout the entire process. This functionality minimizes re-evaporation of the precursors and cross-contamination between sources, ensuring fine control over the evaporation rates and high reproducibility.
For the $FAPbI_3$ deposition, $PbI_2$ (TCI >98 % metal trace based purity) and FAI (GreatCell Solar) were filled into two separate crucibles. For both the $PbI_2$ and FAI, we used fresh powders for every deposition. The tooling factor of each chemical was calibrated by checking the film thicknesses by profilometry inside a $N_2$-filled glovebox (Bruker DEKTAK XT profilometer). Two quartz crystal microbalances (QCMs) mounted on the top of vapor sources allowed us to monitor the deposition rate of each source to control the composition.

The rate of $PbI_2$ and FAI was deposited between 0.34-0.60 Å/s and 0.68-1.50 Å/s, respectively. The substrate temperature was maintained at around 18 °C. The distance between evaporator sources and substrate holder was approximately 0.35 m. We observed minimal change in the substrate temperature (<1 °C), and the chamber was typically at a pressure of less than 2.0 $\times 10^{-6}$ mbar during the entire duration of the depositions. Once the films were removed from the evaporator, they were immediately annealed on a hotplate within the same $N_2$-filled glovebox at 150 °C for 20 mins.

40. Francis, C. & Voyles, P. M. pyxem: A Scalable Mature Python Package for Analyzing 4-D STEM Data. *Microsc. Microanal.* **29**, 685–686 (2023).
41. Cautaerts, N. *et al.* Free, flexible and fast: Orientation mapping using the multi-core and GPU-accelerated template matching capabilities in the Python-based open source 4D-STEM analysis toolbox Pyxem. *Ultramicroscopy* **237**, 113517 (2022).
42. Jones, T. W. *et al.* Lattice strain causes non-radiative losses in halide perovskites. *Energy Environ. Sci.* **12**, 596–606 (2019).

**Funding:**
European Union's Horizon 2020 Research and Innovation Program, European Research Council, HYPERION, 756962 and PEROVSCI, 957513 (SDS, LD)
EPSRC EP/R023980/1, EP/V012932/1 and EP/T02030X/1 (SDS)
Sir Henry Royce Institute grant EP/R00661X/1 and EP/P024947/1 (SDS)
CAM-IES grant EP/P007767/1 (SDS)
EPSRC Cambridge NanoDTC, EP/L015978/1 (TAS)
Early Career Fellowship supported by the Leverhulme Trust (ECF-2022-593) and the Isaac Newton Trust (22.08(i)) (SK)
European Research Council under the European Union's Horizon 2020 research and innovation programme Grant Agreement No. SCORS – 101020167 (SG)
MSCA Individual Fellowship from the European Union's Horizon 2020 PeTSoC, No. 891205 (TCJY)
Marie Sklodowska-Curie Postdoctoral Fellowships via UKRI Horizon Europe Guarantee EP/X025764/1 (JM)
Royal Society University Research Fellowship URF/R1/231287 (IEJ)
Oppenheimer Research Fellowship and a Schmidt Science Fellowship (TASD)
EPSRC for funding under grant numbers EP/V007785/1, and EP/R008779/1 (PAM)
Royal Society and Tata Group (grant no. UF150033) (SDS)

**Author contributions:**

Conceptualization: D.G., S.D.S.

Big data processing and nano-structure analysis: T.A.S.

Cryo-TA and quantum transient interpretation: D.G., S.G., L.D.

Cryo-UVVIS: T.M.; I.E.J., D.G.

Correlative optical and electron microscopy: T.A.S., S.K., S.M.F, D.G., M.D., C.M.

Nanotwinning discovery: T.A.S., T.A.S.D.,

Samples: T.C.J.Y., S.N., J.M., T.K.,

Cryo-WAX and SAX: B.W. R.G.

XRD: C.L., Y.L.

Visualization: D.G., T.A.S., S.D.S., S.K., K.Y.J.

Funding acquisition: S.D.S.

Project administration: S.D.S.

Supervision: S.D.S., P.A.M., T.A.S.D., P.M.B., H.S.

Writing–original draft: D.G., T.A.S., S.D.S.

**Competing interests:**

S.D.S. is a co-founder of Swift Solar Inc.

**Data and materials availability:**

The data that support the findings of this study is available to download at the University of Cambridge's Apollo Repository [DOI to be added at acceptance].

**Code availability:**

SLIC clustering code can be found at github.com/TomSelby/hyperSLIC

Stitching code can be found at github.com/TomSelby/hyper_stitch

# Supplementary information for

## Visualizing nanodomain superlattices in halide perovskites giving picosecond quantum transients

Dengyang Guo[1,2]*, Thomas A. Selby[1]*, Simon Kahmann[1†], Sebastian Gorgon[2], Linjie Dai[1,2], Milos Dubajic[1], Terry Chien-Jen Yang[1], Simon M. Fairclough[3], Thomas Marsh[2], Ian E. Jacobs[2], Baohu Wu[4], Renjun Guo[5], Satyawan Nagane[1], Tiarnan A. S. Doherty[1,3], Kangyu Ji[2], Cheng Liu[2], Yang Lu [1,2], Taeheon Kang[1], Capucine Mamak[1], Jian Mao[1], Peter Müller-Buschbaum[4,5], Henning Sirringhaus[2], Paul A. Midgley[3], Samuel D. Stranks[1,2]✉

Corresponding author: sds65@cam.ac.uk

**The PDF file includes:**

Supplementary Text

**Other Supplementary Materials for this manuscript include the following:**

Supplementary Movie 1: Wavelength dependent emergence of bright spots in the hyperspectral photoluminescence map

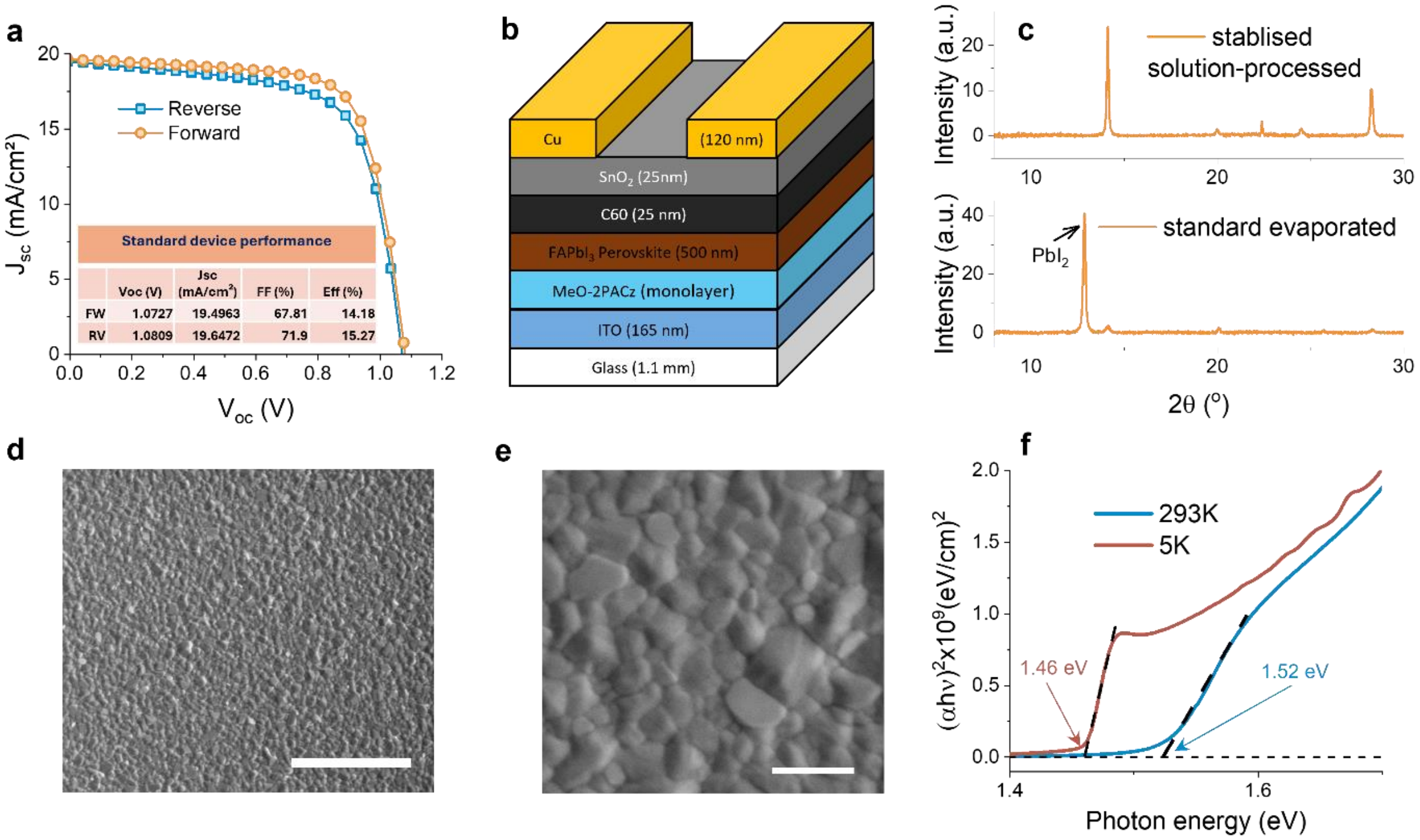

**Supplementary Fig. 1 | Evaporated $FAPbI_3$ films exhibit good quality. (a)** A representative J-V curve of the evaporated $FAPbI_3$ as a solar cell candidate. The inset provides details of the device performance, including forward and reverse scans. **(b)** Solar cell device structure. Commonly available transport and contact layers are utilized for the standard demonstration. **(c)** XRD patterns of both standard evaporated and stabilized solution-processed $FAPbI_3$ films. The signal of $PbI_2$ is from the slight extra proportion (20%) applied to stabilize the evaporated $FAPbI_3$ films.[1] In Fig. 1 and Supplementary Fig. 2, we find the quantum feature of $FAPbI_3$ films is not affected by such a slight $PbI_2$ addition. **(d)** and **(e)** display scanning electron microscopy micrographs of the evaporated $FAPbI_3$ films with scale bars of 4 µm and 500 nm, respectively. **(f)** The Tauc plot absorbance of $FAPbI_3$ films recorded at 293 K and 5 K, which show the position of the bandgap at 1.52 eV (817 nm) and 1.46 eV (849 nm), respectively. The continuous film of perovskite grains on the order of hundreds of nanometers in size can be also directly observed from the SED measurement, which can be seen in Fig. 3.

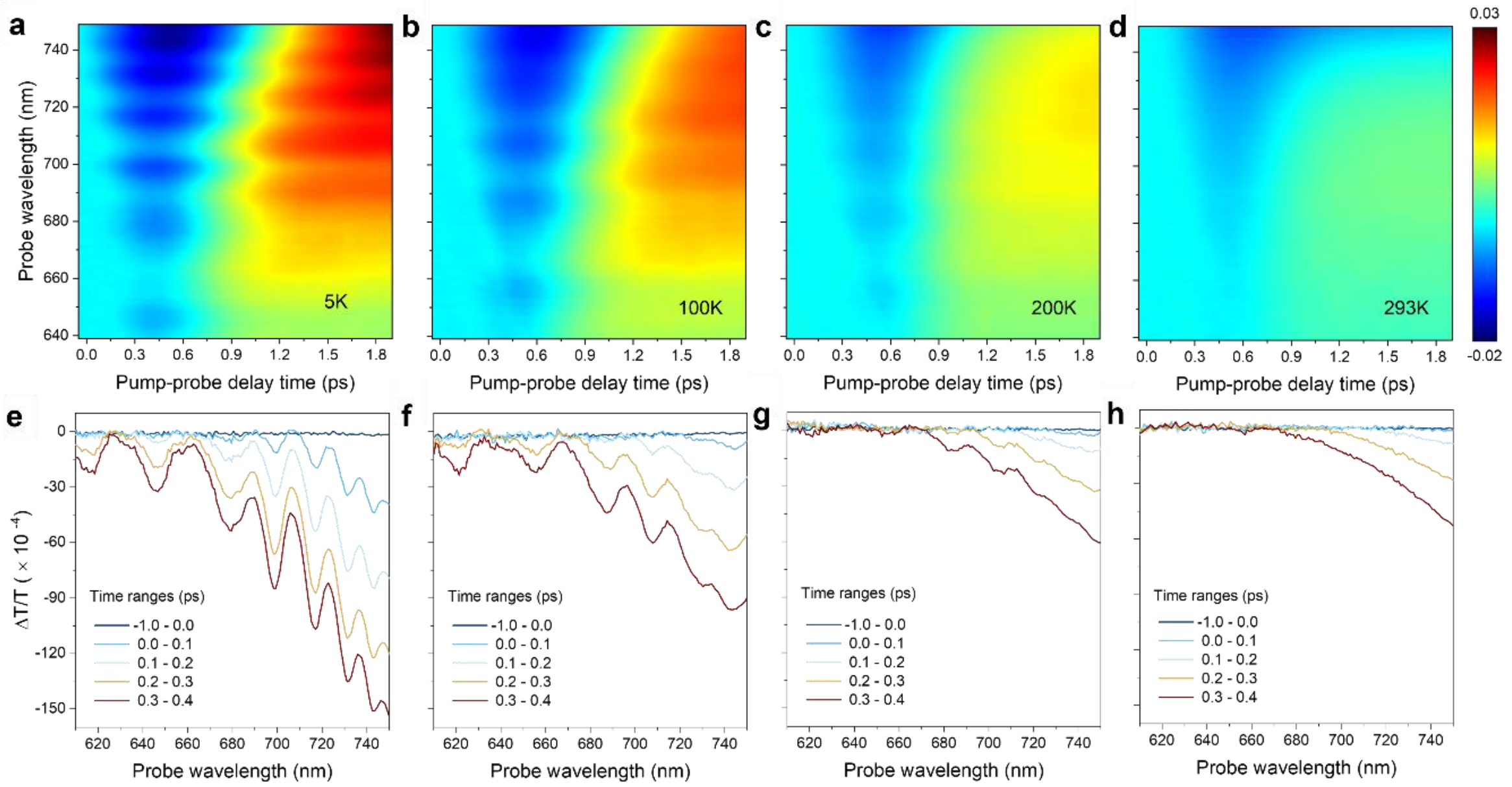

**Supplementary Fig.2 | Temperature-dependent TA maps of solution-processed $FAPbI_3$. (a-d)** Temperature-dependent TA maps of the stabilized solution-processed $FAPbI_3$ show that discrete absorption peaks fade out as the temperature increases, suggesting the quantum confinement is intrinsic and present regardless of the material deposition method. A different colour set is applied to distinguish the films from the two deposition methods. **(e-h)** Absorption spectra at initial time points, vertically extracted from the maps.

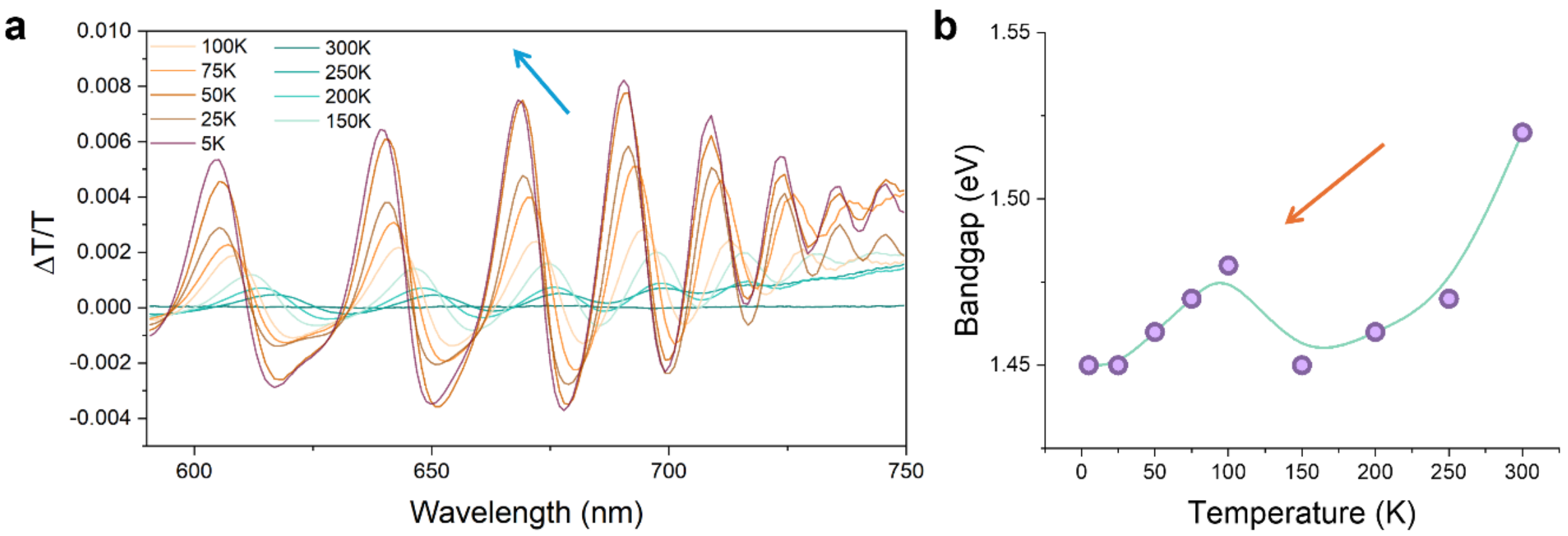

**Supplementary Fig.3 | Temperature dependence of the quantum peaks and bandgap. (a)** The blue shift in the quantum energy levels with temperature, which is independent of the shift in the bandgap (a two-step red shift breaks at phase transition temperature around 140 K) **(b)**. The bandgap values are obtained from the Tauc plot absorbance of $FAPbI_3$ films at each temperature, as exampled in Supplementary Fig. 1f.

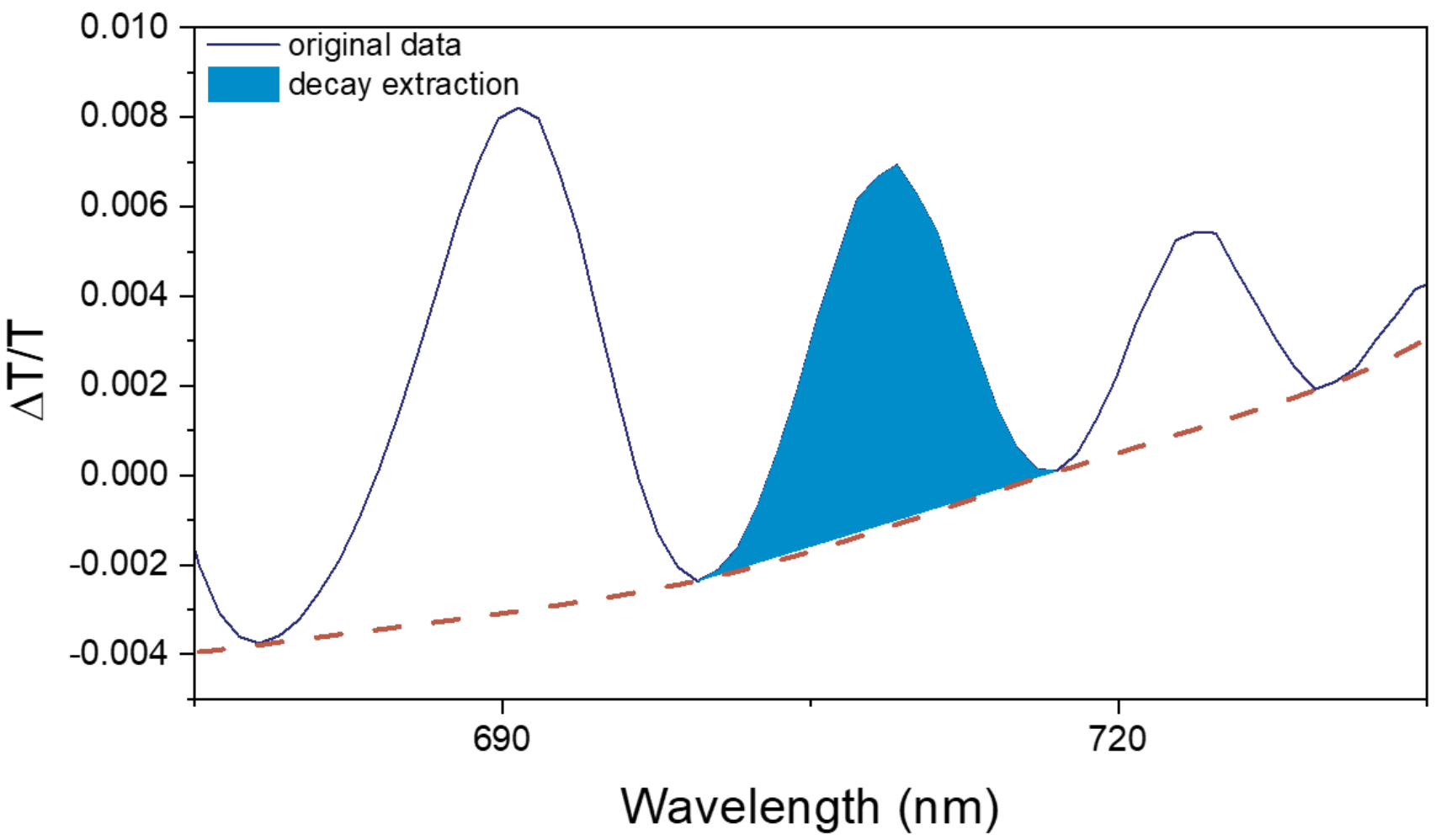

**Supplementary Fig.4 | Extraction method for the transient peaks from TA data.** By utilizing the smooth curve as the baseline, we extract the transient dynamics of the excited states at each quantum energy level, i.e. connecting the coloured area by time. The background arises from the difference in dielectric constant between the positions of the sample where the identical probe and reference beams are located, with only the former being overlapped by the pump.[2]

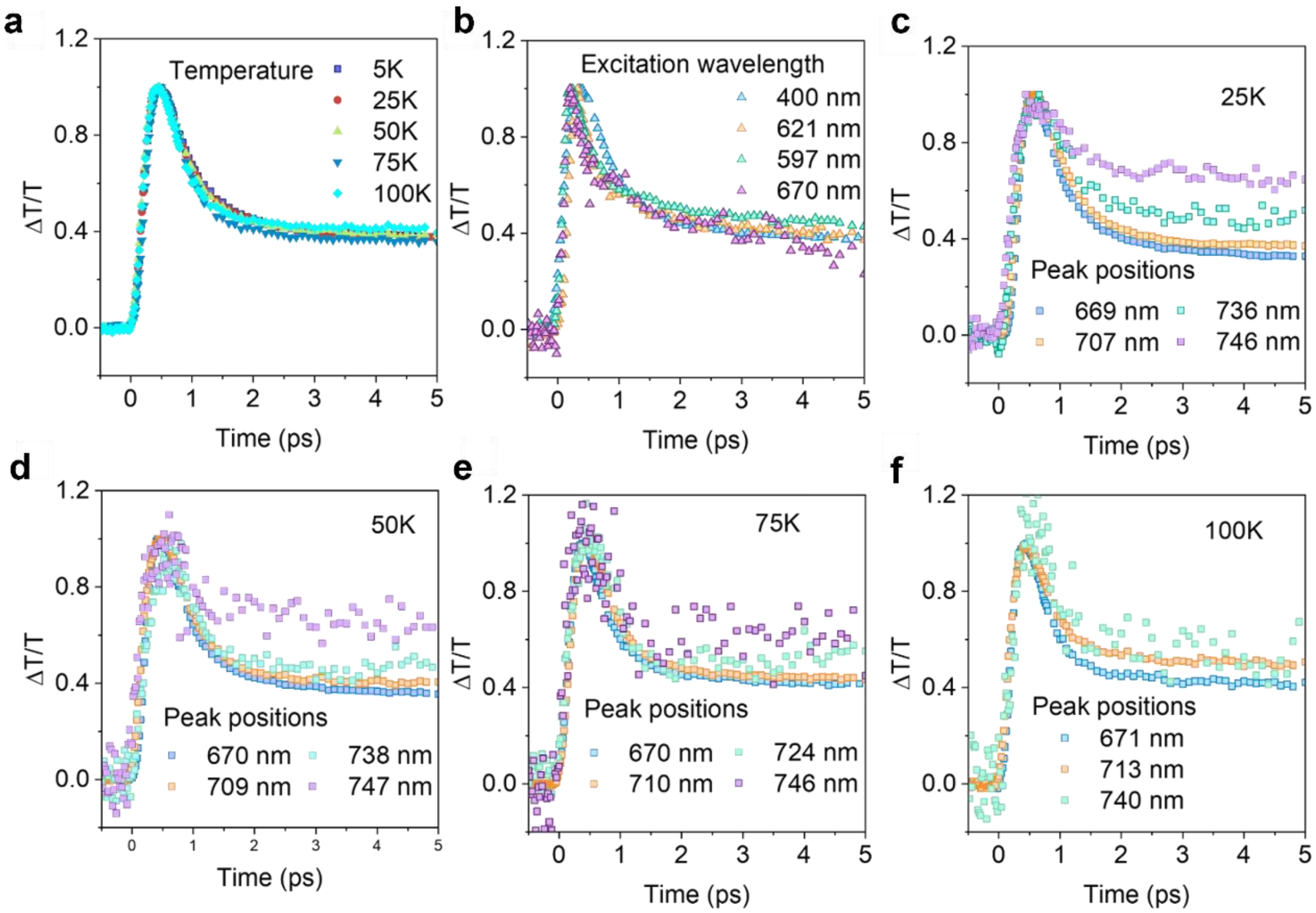

**Supplementary Fig.5 | Temperature-dependent TA transients of evaporated $FAPbI_3$ thin film, complementary to Fig.1. (a)** transients at 691 nm at different temperatures, complementary to Fig.1e with transients at 723 nm. **(b)** transients at 706 nm under different excitation wavelengths, complementary to Fig.1d with transients at 723 nm. **(c-f)** decay characteristics within each level under identical excitation wavelengths from 25 K to 100 K.

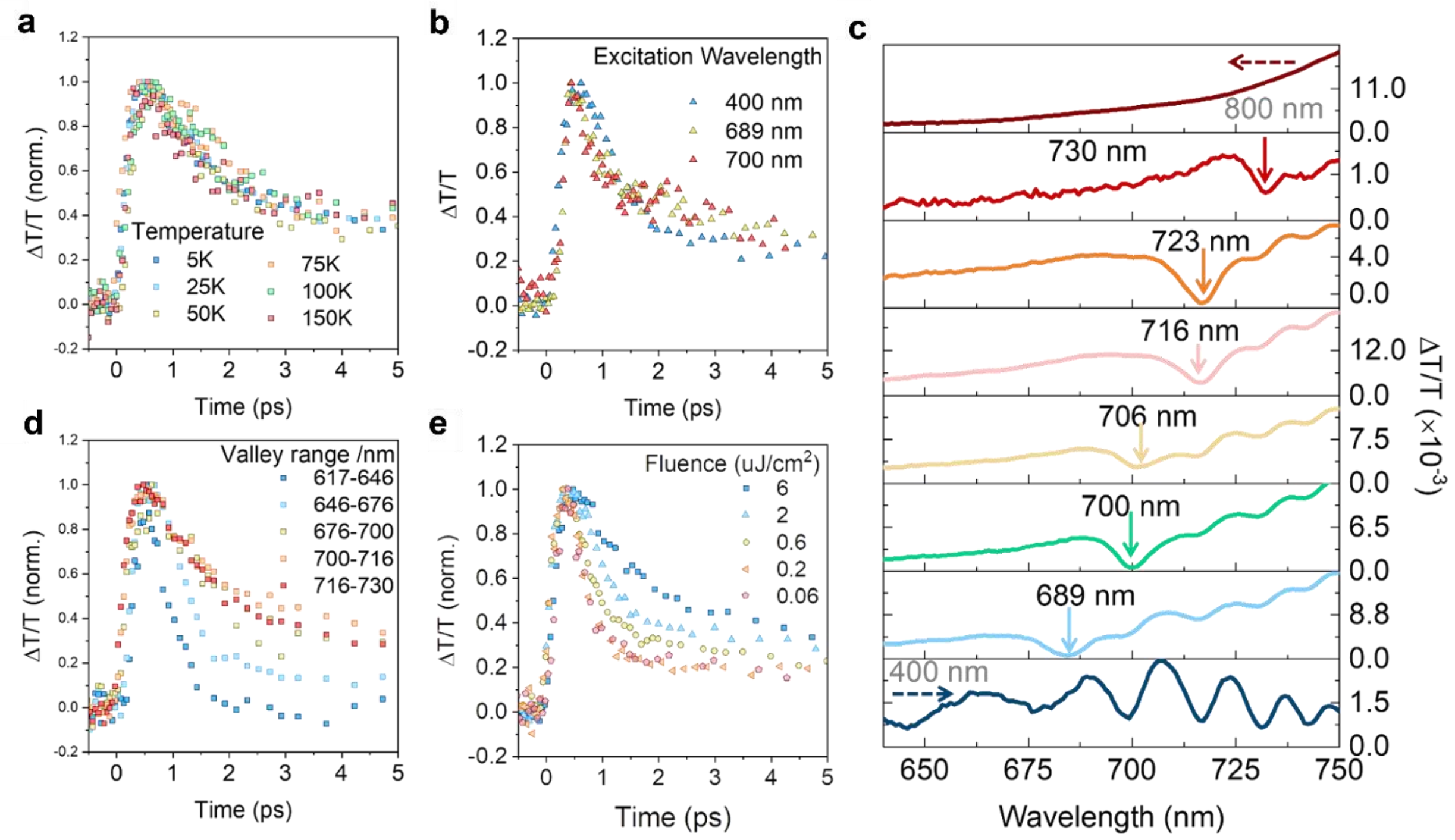

**Supplementary Fig.6 | TA Recordings of the stabilized solution-processed $FAPbI_3$ thin film.** Temperature-dependent **(a)** and excitation wavelength dependent **(b)** transients at peak position 723 nm. **(c)** Excitation wavelength dependent absorption spectra within a consistent time window. **(d)** Transients at different quantum peaks, excited by a fixed 400 nm laser beam with fluence 5.7 uJ/$cm^2$. **(e)** Power dependent transients at peak 723 nm, excited by laser at 400 nm.

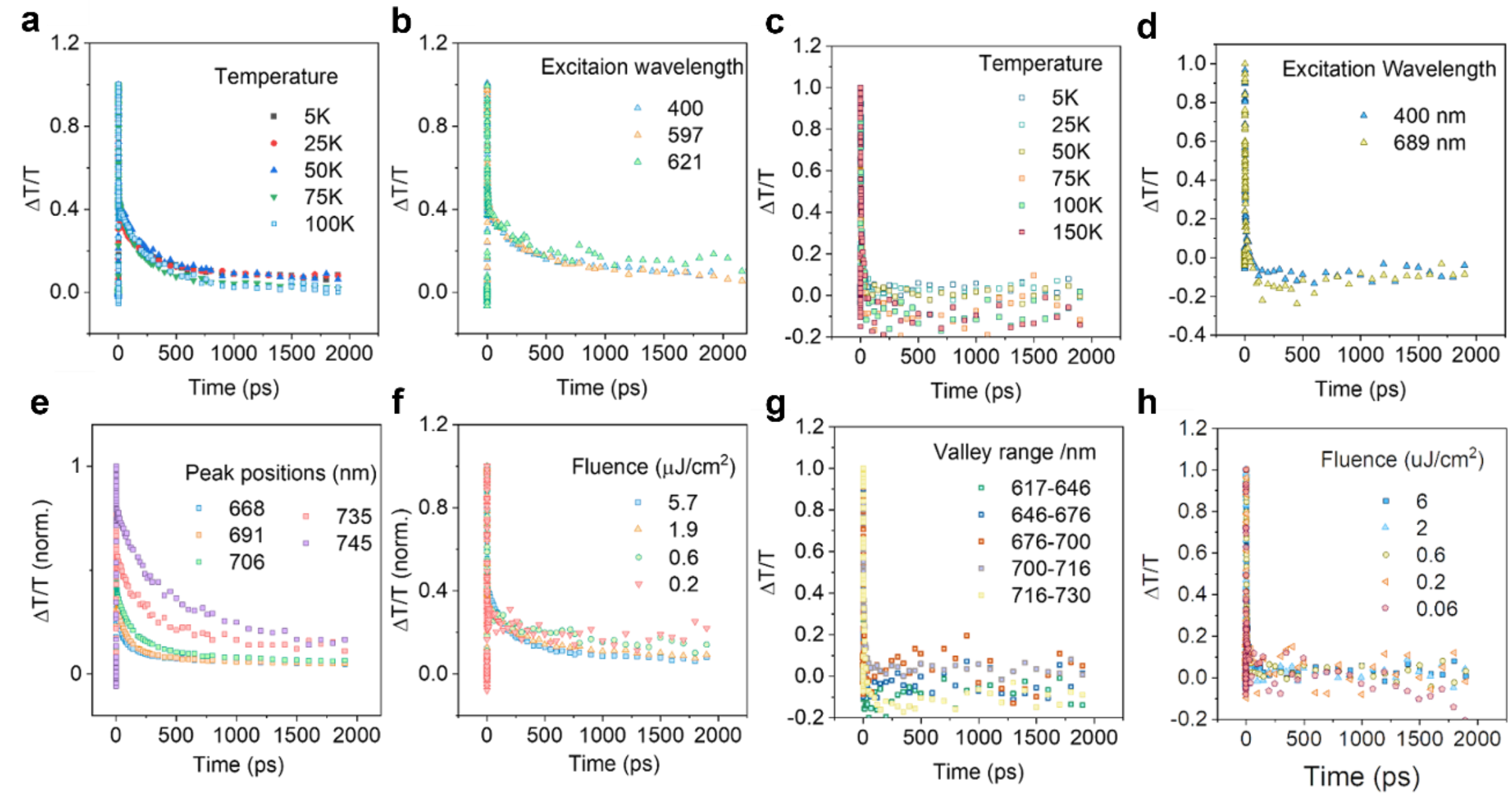

**Supplementary Fig.7 | Full TA decays at the timescale of 2ns** for the evaporated samples **(a-d)** and solution-processed samples **(e-h)**. Both samples show an ultrafast majority (1/e) decay, as clearly resolved in Fig. 1g with the timescales of 5 ps and 50 ps. Here, with the longer timescale of 2 ps, the evaporated samples exhibit a long-lasting signal pointing to 1 ns and marginally extending to 2 ns. In contrast, the solution-processed samples show no such tail at all, suggesting that the long signals are not intrinsic property of the material, and can be extrinsically controlled.

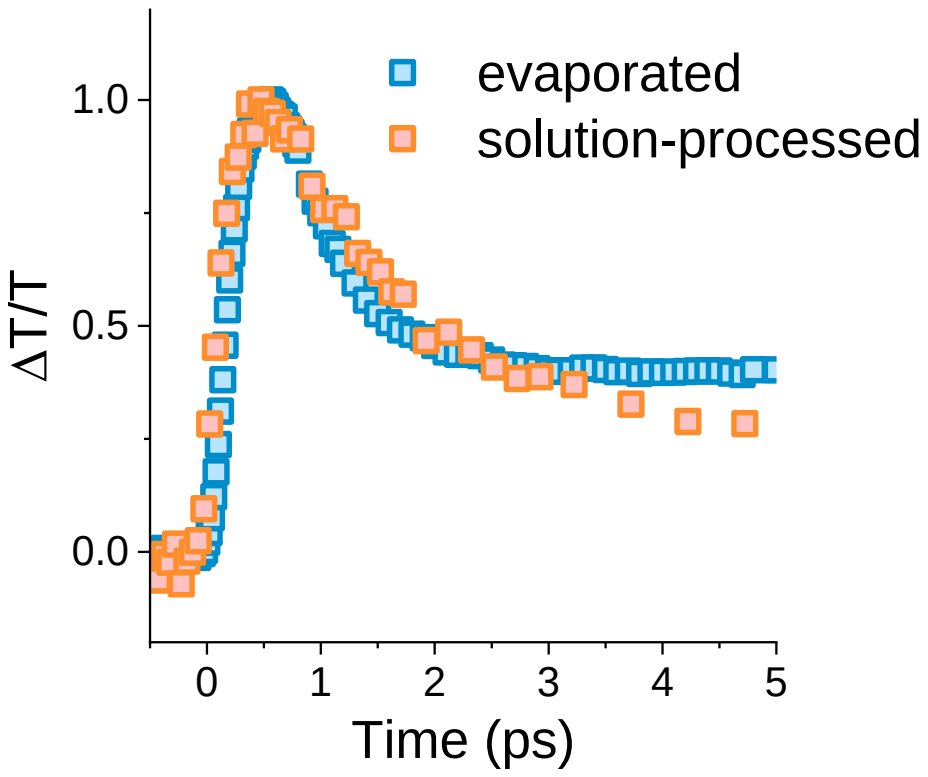

**Supplementary Fig.8 | The TA transient comparison** between the solution-processed and evaporated $FAPbI_3$ samples over the timescale of 5 ps.

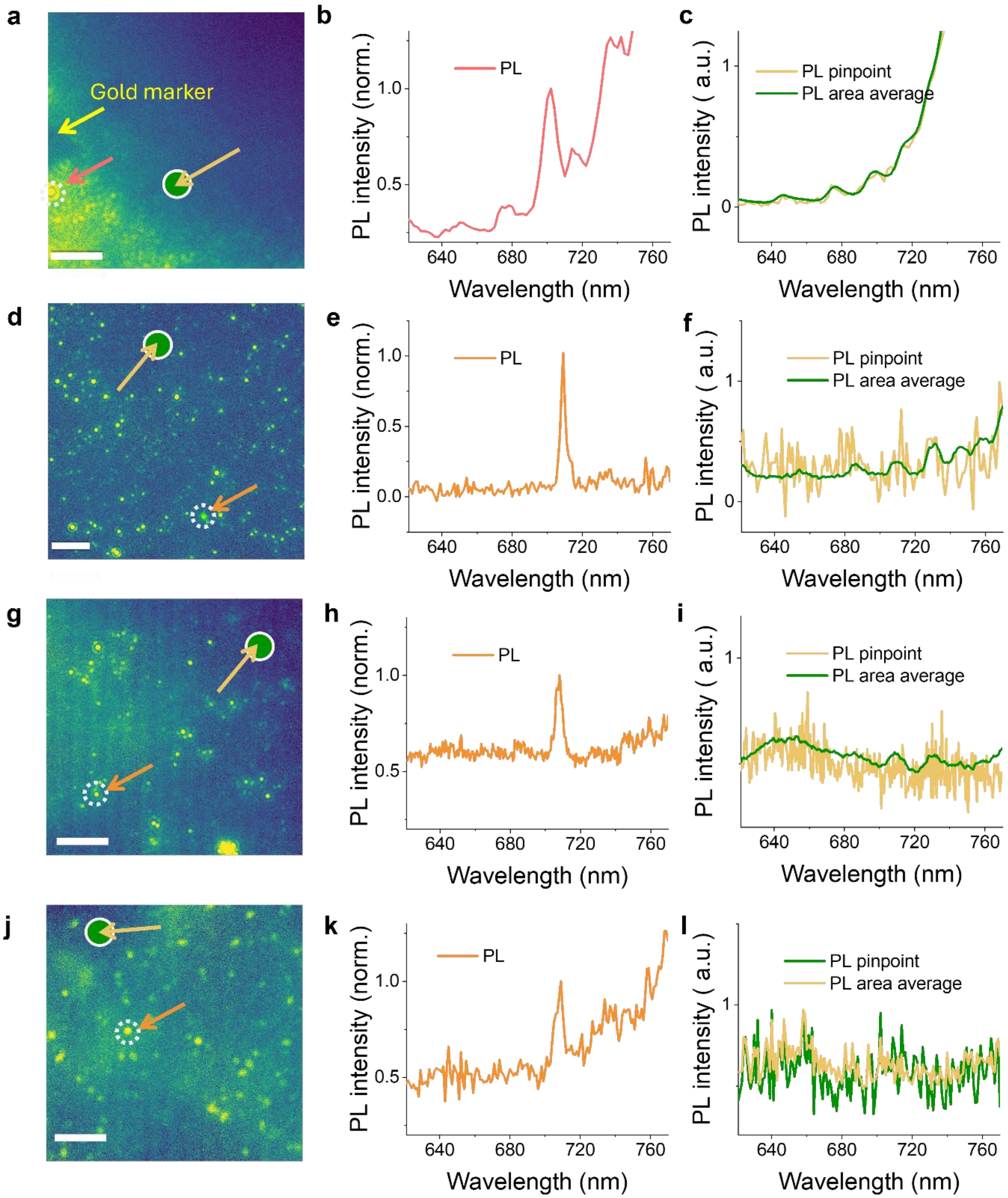

**Supplementary Fig.9 | PL maps and local emission spectra of evaporated and solution-processed $FAPbI_3$ samples under different post-treatment conditions**. PL maps show spatial dependence of PL intensity at 706nm and are derived from hyperspectral cubes. The full PL spectra from local bright spots on the maps are shown. For reference, the background spectra are obtained by both pinpointing a single pixel and averaging over a circular area. **(a-c)**: Evaporated $FAPbI_3$ thin film. The pointed dark triangle is the gold marker, used as a locator for spatial correlation (**a**). The marked bright emitter shows an outstanding high intensity PL peak at 706 nm along with other

above bandgap peaks (**b**). The background PL peaks are also observed at dim areas in PL maps where no clear bright spots are observed (**c**). (**d-f**): solution-processed film without any additional treatment but aged in glovebox for one year. The background multi-peaks are in line with the PL recorded from evaporated samples and in accordance with the conclusion of ubiquitous spread of nanotwinning domains from the structural analysis in section 3 of the main text. (**g-i**) fresh solution-processed film without any additional treatment. In this case, clear but less background peaks are observed. (**j-l**): solution-processed film with an additional annealing treatment (150$^{o}$C in air for 30 minutes) and aged in glovebox for one year. This sample, after post-treatment (**l**), does not show clear background multi-peaks, meaning much less overlapping or pile-up of different single emission peaks. The suppressed appearance of background peaks suggests that the overall spread distribution of single peaks can be modulated by post-treatment. Such suppression lasts after aging implies that the annealing post-treatment does not only improve PL of the whole film as we previously reported[14] but also has a long-term structural impact. Despite the suppression of background peaks, the single emission peak could still be found (**k**), suggesting an intrinsic formation of the local nanoscale structural source within the $FAPbI_3$ samples.

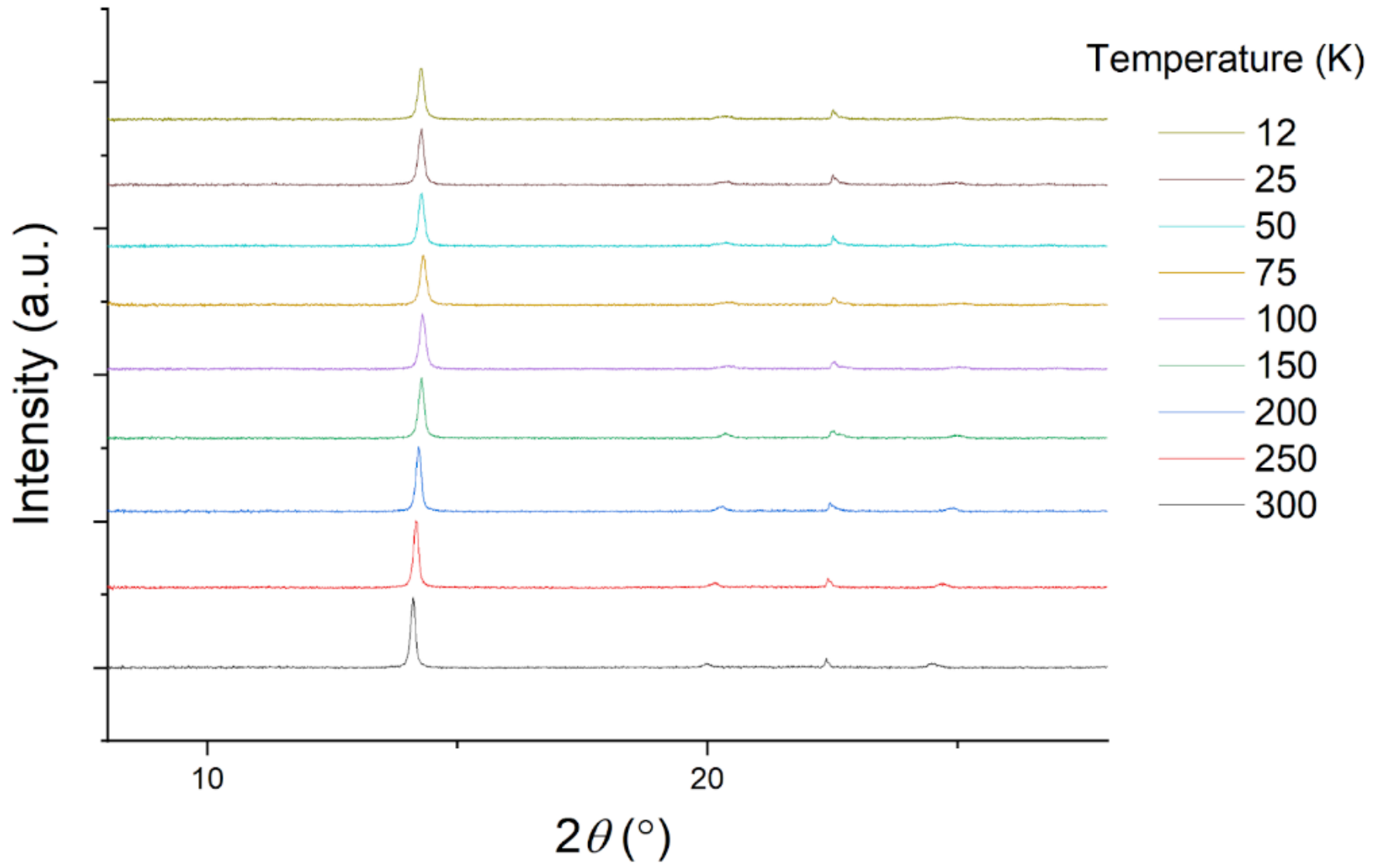

**Supplementary Fig.10 | Temperature dependent X-ray diffraction** of the solution-processed $FAPbI_3$ thin film. The $PbI_2$ which peaks at 12.8 is not observed through all temperatures.

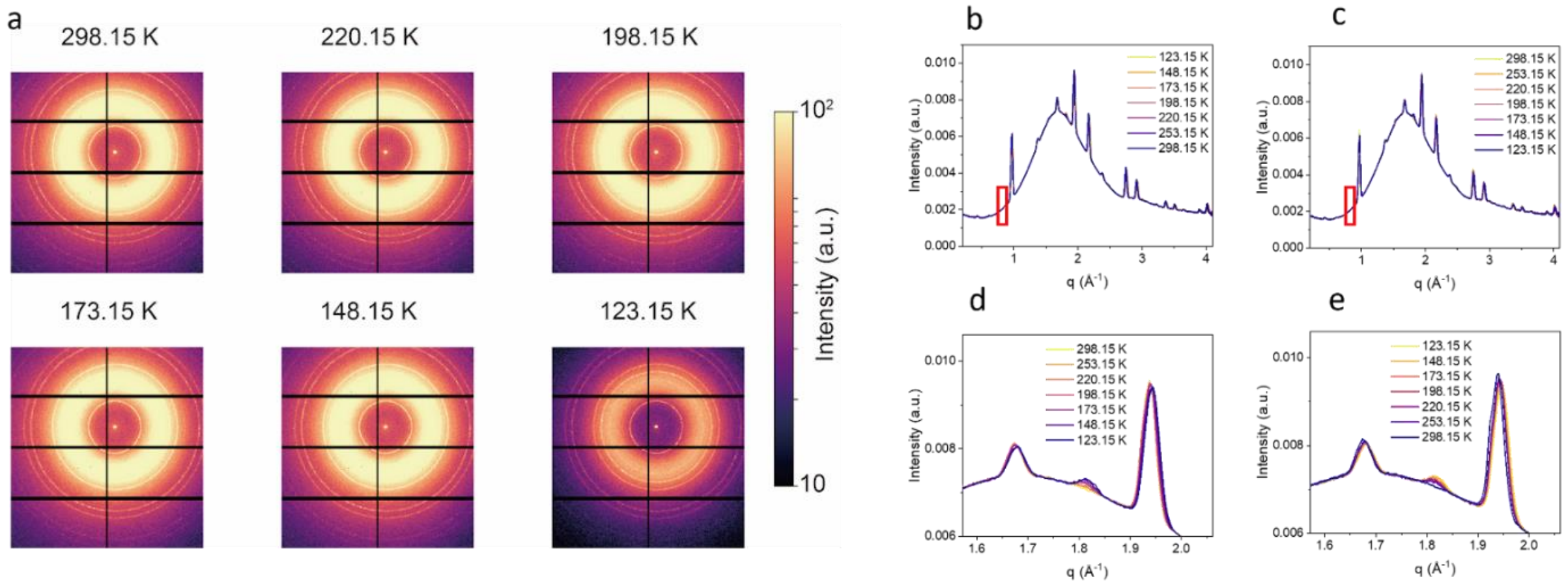

**Supplementary Fig.11 | Temperature-dependent wide-angle X-ray scattering** (cryo-WAXS) of a solution-processed $FAPbI_3$ thin films. The $PbI_2$ peak is located at 0.89 $Å^{-1}$, and there is no such peak as shown in **b** and **c**. This is further confirmation of non-existent $PbI_2$ in our fabricated materials at any measured temperature. Therefore, the excess $PbI_2$ in the evaporated $FAPbI_3$ is not related to the quantum signals.

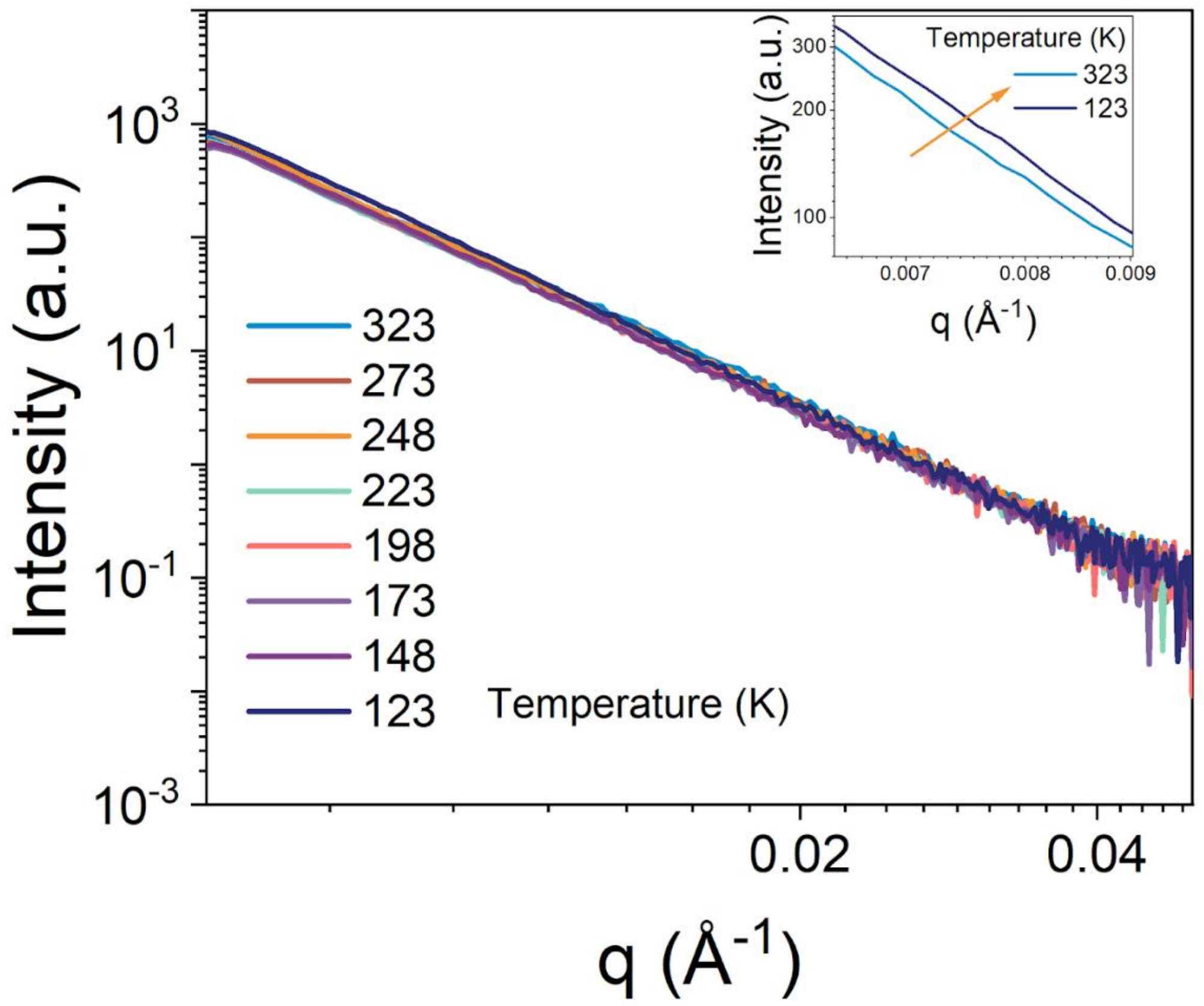

**Supplementary Fig.12 | Temperature dependent small-angle X-ray scattering** (SAXS) results reveal structure information on the evolution of the structure from 1 to 50 nanometers. The results show that during the cooling process of the solution-processed $FAPbI_3$ sample, the entire sample exhibits a dense bulk-like structure (low q scattering following the power law $I = q^{-4}$). Slight roughness interface appears from 323K due to the shrinkage of the film (Supplementary Fig.11 inset).

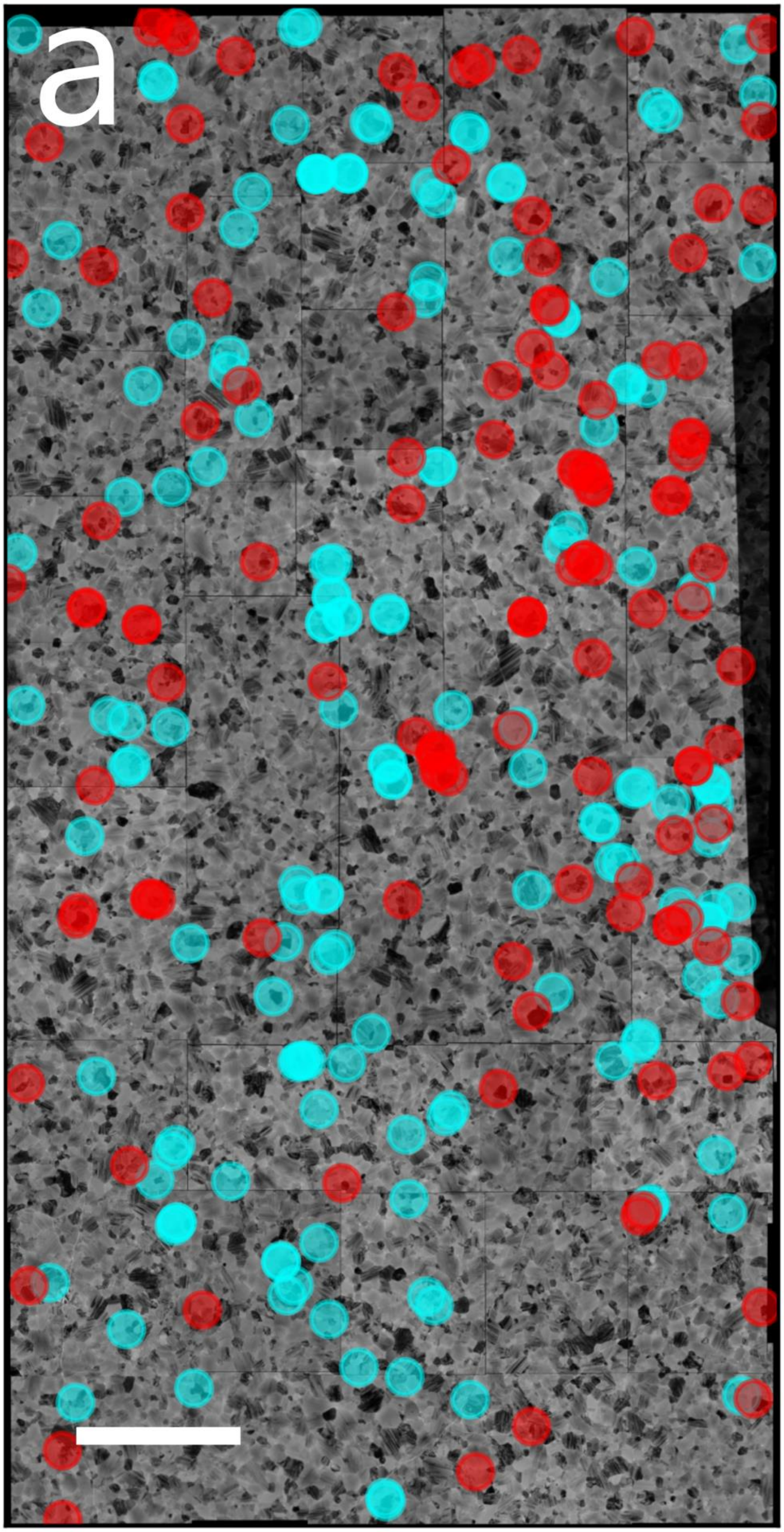
a

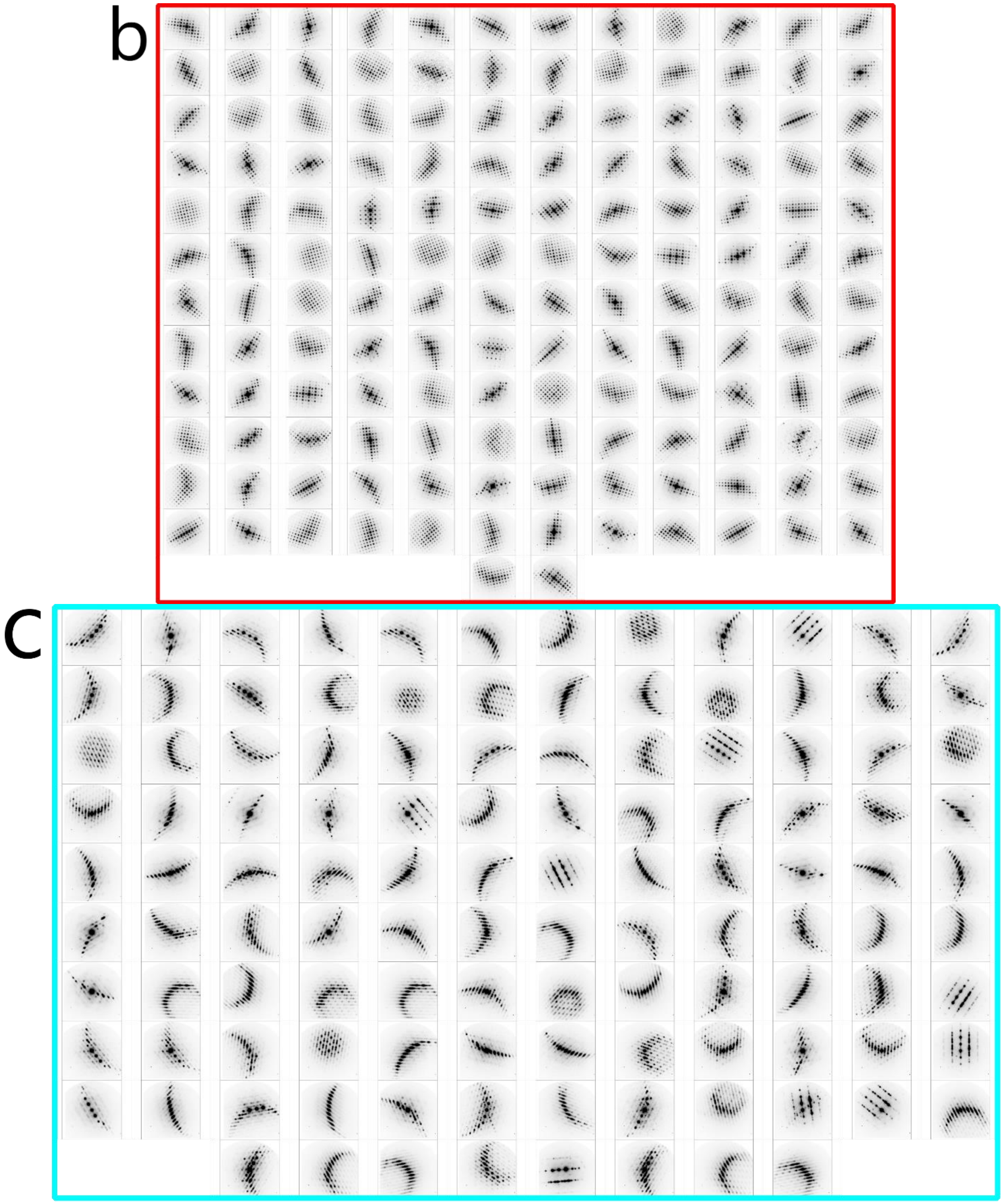

**Supplementary Fig.13 | Stitched SED data recorded at ambient temperature. (a)** The stitched SED data of thermally evaporated $FAPbI_3$ on a $SiN_x$ membrane at ambient temperature. **(b)** Patterns which are oriented close to a <001>$_c$ zone axis are marked in red. **(c)** Patterns which show characteristic $\{111\}_c$ twinning are marked in cyan. Scale bar: 3μm.

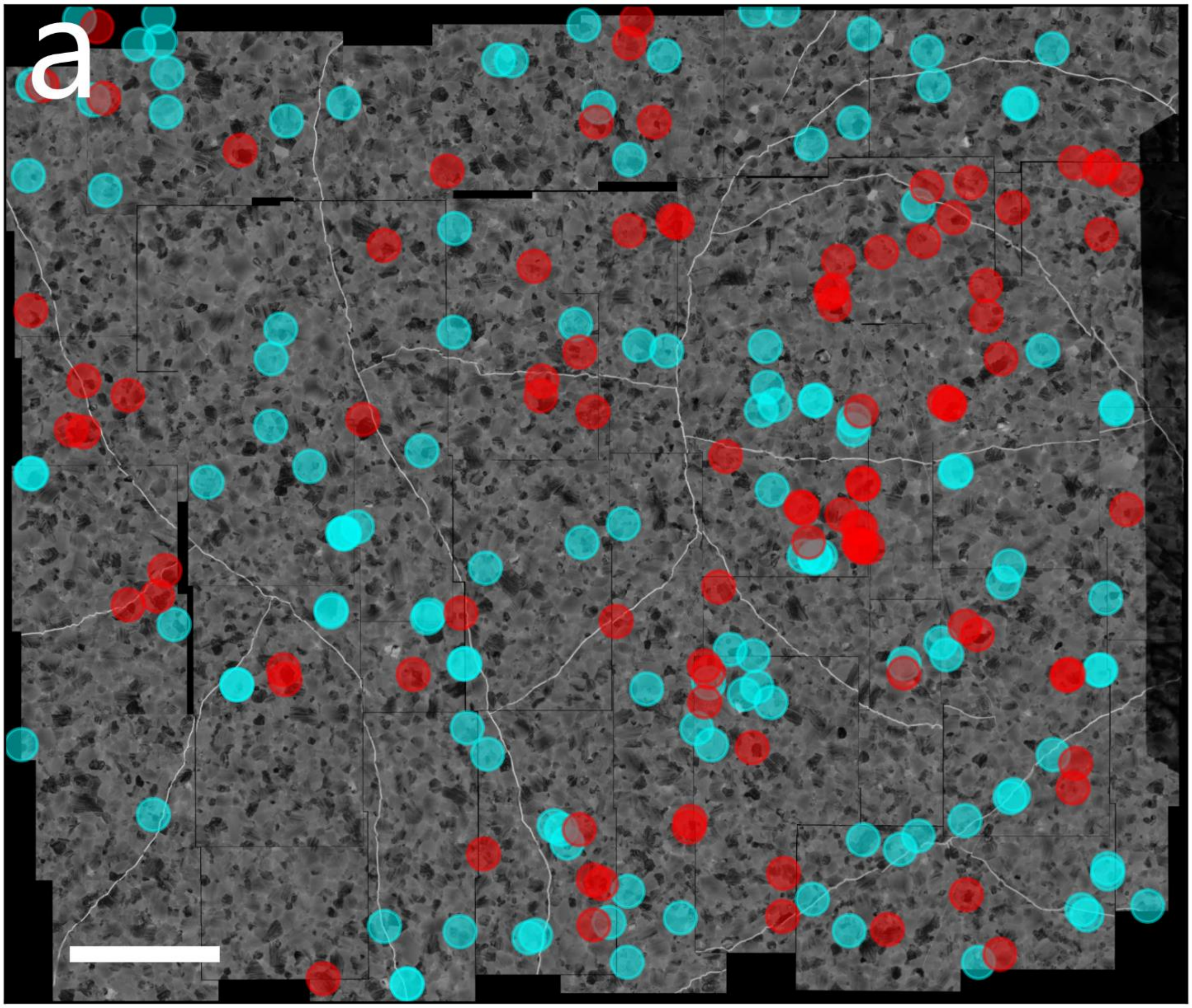
a

b

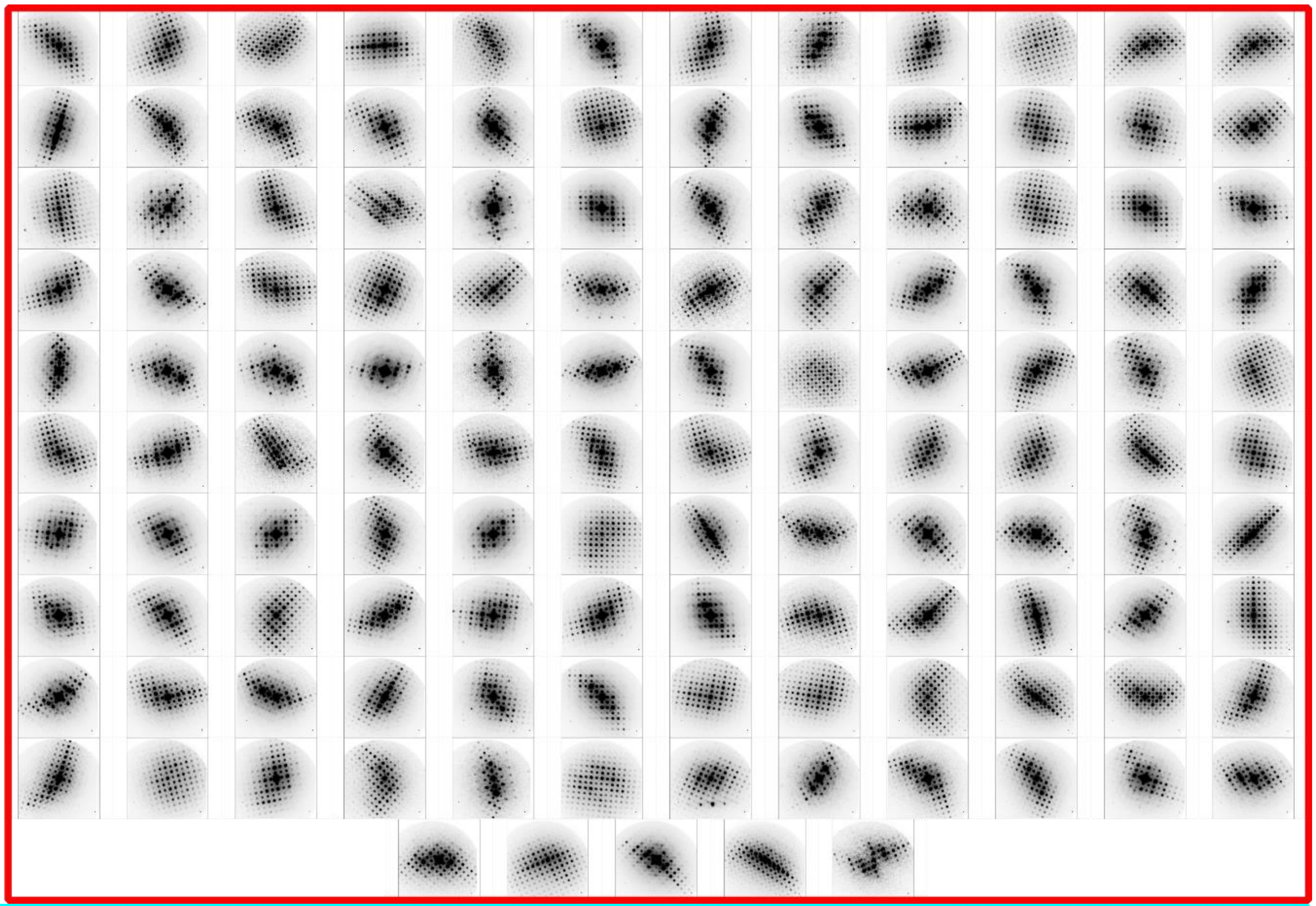

c

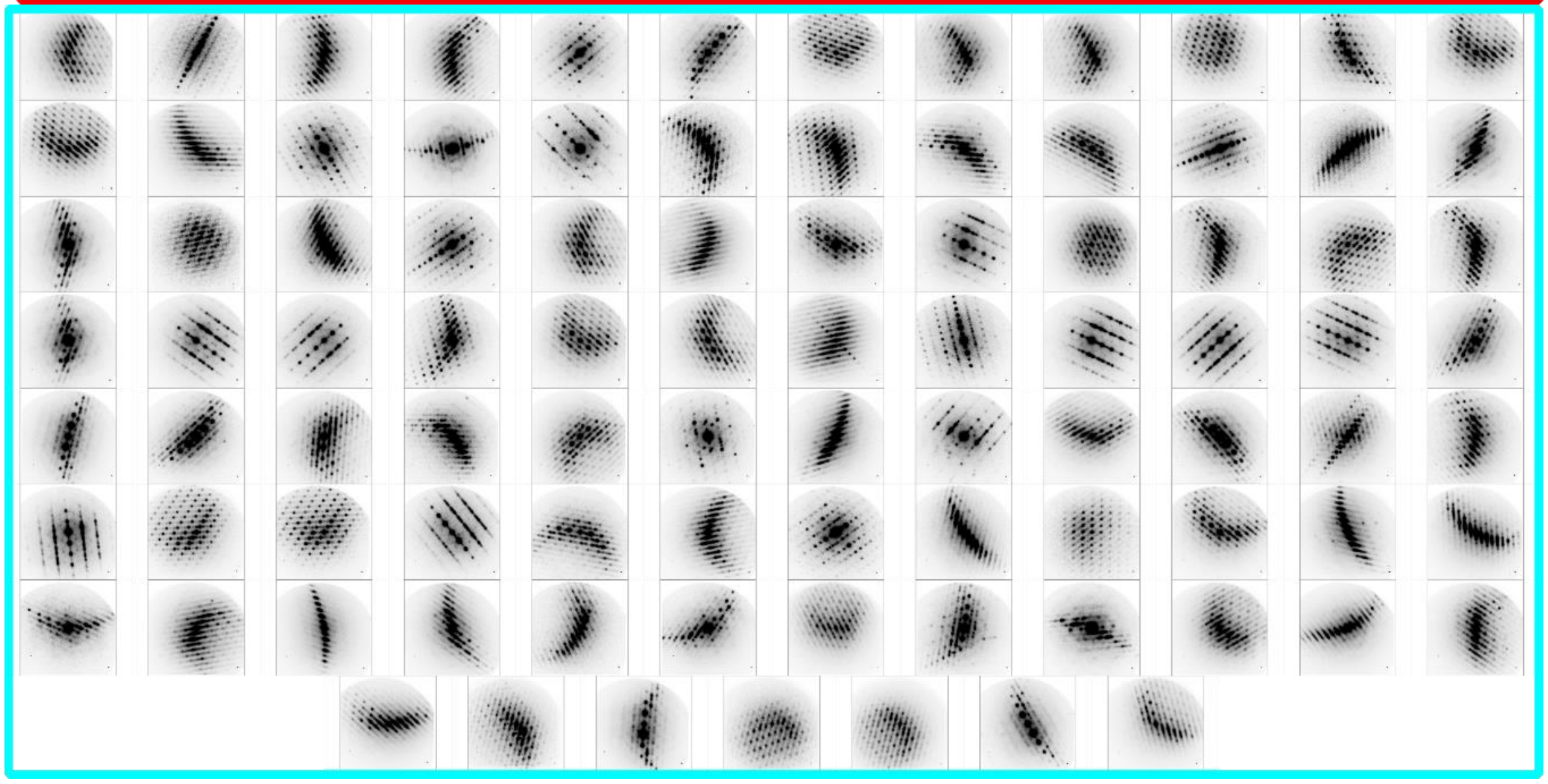

**Supplementary Fig.14 | Stitched SED data recorded at ~90K.(a)** The stitched SED data of thermally evaporated $FAPbI_3$ on a $SiN_x$ membrane at ~90K. Note the stitching is not as robust as the ambient equivalent due to the slight drift experienced at cryogenic temperatures **(b)** Patterns which are oriented close to a <001>$_c$ zone axis are marked in red. **(c)** Patterns which show characteristic {111}$_c$ twinning are marked in cyan. Scale bar: 3μm.

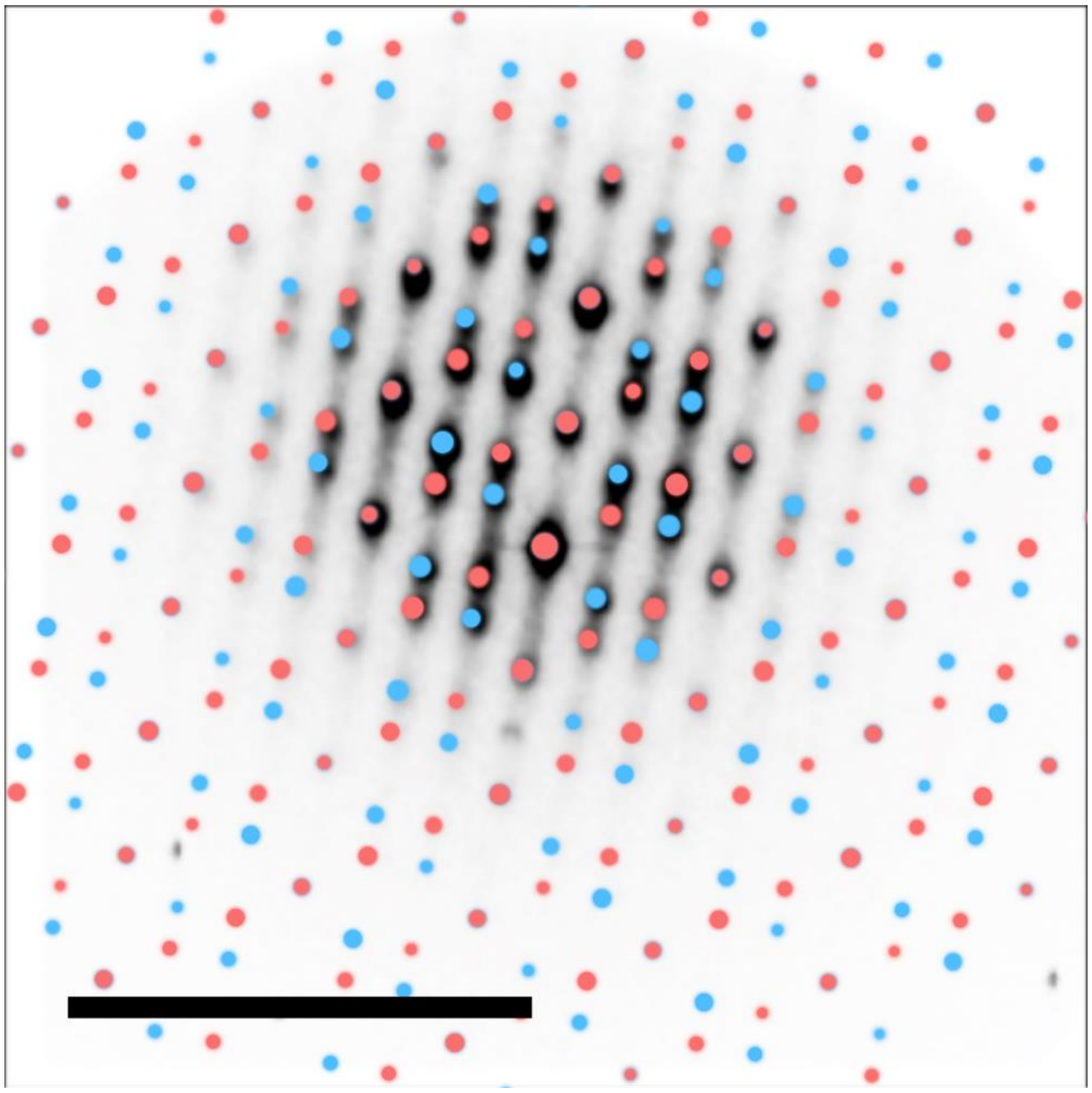

**Supplementary Fig.15 | A kinematical simulation of two [110]$_c$ zone axis patterns** rotated such that the $(\bar{1}11)_c$ and $(1\bar{1}1)_c$ reflections are shared and overlaid onto an experimental pattern taken at ambient temperature *.cif* files for the simulation were based on crystallographic data.[3] Scale bar is 1Å$^{-1}$.

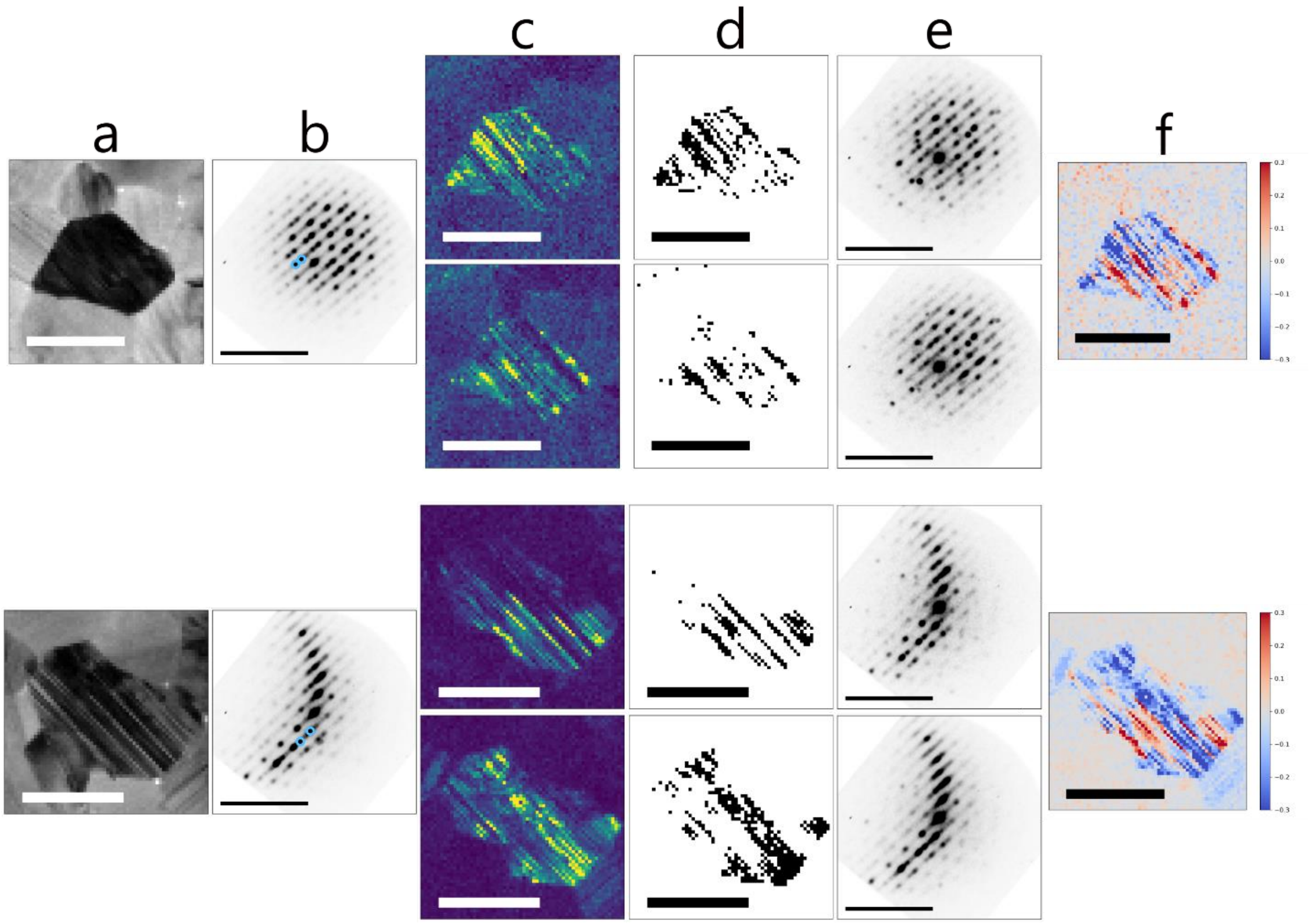

**Supplementary Fig.16 | SED data of thermally evaporated $FAPbI_3$ on SiNx grids** recorded at ambient temperature. **(a)** Virtual bright field images of grains oriented near the $\{110\}_c$ zone axis recorded at ambient temperature. **(b)** Average diffraction patterns extracted from the grain marked in (a). **(c)** VDF images formed by placing an aperture over Bragg spots of complementary twins. **(d)** Binary masks formed from the virtual darkfield images in (c). **(e)** Diffraction patterns extracted when the binary masks are applied to the SED data in (a). Importantly it is observed that each of these patterns is still indicative of nanotwinning and that there is not a great change in the position of the Ewald sphere or Laue circle between the images. **(f)** Image formed when the two complementary VDF images are subtracted from each other showing the contrast approximately anti-correlates. Scale bars a,c,d&f: 200 nm; b&e: $1Å^{-1}$.

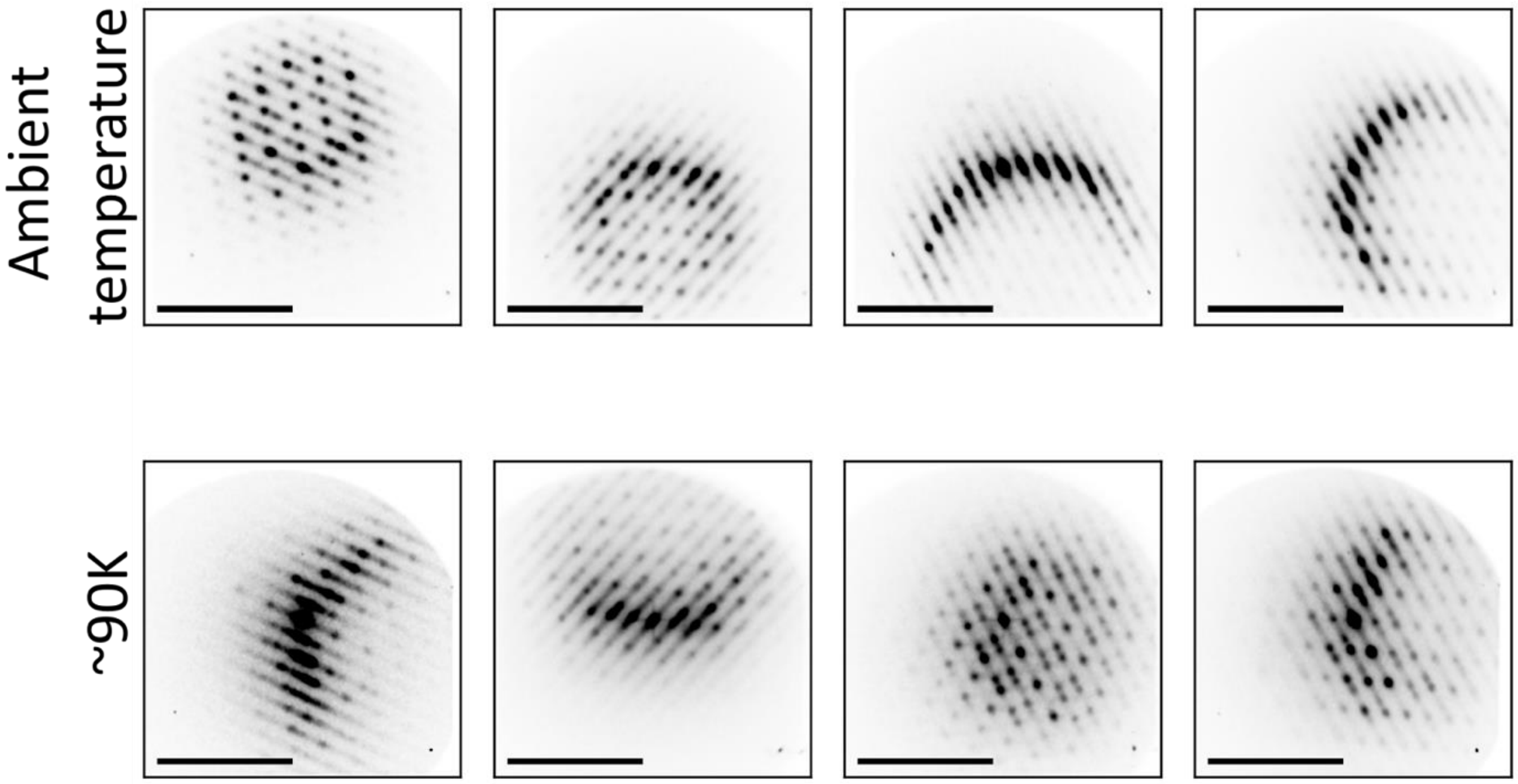

**Supplementary Fig.17 | Comparison of twinned <110>$_c$ zone axis patterns found at ambient and cryogenic temperature.** Averaged diffraction patterns from SED of thermally evaporated $FAPbI_3$ on a $SiN_x$ membrane showing $\{111\}_c$ type nanotwinning between ambient and cryogenic temperatures with patterns from the same grain shown in each column. No great change is observed in the nanotwinning between ambient and cryogenic temperatures but a small change in orientation is often observed upon cooling. Scale bars 1Å$^{-1}$.

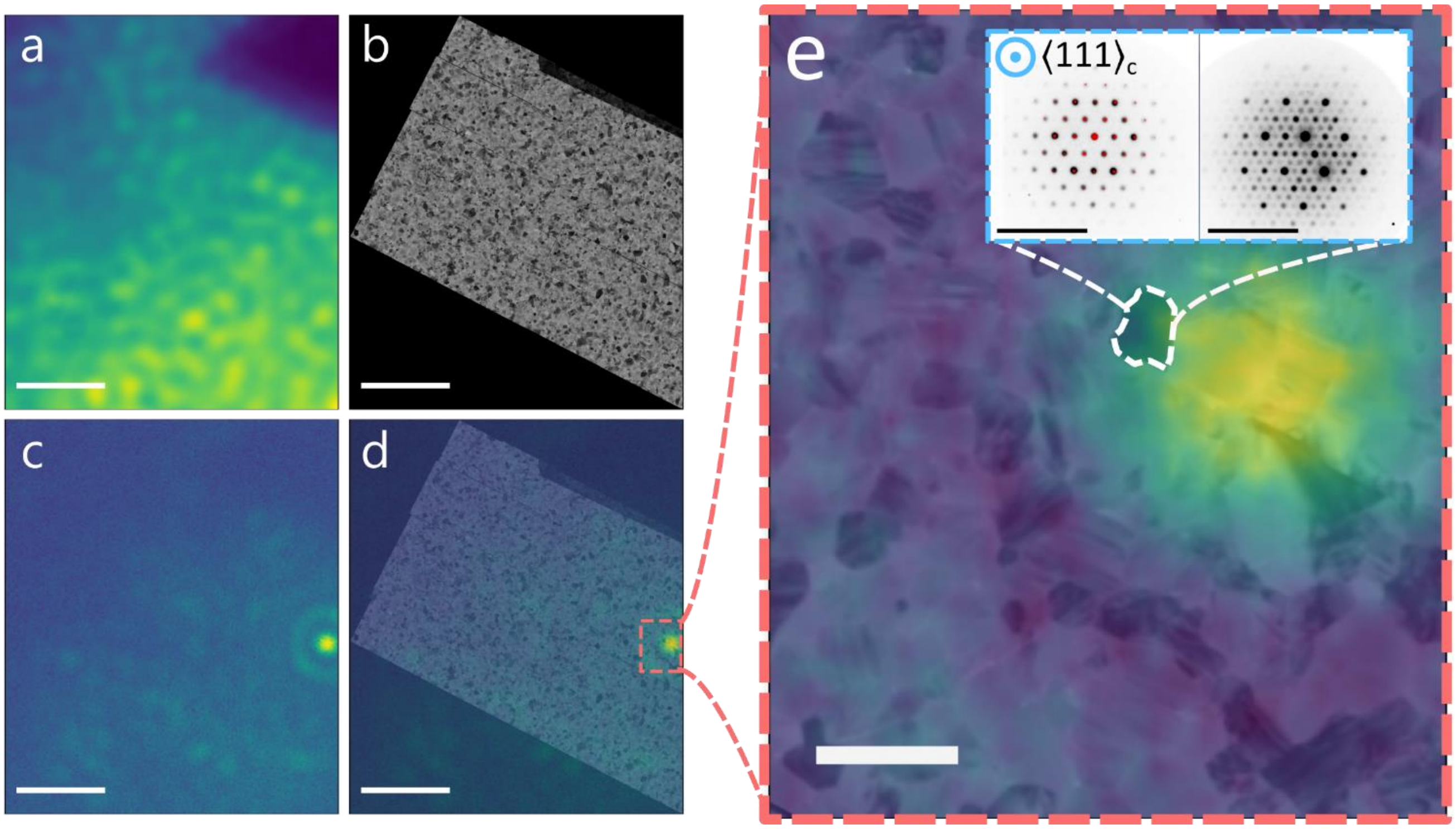

**Supplementary Fig.18 | Additional spatial correlations between local photophysics (PL) and structure (SED) reveal bright isolated emission arising from a grain oriented close to the <111>$_c$/<001>$_h$ zone axis where nanotwins are oriented in the plane of the film**. (**a**) Hyperspectral data summed over all wavelengths taken at 80K. (**b**) Stitched SED (ambient temperature) once correlated to the hyperspectral PL (80K) via the use of a Au fiducial marker. (**c**) The frame at 706 nm in the hyperspectral data. (**d**) Overlaid SED and hyperspectral PL data. (**e**) Zoom-in of the panel marked in (d), inset showing the diffraction pattern from the grain marked at both ambient and cryogenic temperatures (~90K) respectively (overlaid red pattern shows kinematical simulation of diffraction from cubic $FAPbI_3$ oriented along the <111>$_c$ zone axis.[35]

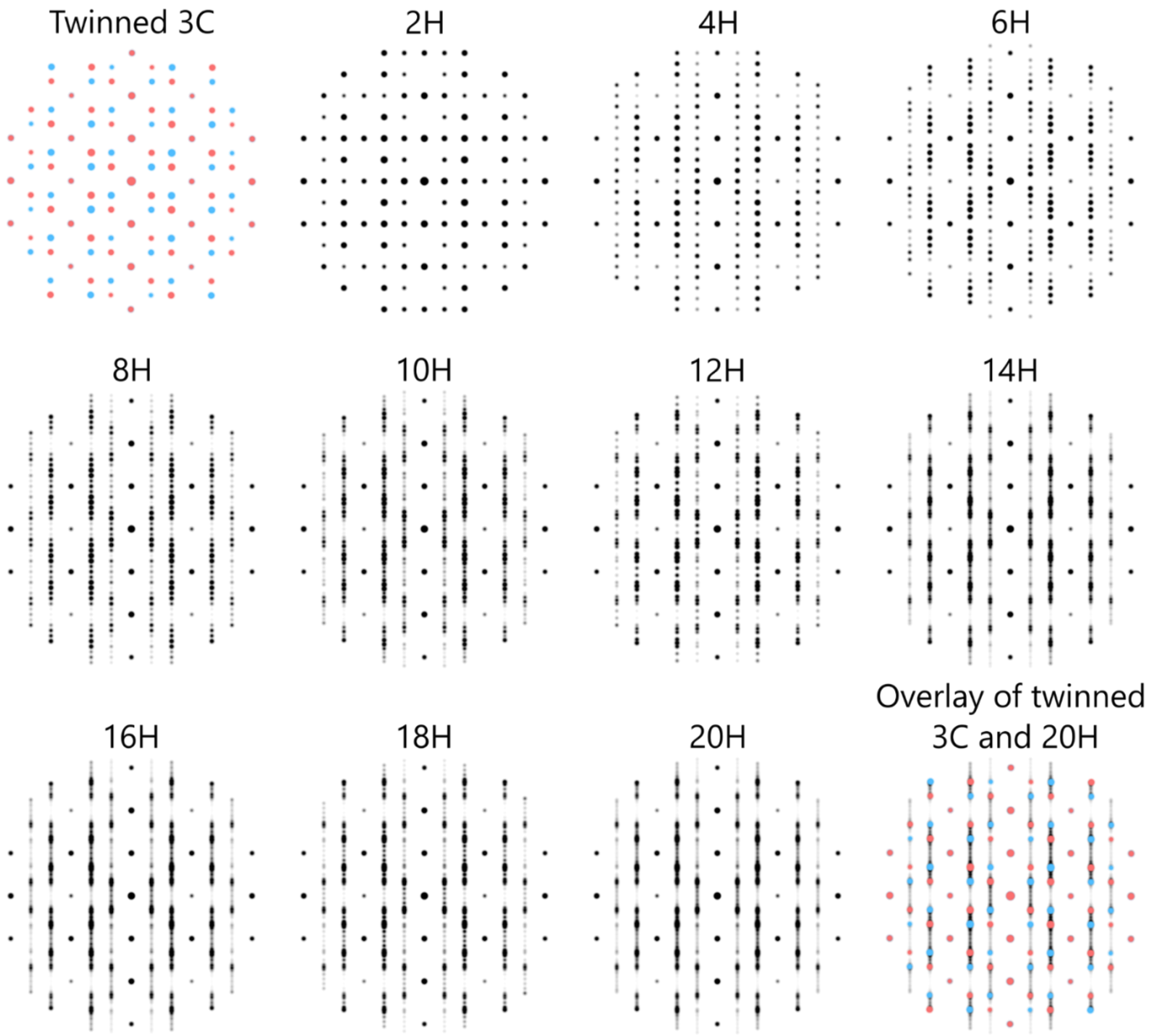

**Supplementary Fig.19 | Kinematical simulations of diffraction patterns with increasing polytype order.** Kinematical simulations showing that when the order of the hexagonal polytype increases the $[100]_h$ zone axis pattern resembles the twinned 3C phase with streaking along the $<111>_c$ direction. .*cif* files were created similarly to.[4,5]

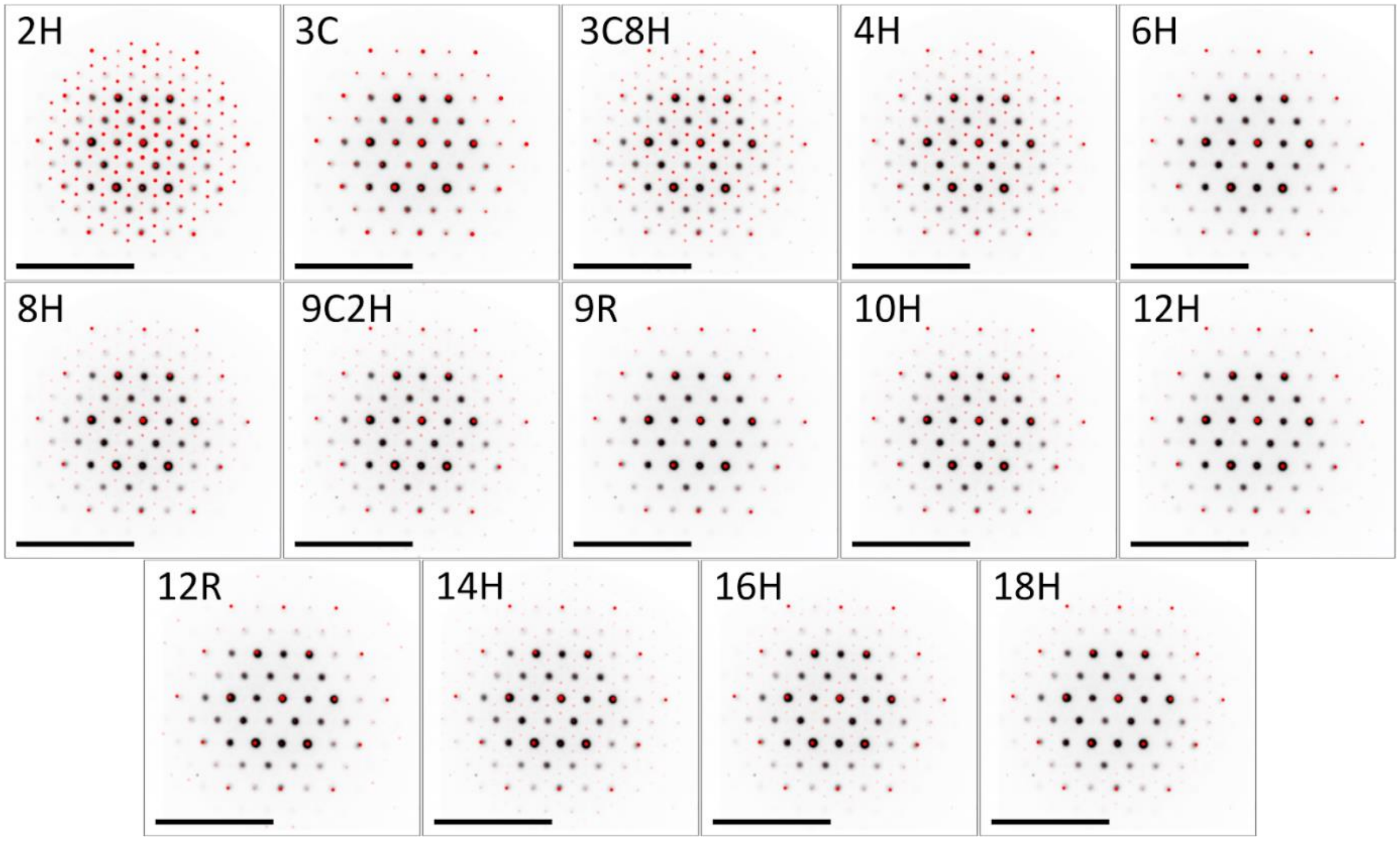

**Supplementary Fig.20 | Kinematically simulated patterns of the hexagonal polytypes** of $FAPbI_3$ with increasing corner sharing layers viewed along $[001]_h$ direction (red) overlaid with the experimental $<111>_c$ /$<001>_h$ zone axis pattern (black) recorded at ambient temperature. Notice that extra reflections are absent for the 3C, 6H, 9R, 12H and 12R polytype. All scale bars are $1Å^{-1}$.

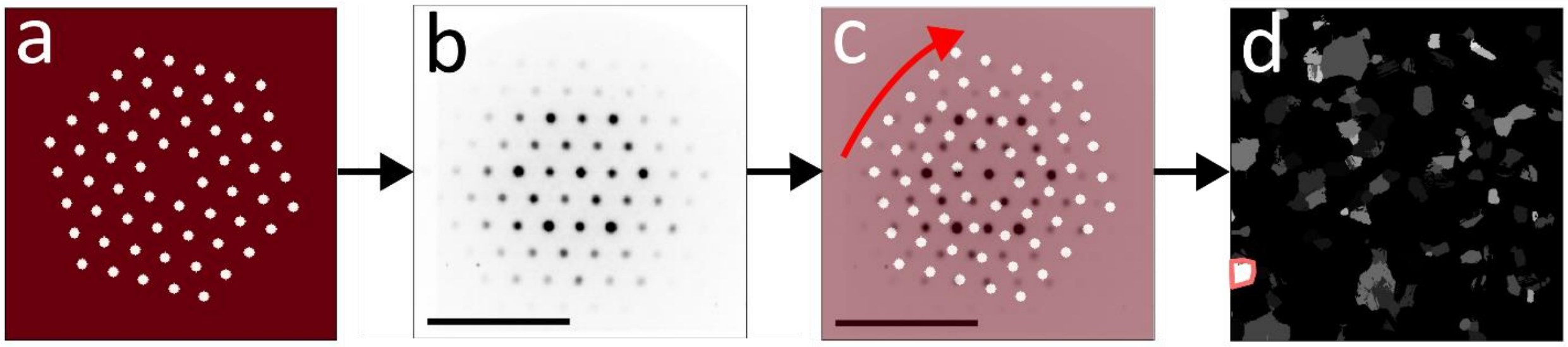

**Supplementary Fig.21 | An example workflow to quickly identify grains oriented near the <111>$_c$/<001>$_h$ zone axis** shown for thermally evaporated $FAPbI_3$ on $SiN_x$ grids. **(a)** A mask created from a simulation of a <111>$_c$ zone axis pattern. **(b)** Experimental pattern found via the clustering outlined in Supplementary Note 2. **(c)** The mask is overlaid on the experimental patterns and rotated 360°; a VDF image is then formed where the intensity is maximal throughout this rotation. **(d)** Image found where intensity is given by the maximum from step **(c)** for each cluster. Notice that the bright grain marked (pink) is oriented along the <111>$_c$ direction as this matches well with the mask Scale bars 1Å$^{-1}$.

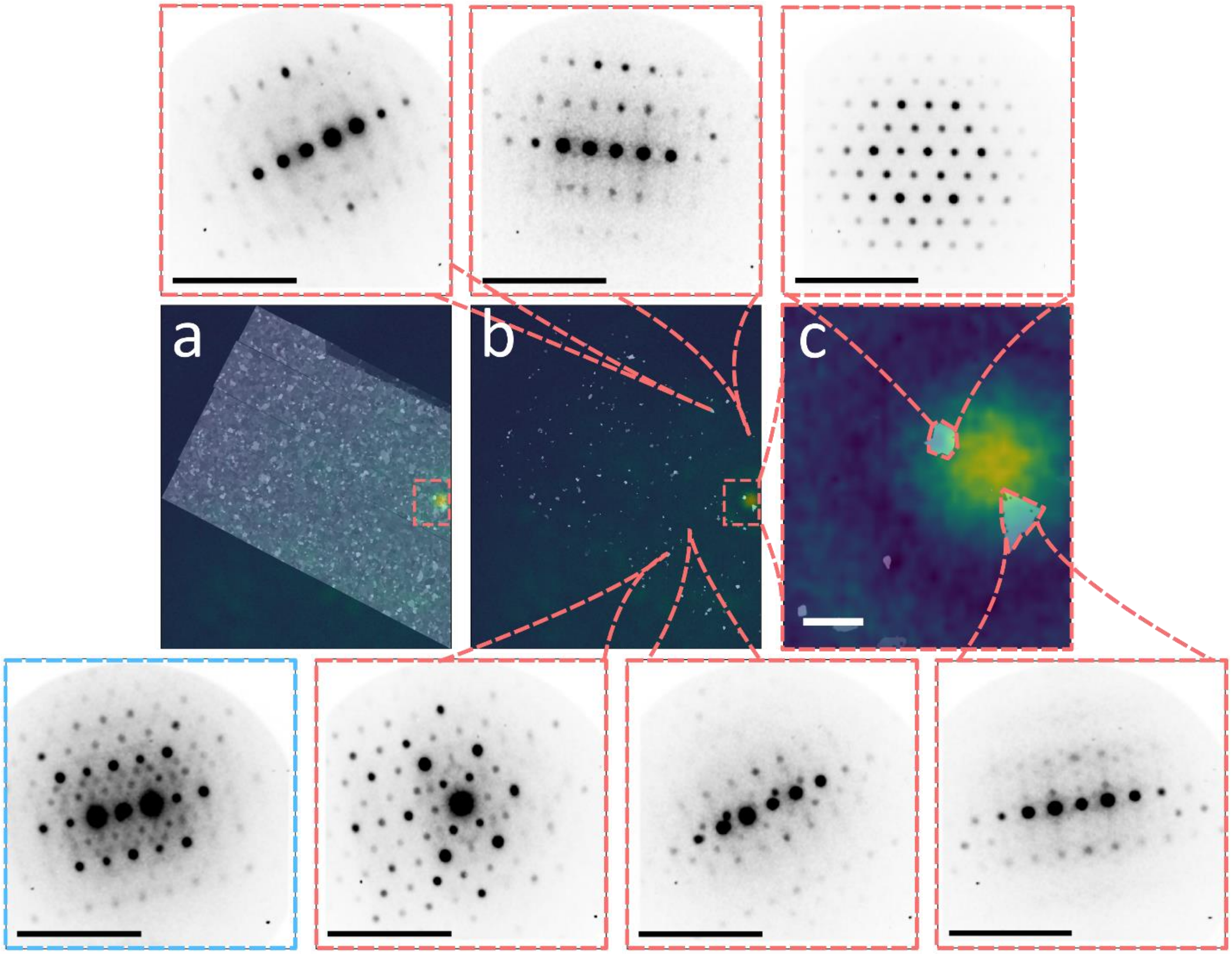

**Supplementary Fig.22 | Overlay of '<111>$_c$ map' and hyperspectral PL data (80K) (a)** The '<111>$_c$ zone axis map' for thermally evaporated $FAPbI_3$ deposited on $SiN_x$ grids, constructed using the method described in Supplementary Fig. 21 overlaid with the hyperspectral PL data showing the isolated emitter (80K). **(b)** The overlay shown in (a) with a threshold so only the most intense features in the '<111>$_c$ zone axis map' are shown. **(c)** Expansion from the area marked in (b). Surrounding diffraction patterns which show the <111>$_c$ zone axis pattern discussed is a distinct structural feature, especially in the vicinity of the isolated emitter, with very few grains being observed oriented along this zone axis and many of the other bright grains being false positives. An instance of another grain oriented close to the <111>$_c$ zone axis is observed in this sample with superstructure reflections appearing at low temperature (shown in blue and discussed in Supplementary note 3). Scale bars: central row: 500 nm and top/bottom row: 1Å$^{-1}$.

**Supplementary Text 1: Acquisition and preprocessing of the SED data**

The contiguous datasets were acquired as outlined in the Materials and Methods section however the small note on electron beam damage should be expanded on from the main text. Although we are confident that damage was mitigated sufficiently and no observable changes in the SED data was observed between scans, damage mitigation is complicated by the relative cross sections between radiolysis and knock-on damage being largely unknown for this class of material. This means that measurements performed at cryogenic temperatures may either improve or worsen the effect of beam induced damage.[6–8]

Analysis of the SED dataset was performed with the use of python packages pyxem 0.16 and py4DSTEM 0.14.[9,10] Firstly, during data processing the direct beam was aligned to the central pixel of the detector and the ellipticity of the polycrystalline disk confirmed to be minimal by summing all diffraction patterns across an entire SED scan. Virtual bright field images were formed by taking intensity from a virtual aperture 0.04 $Å^{-1}$ in radius placed around the central spot. Virtual dark field images were similarly formed however a virtual aperture was placed around a diffracted reflection with a radius of 0.04 $Å^{-1}$.

**Supplementary Text 2: Simple linear iterative clustering (SLIC)**

The clustering methodology is adapted from simple linear iterative clustering (SLIC) used widely in the field of remote sensing and is applied to high dimensional microscopy datasets herein, to our knowledge for the first time.[11,12] SLIC can be thought of as a variant to k-means clustering and provides a general, intuitive, robust, and computationally inexpensive methodology to cluster SED data (Supplementary Fig. 23). During this procedure detector pixels that have a low dynamic range or variance over the SED scan are discarded. Typically, 98% of the pixels are removed meaning only the important data points for clustering are retained. Next 700 centroid seeds ($N$=700 in pseudocode below) are initialized randomly or as a regular grid into the SED scan in real space. Data in the vicinity of the centroid (defined by $S$) is then considered and assigned to a cluster depending on the similarity to each centroid; defined via the Euclidean distance. The spatial and channel distances are then combined into a single measure via a weighting factor $m$, this encodes intuition that 'pixels' close together likely belong to the same cluster. Once this step is complete for all the data the centroids are updated and the process repeats iteratively. After the iterative process a final step which combines clusters that correlate highly with each other is performed. The mean diffraction patterns from each cluster are then computed from the original data. As much of the data prior to clustering is discarded and each centroid only considers data in its vicinity this process is surprisingly efficient with it taking 99s on a standard desktop machine (11th Gen Intel(R) Core (TM) i5-11400 CPU and 32Gb of RAM), this could likely be further improved with by performing the calculations on a GPU.

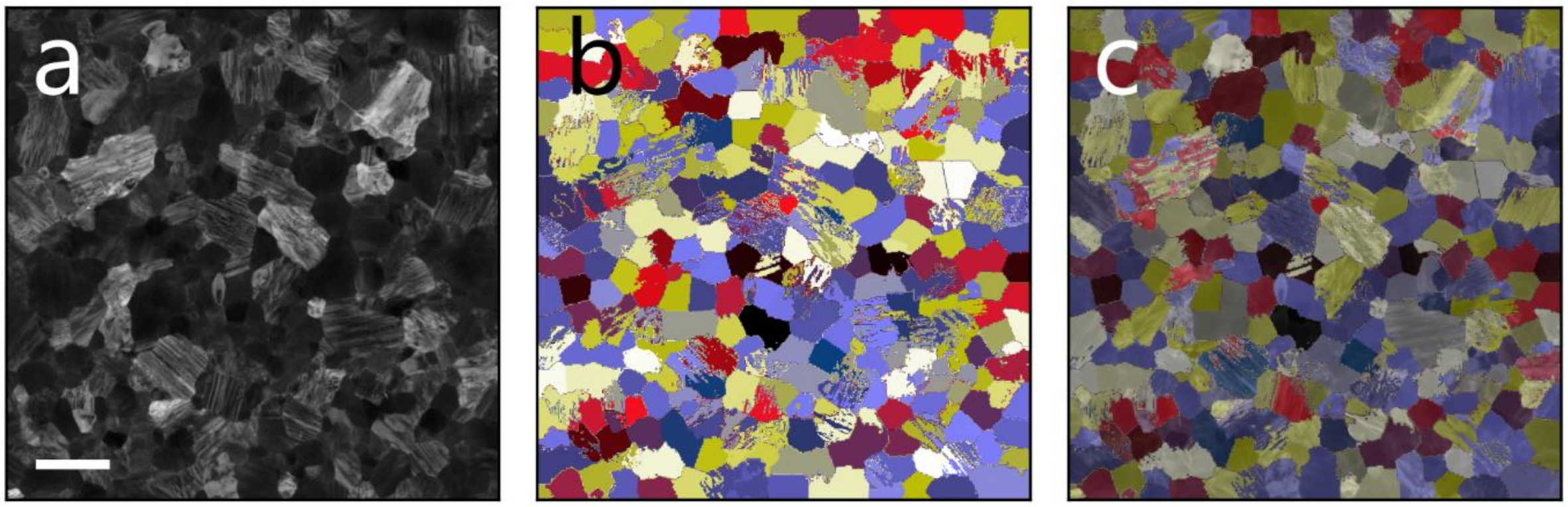

**Supplementary Fig.23 | The SLIC methodology applied to an SED scan. (a)** Shows a dark field image formed from the SED scan. **(b)** Shows the resulting clusters after the SLIC clustering. **(c)** An overlay of (a) & (b). Scale bar in (a): 500 nm.

When compared to other clustering methods, such as principal component analysis (PCA) and traditional k-means clustering, the SLIC approach is confirmed to be an improvement in our case. If PCA is performed on the same SED scan as above, (after pixels which have variance below the 98th percentile are removed so that a direct comparison to SLIC can be made) the compute time is improved with the entire process taking only 33s. However, this comes at the detriment of the clustering quality. During this approach PCA is applied to the data and the components which retain up to 1% of the explained variance ratio are used as a cut-off similarly to Duran *et al.*; this leaves 28 principle components to be included for subsequent k-means clustering (Supplementary Fig. 24).[13] Once clustered we observe a drop in accuracy due to this approach not encoding the intuition that pixels in close proximity are likely to belong to the same cluster, as SLIC does. Ultimately this leads to diffraction patterns which are not from the same grain being clustered together, giving the appearance of overlapping patterns in the resulting clusters.

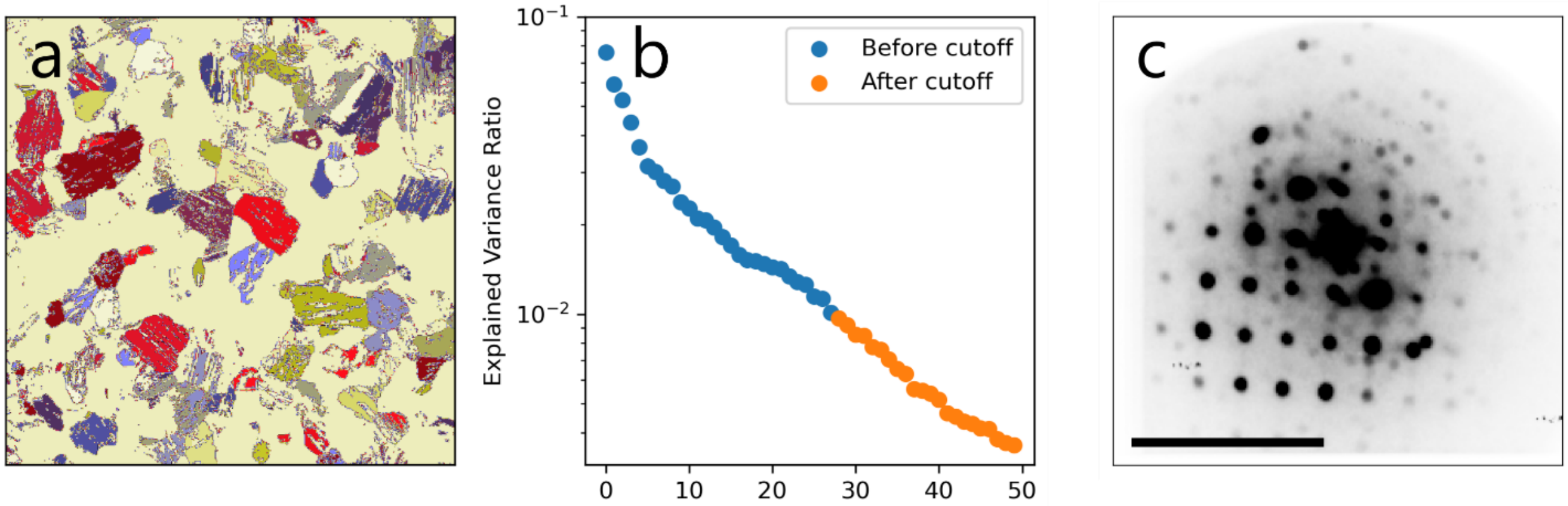

**Supplementary Fig.24 | PCA clustering results. (a)** The results when PCA clustering is performed. **(b)** A scree plot showing the cutoff when explained variance ratio is below 1%. **(c)** A

mean diffraction pattern taken from a cluster in (a) where multiple crystallites are contributing to the signal. Scale bar in (c): $1 Å^{-1}$.

It is important to note, performing PCA as a dimensionality reduction technique and related unsupervised methods are undoubtedly valuable tools that are likely to outperform SLIC on different problems due to SLIC in part relying on the crystallinity and polycrystalline nature of the perovskite film.

Code used to perform SLIC can be found on Github. In pseudo code the adapted SLIC algorithm is as follow:

1. Unravel each diffraction pattern into a 1D array.
2. **Require:** dynamic range or variance of channels > 95th percentile
3. Normalise the channels between [0,1]
4. Initialize N centroids
5. **while** converged = False **do**
6. **for all** centroids $\in \{1, . . . , N\}$ **do**
7. **for all** 'pixels' in the vicinity (S) of each centroid **do**
8. Calculate the distance measure ($D_{new}$) between each 'pixel' and centroid given weighting factor (m)
9. **if** $D_{new} < D_{current}$ **then**
10. Reassign the 'pixel' to the new centroid
11. Combine clusters which are highly correlated
12. Calculate mean diffraction patterns from the original data

**Supplementary Text 3: Determination of the Low Temperature Phase**

The appearance of superstructure reflections upon cooling in electron diffraction patterns can be used to inform the lower symmetry space group of perovskites[14]. The methodology used in this work follows similarly to that used by Woodward *et al.*[15] and we assume $O_h$ tilting is the only structural distortion present. Throughout the discussion we will use the Glazer notation as well as noting the space group; Glazer notation considers tilting independently about the pseudocubic $<001>_c$ directions with a superscript 0, + and - denoting no tilt, inphase and antiphase tilting respectively[16,17].

Firstly, considering the literature to date there already exists reports of the phase behavior of $FAPbI_3$ at low temperature, but a consistent understanding of what the space group the γ phase occupies is lacking[18–20]. This debate is undoubtedly due to the many possible lower symmetry space groups perovskites can occupy due to structural distortions of the cubic aristotype. Indeed, common causes of superstructure reflections come from octahedral tilting or cation/vacancy ordering; or indeed a combination of both. Furthermore, as prior work reported experiments either on single crystals, which are not representative of the thin films; or use powder-based techniques, which use spatial averaging and azimuthally integrated diffraction patterns, an accurate identification of the phase behavior is especially challenging. From this it is evident that scanning electron diffraction is one of only a few techniques which can achieve a genuine understanding of the phase behavior of $FAPbI_3$ thin films at low temperature.

It is well established that upon cooling at 285K the cubic aristotype (α phase, $Pm\bar{3}m$, $a^0a^0a^0$) continually converts to a lower symmetry space group with in–phase tilting occurring about a single pseudocubic axis via a second order phase transition. This gives the β phase ($P4/mbm$, $a^0a^0c^+$). Upon further cooling it is known there is a first order phase transition at ~140K to the γ phase but the identity of this phase is not entirely understood. Fabini *et al.* have found the γ phase still possesses a $P4/mbm$ symmetry using high resolution powder X-ray diffraction (P-XRD).[19] In that work the $Immm$ ($a^+b^+c^+$); $I4/mmm$ ($a^0b^+b^+$); and $Im\bar{3}$ ($a^+a^+a^+$) space groups are considered but rejected from the fact that no low angle Bragg peaks are detected as would be expected.

In other reports it was commented that the "synthesis conditions, sample environment, and cycling temperature all play a role in the observed phases of the material"[18] and showed that below 140K the motion of the FA cation is restricted, forming a glassy state. However, it is also noted that additional Bragg reflections are observed but the origin of these and therefore the space group of the γ phase was not determined. The presumption of complexity is further reinforced by work considering the cubic (α phase) to hexagonal (δ phase) upon cooling which shows kinetic trapping of the pseudocubic phase occurs depending on cooling rates[20].

Now considering the electron diffraction patterns obtained from our data. Firstly, by collating $<001>_c$ zone axis patterns we confirm no superstructure reflections are present at room temperature as expected for the cubic $Pm\bar{3}m$, $a^0a^0a^0$ structure (as shown in Supplementary Fig.

13b). Upon cooling to 90K superstructure peaks appear consistently at ½{ooe}$_c$ positions with no absences. Where *o* and *e* denote odd and even miller indices as described by Woodward *et al.* [15] It is important to note that if a *P*4/*mbm*, $a^0a^0c^+$ space group was retained at 90K it is expected to only see superstructure reflections in ⅓ of the patterns sampled and systematic absences are also expected. Next, we consider the <111>$_c$/<001>$_h$ zone axis patterns, we again see the appearance of superstructure reflections upon cooling, at ½{ooe}$_c$ positions. This is consistent with in-phase tilting about multiple pseudocubic axes and importantly inconsistent with *P*4/*mbm*, $a^0a^0c^+$ even if dynamical diffraction is considered (Supplementary Fig. 25)[15].

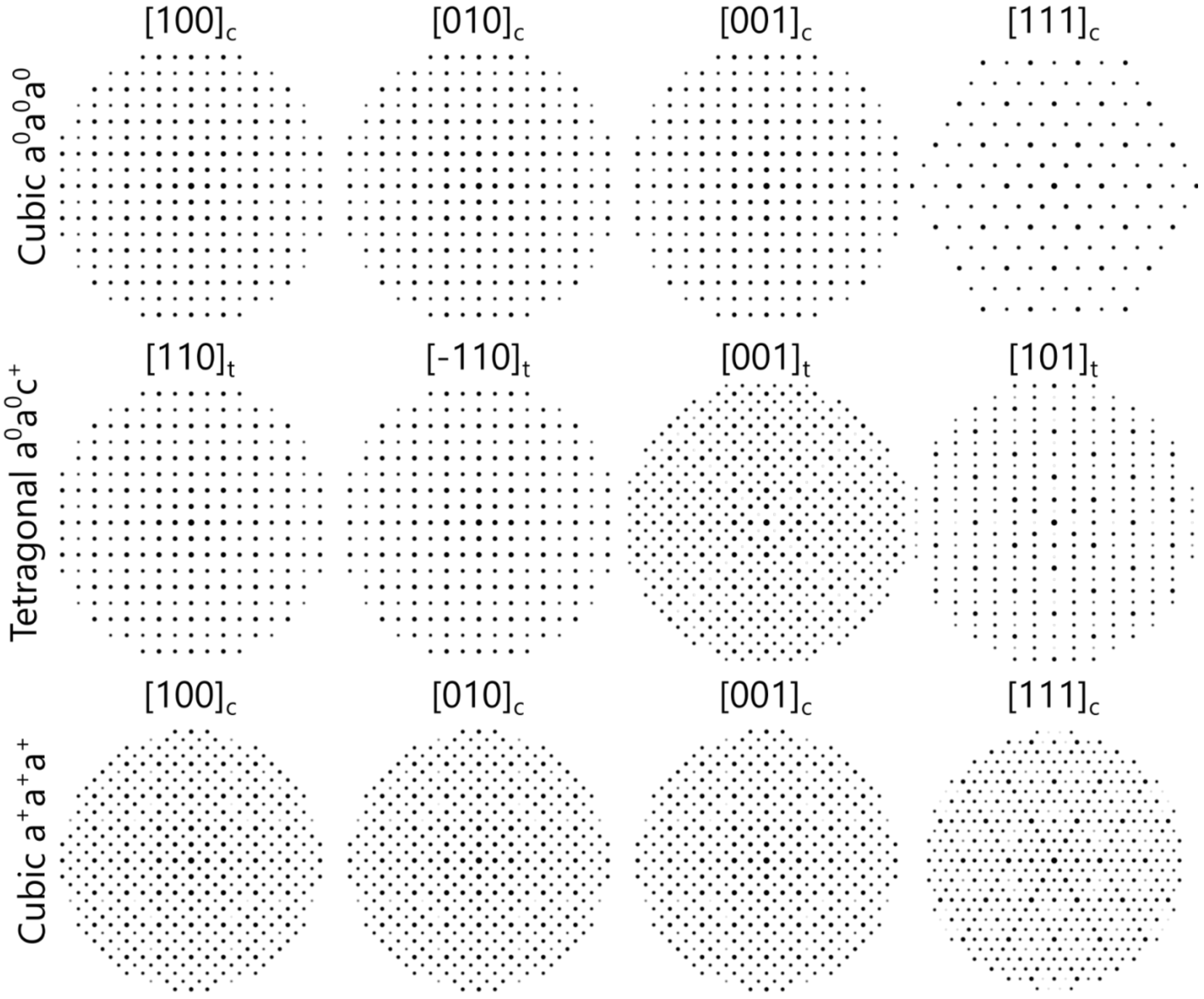

**Supplementary Fig.25 | Kinematically simulated zone axis patterns** along [100]$_c$, [010]$_c$ and [001]$_c$ and [111]$_c$ directions in the pseudo cubic unit cell. The cubic $Im\bar{3}$ ($a^+a^+a^+$) space group is used to illustrate the appearance of superstructure peaks, this is equivalent in the *Immm* ($a^+b^+c^+$) and *I*4/*mmm* ($a^0b^+b^+$) tilt systems.

We attempt to see superstructure reflections consistent with our proposed structure along a <110>$_c$ ZA however do not observe any additional spots upon cooling. This is attributed to the fact that the superstructure reflections are weak compared to the background diffuse scattering observed from nanotwinning in <110>$_c$ ZA patterns. We can however consider patterns oriented along a <123>$_c$ ZA where the Bragg spots are further apart, allowing for the appearance of superstructure reflections to be observed. This is consistent with our assignment (Supplementary Fig. 26).

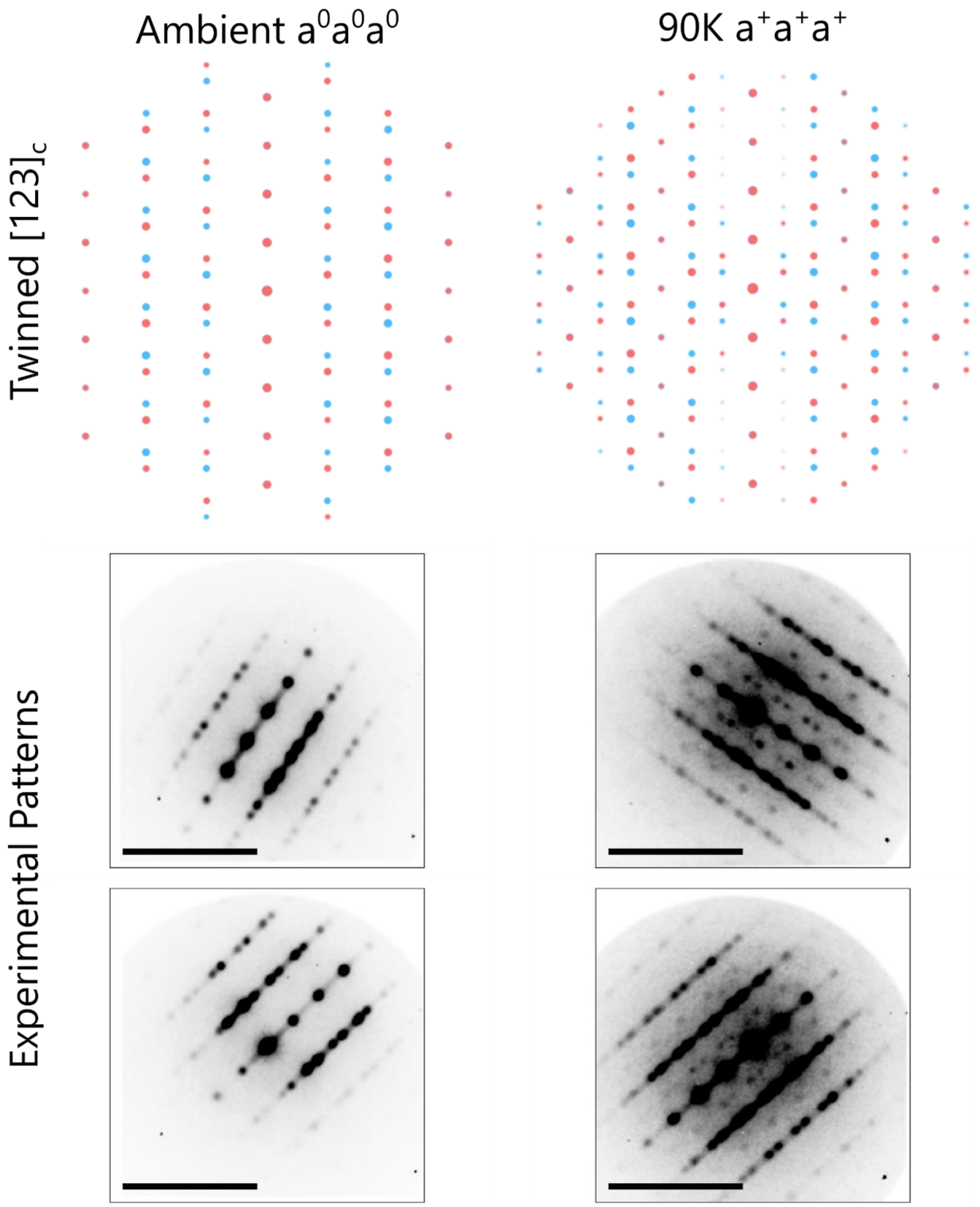

**Supplementary Fig.26 | Showing the superstructure peaks observed upon cooling** if oriented along a $[123]_c$ ZA with nanotwinning. All scale bars are $1Å^{-1}$.

Considering hexagonal polytypes which can be described as a highly ordered array of nanotwins. Upon tilting of the corner sharing layers between $\{111\}_c$ twins in an $a^+a^+a^+$ fashion this results in superstructure peaks similar to those observed experimentally at ½$\{ooe\}_c$ positions (Supplementary Fig. 27). To create .cif files for the tilted hexagonal polytypes firstly the basis vectors of a parent .cif file, whether that be $a^0a^0a^0$, $a^+a^+a^+$ or $a^0a^0c^+$ were changed to possess hexagonal symmetry, with the new lattice parameter along c, denoted $[001]_h$, representing what was previously a pseudocubic $<111>_c$ direction. The order of the polytype is then defined by altering the number of corner sharing layers present and a face sharing twin boundary layer created. As we propose isolated emitters are ordered hexagonal polytypes they should not be assigned a space group of *Immm* ($a^+b^+c^+$)*; I*4/*mmm* ($a^0b^+b^+$)*; and* $Im\overline{3}$ ($a^+a^+a^+$) despite tilting being consistent with in-phase tilting about multiple pseudocubic axes.

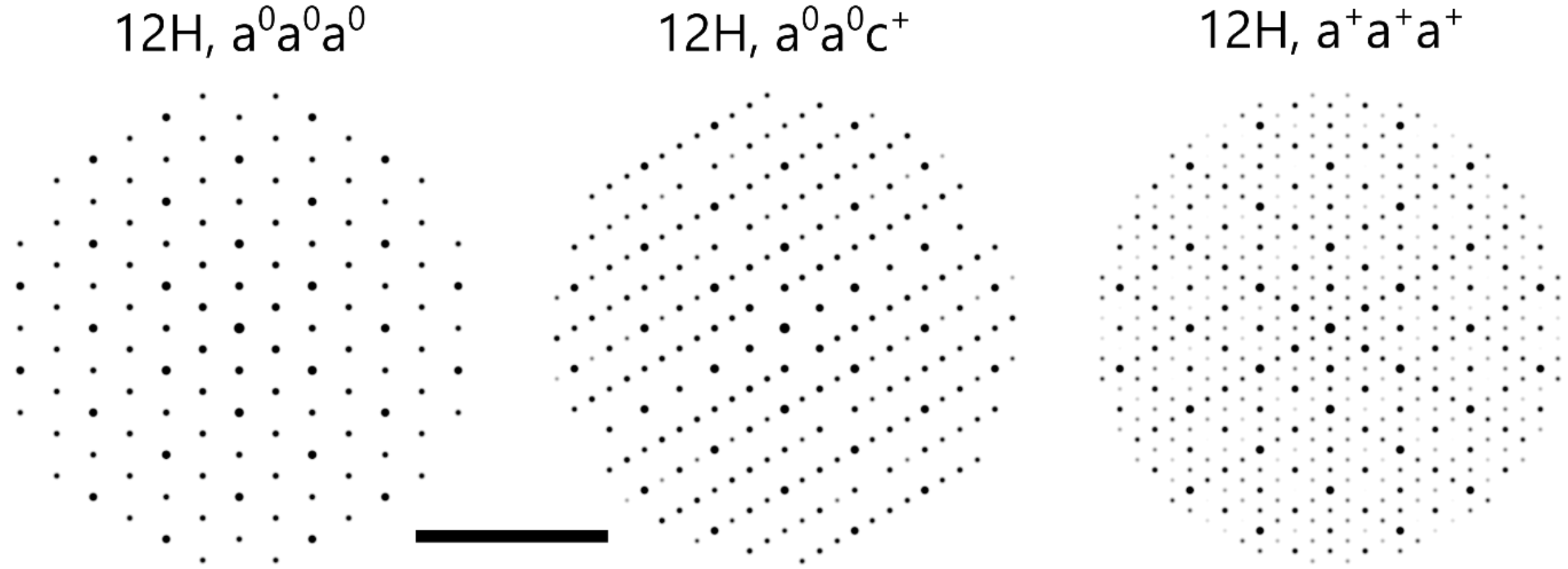

**Supplementary Fig.27 | Kinematically, simulated diffraction patterns of a 12H polytype** with various tiling patterns imposed on the corner sharing layers. Scale bar: 1Å$^{-1}$.

If the position of $\{hh0\}_c$ reflections in the cryogenic temperature $<001>_h/<111>_c$ zone axis pattern are inspected closely it can be seen they do not noticeably distort away from the ambient equivalents. This suggests that the tilting is equal along all three pseudocubic directions within the limits of experimental error. To measure any distortion accurately a line is fit across complementary $\{hh0\}_c$ Bragg reflections and the angles between these calculated. To gain an estimate of the uncertainty of the angles we consider the case where a $\{110\}_c$, $\{220\}_c$ or $\{330\}_c$ reflection moves position by a single pixel normal to the relevant g vector. This gives shifts of 2.3°, 1.8° and 0.8° for $\{110\}_c$, $\{220\}_c$ and $\{330\}_c$ reflections respectively, as the differences in angle we observe experimentally are significantly smaller than these values we are propose 6-fold symmetry is retained within experimental error. (Supplementary Fig. 28).

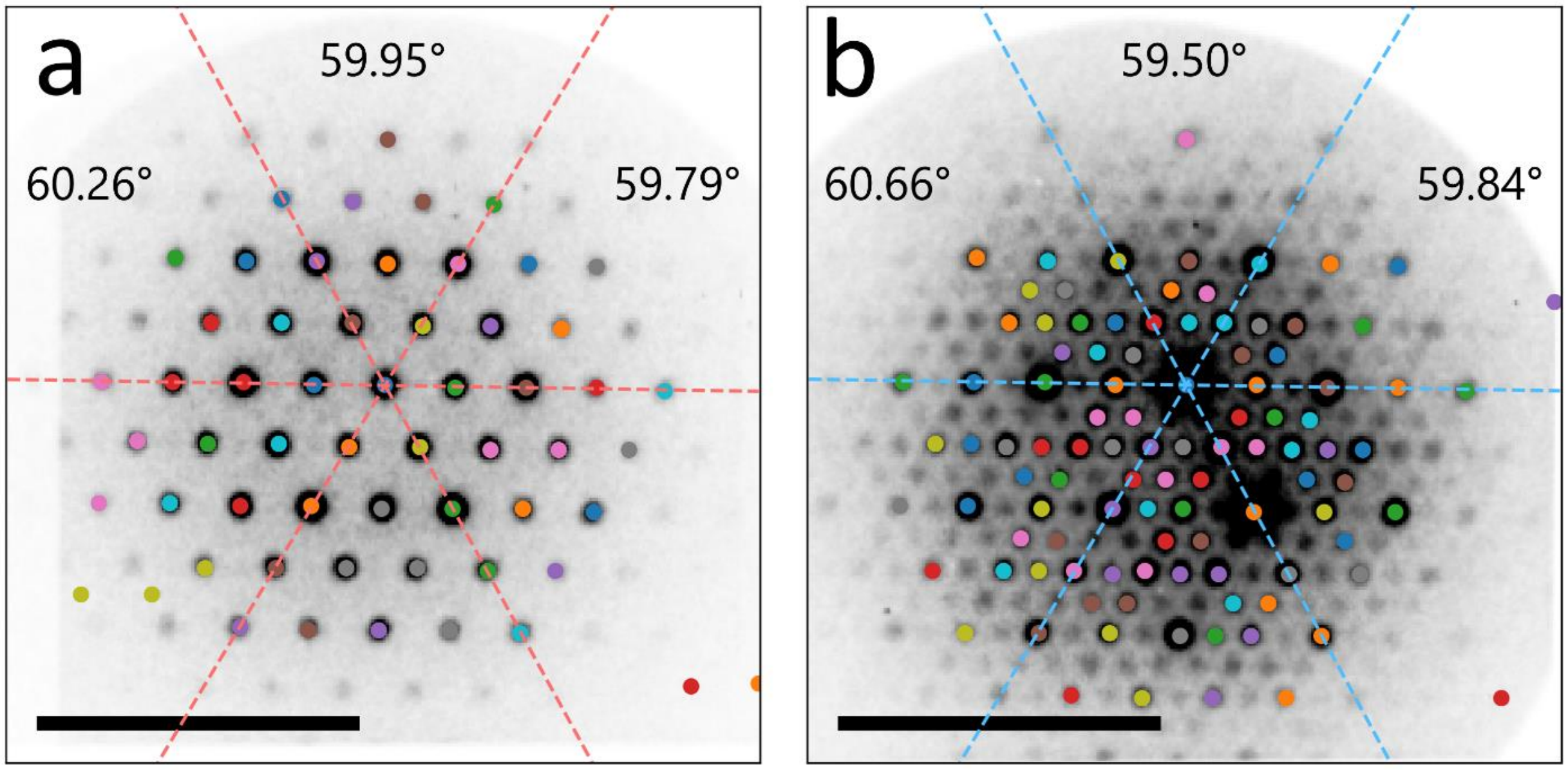

**Supplementary Fig.28 | The experimental <111>$_c$ zone axis pattern** at **(a)** ambient temperature and **(b)** cryogenic temperature, with peaks found via a difference of gaussian method[9]. Overlaid over each pattern are lines fit through the peaks belonging to the {hh0}$_c$ family, the angles between these lines show minimal distortion of the 6-fold rotational symmetry upon cooling. Scale bars: 1Å$^{-1}$.

Symmetry breaking upon cooling is also observed in patterns which are originally indexed to a hexagonal polytype at ambient conditions. It is understandable that there are numerous hexagonal polytypes present in the film due to the ubiquitous nature of the {111}$_c$ nanotwinning present. This behaviour further reinforces the fact that in high order hexagonal polytypes the majority of octahedra are still corner sharing (except for the layers which can be considered a {111}$_c$ twin) and will similarly distort to the discussion above (Supplementary Fig. 29).

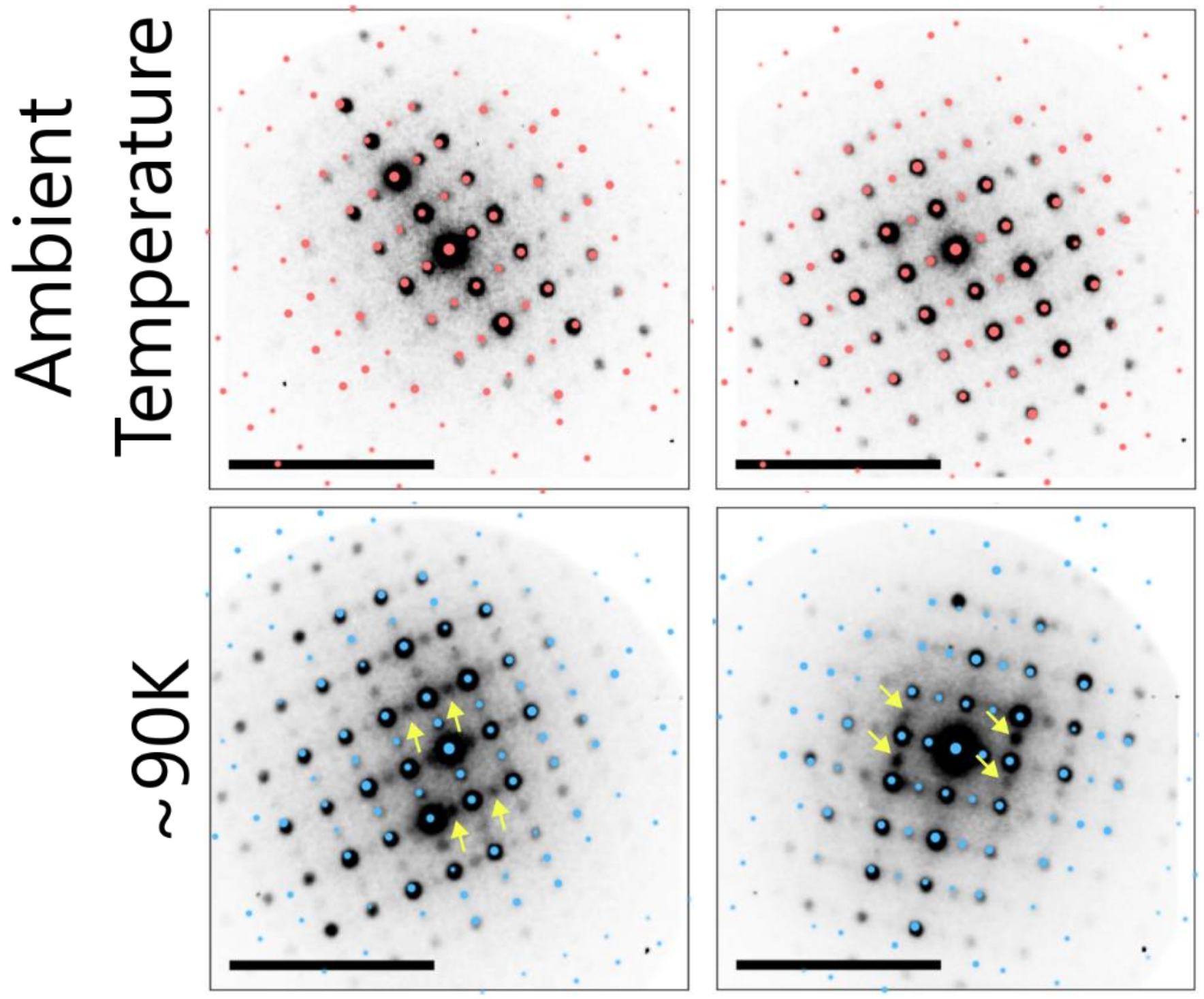

**Supplementary Fig.29 | The appearance of extra spots at low temperature** (blue) compared with ambient temperature (pink) are marked by arrows and attributed to octahedral tilting of the corner sharing layers. Diffraction patterns which can be indexed to $[112]_h$ zone axis of a 6H polytype are shown.

There are several complications which mean our assignment should be considered cautiously. Firstly, dynamical diffraction will excite reflections which are nominally systematically absent, especially when close to a zone axis pattern. Secondly, due to the dynamic nature and structural heterogeneity of hybrid perovskites, defining a unifying space group may not be appropriate with phase transitions being highly dependent on the local environment[21–23], particularly with the high density of extended defects present. Despite these caveats a consistent structural model for the SED data collected herein has been proposed.

**Supplementary Text 4: Stitching of the SED data and spatial correlation with hyperspectral PL**

For brevity we demonstrate the stitching and correlation methodology for one of the two correlative SED datasets recorded as it is similar in both cases. In this sample we first correlated the SED data at ambient temperature before finding the corresponding structural features in the cryogenic dataset; this was done to prevent the slight drift experienced at ~90 K between scans affecting the registration.

To stitch many SED scans together we record multiple contiguous scans and form virtual bright or dark field (VBF and VDF) images which act as a proxy for the 4D dataset. A key-point detection and matching algorithm is then applied between the images; common choices for this include the scale invariant feature transform (SIFT) or binary robust invariant scalable key-points (BRISK) algorithms, implemented using OpenCV[24–26]. Although these are undoubtedly valuable tools, it has been empirically found that pretrained neural network-based approaches such as Key-Net-AdaLAM, implemented using Kornia provide improved mappings[27]. Once the keypoints are detected the random sample consensus algorithm (RANSAC), or variants thereof, is used to define an affine transform which stitches one image onto the other (Supplementary Fig. 30)[28]. As an affine transform is utilized differences in scaling, rotation, shear, and translation can be accounted for between the images. To make these tools accessible a Python based GUI has been developed which can be found on Github.

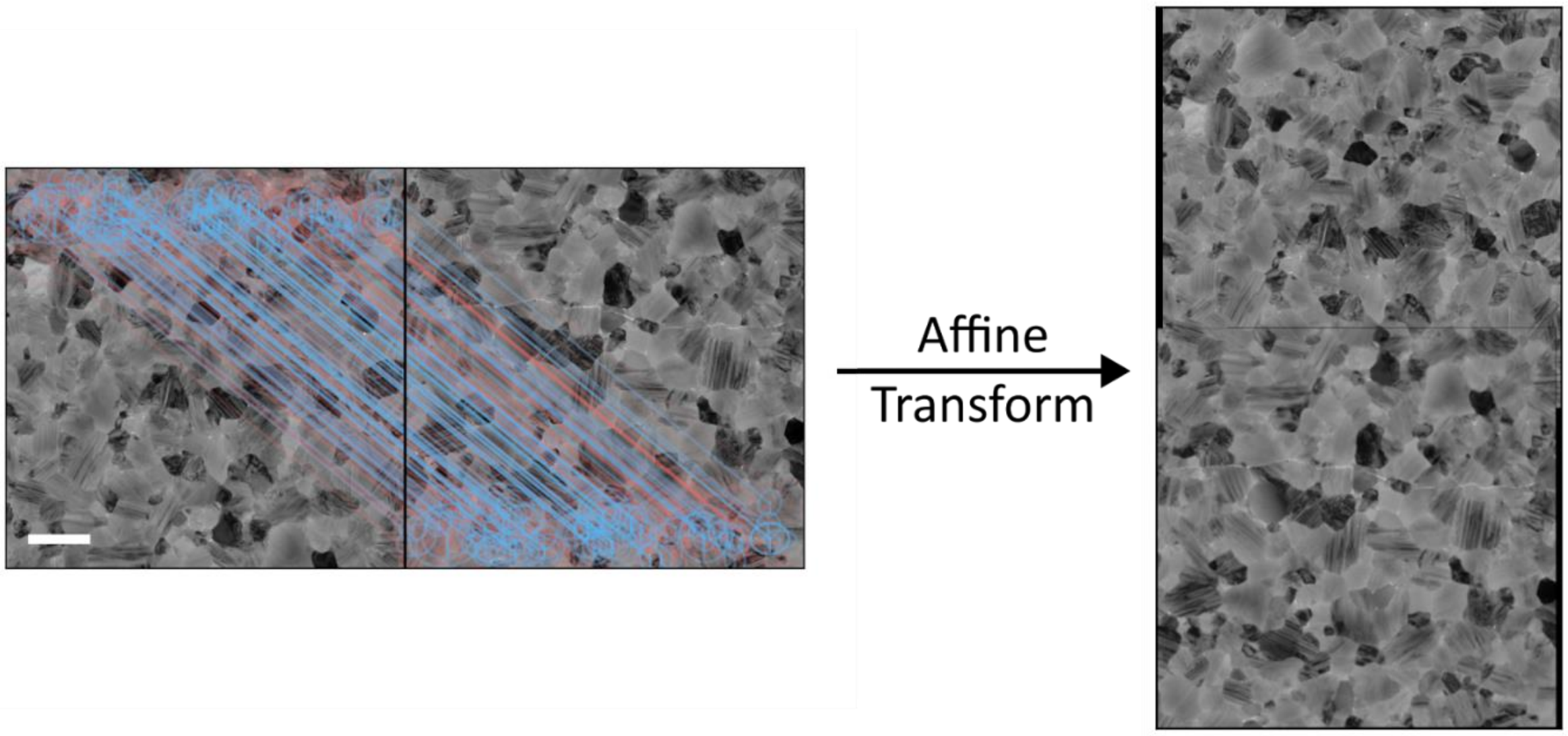

**Supplementary Fig.30 | Keypoints between the two virtual bright field images** detected using AdaLAM[27], with tentative and confident matches shown in pink and blue respectively. The RANSAC algorithm is used to define an affine transform which stitches the images together. Scale bar: 500 nm.

To then perform the spatial correlation of the hyperspectral PL and SED data we use the Au fiducial marker to find a common region between the two imaging modalities. To account for differences in pixel sizes between the hyperspectral PL and SED datasets the hyperspectral PL data is upsampled using linear interpolation. Due to the different contrast mechanisms between the techniques both datasets are then cropped such that only the Au fiducial marker is used during the registration. The resultant transform to overlay the datasets can be found several ways, herein we use two independent methods and compare their output. Firstly, an image transform can be defined by finding where the normalized cross correlation (NCC) is maximal between the two images accounting for both rotation and translation (Supplementary Fig. 31 & 32). Once the transformation has been found the two datasets can then be overlaid (as shown in main text Fig. 4).

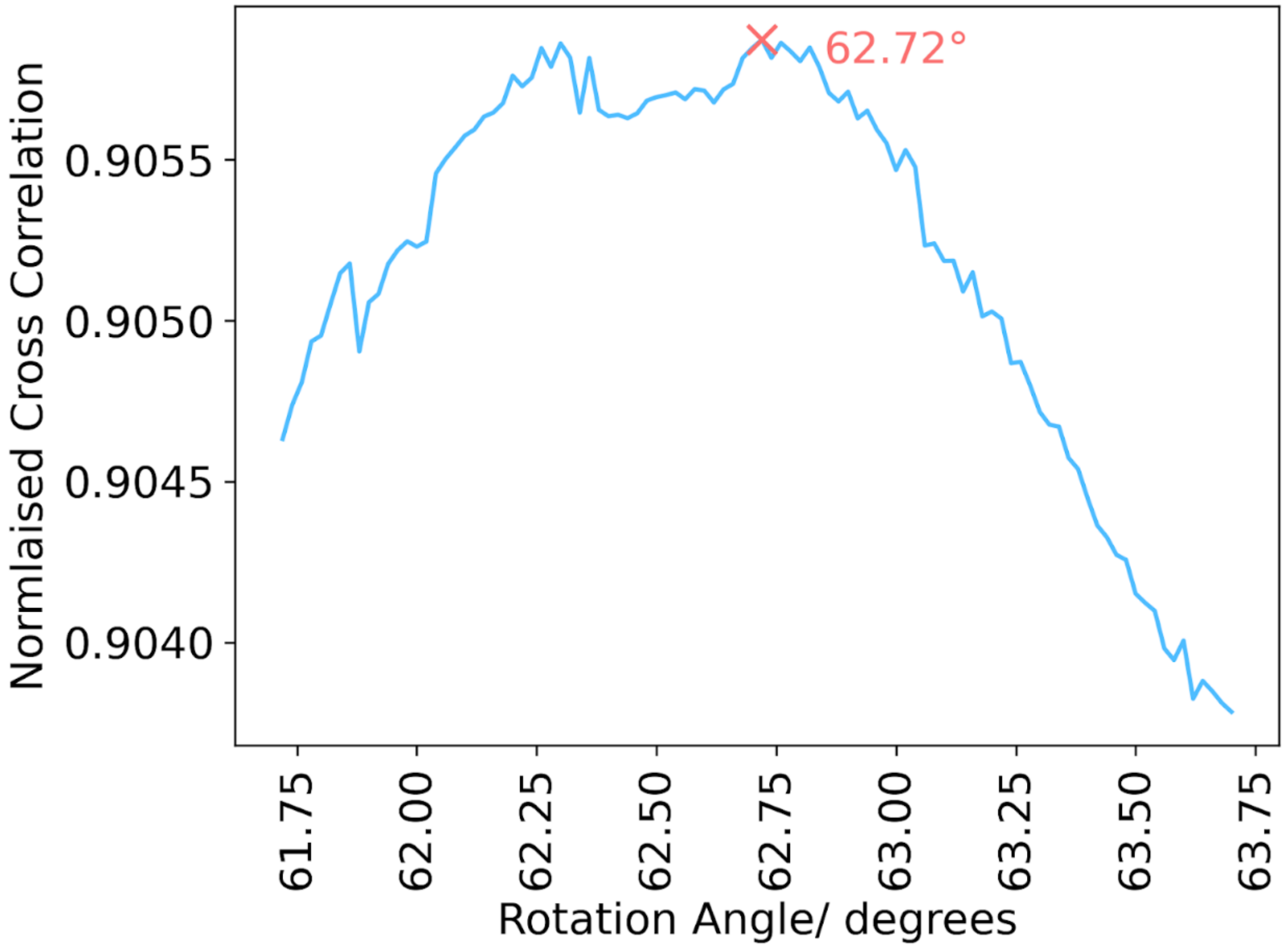

**Supplementary Fig.31 | The optimum rotation** between the two datasets found to be 62.72° where the NCC is maximal

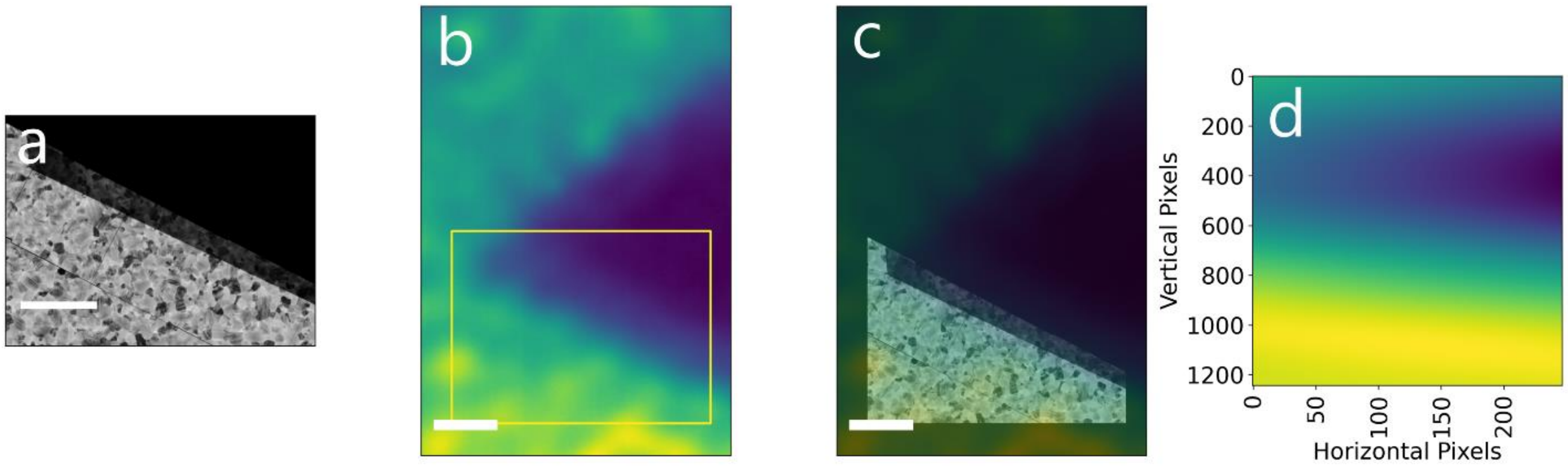

**Supplementary Fig.32 | Finding the optimal translation between the hyperspectral PL and SED. (a)** The stitched and rotated (62.72°) VBF image from the SED data. **(b)** The summed hyperspectral PL data over all channels and an overlaid rectangle showing where the NCC is maximal. **(c)** An overlay of the data shown in a&b. **(d)** The values for the NCC as (a) is rastered over (b) to find the optimal overlap. All scale bars: 2µm

Secondly, the Python package AntsPy, primarily built for the registration of medical images was used to define a mapping[29,30]. AntsPy uses 'multi-resolution gradient descent' and metrics such as the cross correlation (CC) or mutual information (MI) to define the transformation. During this process both the SED and hyperspectral PL were cropped, as above, such that only the Au fiducial marker is used during the correlation and a 'rigid' transform defined so that rotation and translation were accounted for (Supplementary Fig. 33). When the two methods are compared, they are shown to be consistent, with the grain oriented close to a $<111>_c/<001>_h$ zone axis in the SED being close to the bright isolated emitter in the hyperspectral PL.

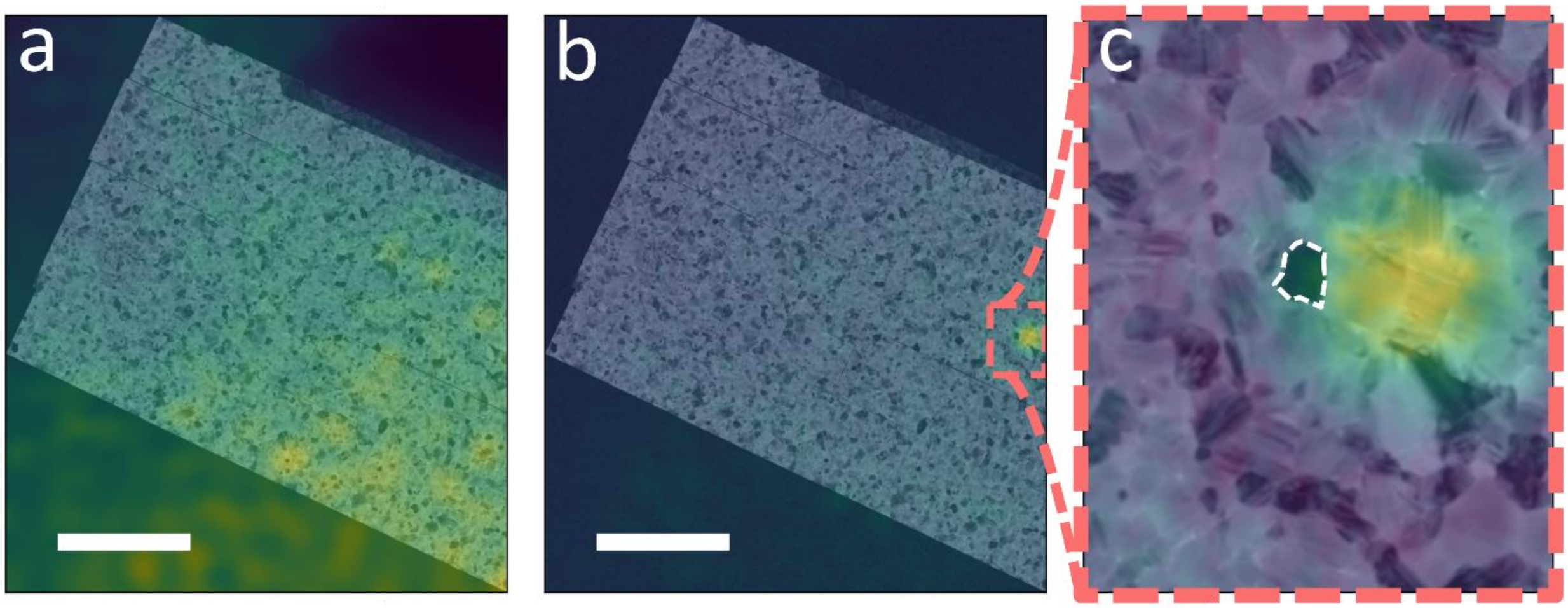

**Supplementary Fig.33 | Overlay of hyperspectral PL and SED**. **(a)** Overlay of hyperspectral PL, summed over all channels, and the stitched VBF image data using the transform defined via AntsPy. **(b)** Overlay of the PL map taken at an emission wavelength of 706 nm overlaid with the stitched VBF image. **(c)** Expansion from (b) showing the grain oriented along the <111>$_c$ zone axis (marked) is in the same locality as the isolated emitter. All scale bars: 5μm.

**Supplementary Text 5: Krönig-Penney superlattice**

As illustrated in the main text Fig. 3j, the Krönig-Penney model describes the quantized behavior of electrons in a periodic potential. The periodical potential is constructed by δ phase as the barrier with width $b$ and α phase as the well with width $a$. The E-k relation can be expressed by[31]:

$$\begin{cases} \cos k(a+b) = \dfrac{1-2\zeta}{2\sqrt{\zeta(1-\zeta)}} \sin\left(\alpha_0 a\sqrt{\zeta}\right)\sinh\left(\alpha_0 b\sqrt{1-\zeta}\right) \\ \qquad\qquad + \cos\left(\alpha_0 a\sqrt{\zeta}\right)\cosh\left(\alpha_0 b\sqrt{1-\zeta}\right) \\ \text{for } 0<\zeta<1 \\ \cos k(a+b) = \dfrac{1-2\zeta}{2\sqrt{\zeta(\zeta-1)}} \sin\left(\alpha_0 a\sqrt{\zeta}\right)\sin\left(\alpha_0 b\sqrt{\zeta-1}\right) \\ \qquad\qquad + \cos\left(\alpha_0 a\sqrt{\zeta}\right)\cos\left(\alpha_0 b\sqrt{\zeta-1}\right) \\ \text{for } 1<\zeta \end{cases} \qquad \text{Eq.S1}$$

where,

$\alpha_0 = \sqrt{\frac{2mU_0}{\hbar^2}}$ and $\zeta = \frac{E}{U_0}$,

In the given equations, the energy $E$ is the sole variable in the functions on the right-hand side, whereas the wavenumber $k$ is the only variable on the left-hand side. Analogous to the scenario of a finite potential well, a solution for $\zeta$ in Eq. S1 enables us to ascertain the energy values and the wavefunctions (once they are normalized). Considering the spacing of octahedral layers along a <111>$_c$ direction, we can get the form of the function on the right-hand side of Eq. S1, referred to as $f(\zeta)$, can be depicted in Supplementary Fig. 34.

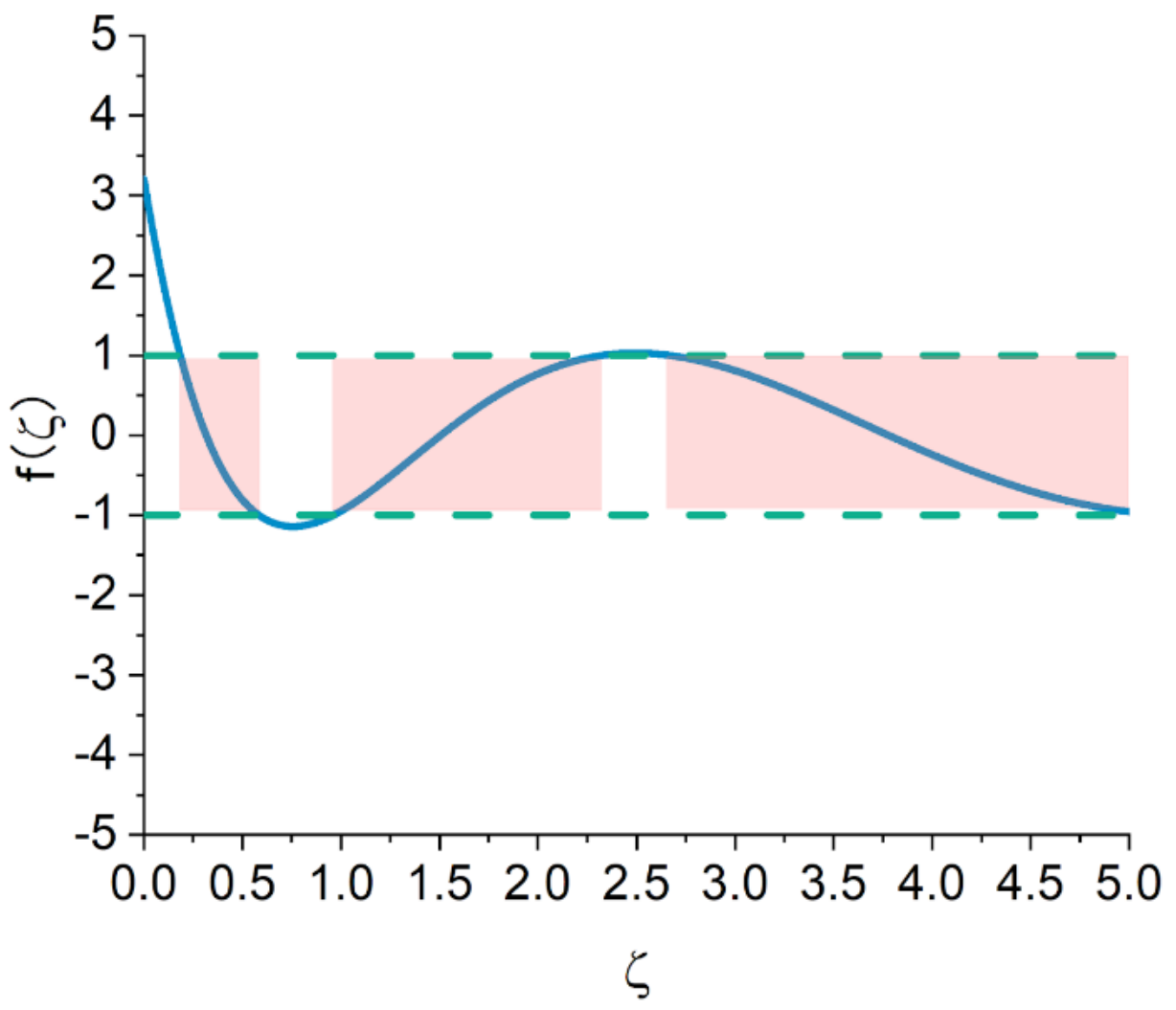

**Supplementary Fig.34 | The bands calculated from the KP model**. The plot of $f$ as a function of $\zeta$ shows the allowed energy bands (coloured area) from the quantum confinement of a superlattice described by a Krönig-Penney model.

In the following calculations, we only consider energy level of n=1, i.e. the first allowed band from the KP model for the following reasons:
1. As visualised in Fig.3, generally well and barrier widths are aperiodic and vary throughout the film. This means that the dimension-determined spacing between quantum energy levels is expected to change from grain to grain, leading to various peak positions for the same quantum number which is contrary to the fixed peak positions at a certain temperature observed experimentally in Fig.1 and 2.
2. As shown in Fig. 1d and Supplementary Fig.6c, the individual quantum levels do not share ground states, suggesting that the quantum levels are not from progression in the same confinement.
3. The independence of quantum levels is also supported by the isolated PL emission at distinct wavelengths, shown in Fig.2.

The KP lattice constructions corresponding to each energy level, collected from the simulation, are listed in table S1 with *u*=0.37 nm. We recognize the well/barrier combination of 4*u*/2*u* and 7*u*/2*u* as 12H and 18H, respectively. In the calculation, the input parameters include bandgaps (2.14eV for δ-phase and 1.52eV for the black absorbing phase), effective mass, and thickness of well and barriers. Owing to the uncertainty of the exact electron affinity of δ-phase and α phase, we consider the upper and lower bound as 0.62 and 0.31 eV. To estimate reasonable values for unknown variables we iterated through the calculation whilst altering unknown parameters in line with the variation reported in the literature. From this we found that the outcome of the KP model is particularly sensitive to effective mass. For example, when we implement the same well and barrier combination, like 12H, but different values of effective mass, m* as $0.27m_0$ or $0.21m_0$, the calculated first discrete energy level varies from 723 nm to 706 nm. This encompasses the emission we observe experimentally but, due to the idealized nature of the KP model and uncertainty in the value m*, variation between experimental data and theoretically calculated values is to be expected. Although this makes it difficult to conclusively assign a specific quantum peak to a specific polytype structure, the model demonstrates that the experimental values can be reproduced using physically meaningful values, and thus the proposal is consistent with experiment, as exampled in Supplementary Table 1.

**Supplementary Table 1 Simulated well and barrier lengths corresponding to each quantum peaks.**

| | $m^*$=0.1$m_0$, $U_0$=0.62 eV | | $m^*$=0.27$m_0$, $U_0$=0.31 eV | | $m^*$=0.21$m_0$, $U_0$=0.31 eV | |
|---|---|---|---|---|---|---|
| Peaks (nm) | Well length ($n_1$*$u$) (nm) | Barrier length ((2 + $n_2$) *$u$) (nm) | Well length ($n_1$*$u$) (nm) | Barrier length ((2 + $n_2$) *$u$) (nm) | Well length ($n_1$*$u$) (nm) | Barrier length ((2 + $n_2$) *$u$) (nm) |
| 632 | 4 | 2+1 | 2 | 2+0 | 2 | 2+1 |
| 662 | 5 | 2+0 | 2 | 2+1 | 3 | 2 |
| 689 | 6 | 2+0 | 3 | 2+0 | 3 | 2+1 |
| 706 | **7** | **2+0** | 3 | 2+1 | **4** | **2+0** |
| 723 | 8 | 2+0 | **4** | **2+0** | 4 | 2+2 |
| 736 | 9 | 2+0 | 4 | 2+2 | 5 | 2+1 |
| 748 | 10 | 2+0 | 5 | 2+1 | 6 | 2+0 |
| 756 | 11 | 2+0 | 6 | 2+0 | **7** | **2+0** |

**Supplementary Text 6: The mean lifetime of the excited carriers in the quantum levels.**

The following calculation is based on a single quantum well under WKB method[32]. The mean lifetime of an electron with energy of 1.756 eV stays at its excited energy level before tunnelling out through the edge of quantum confinement can be estimated by a semi-classical computation. In the first approximation, the ratio of $\psi$ at the outside edge of wall to escape, $x_1$ and at the inside edge of the wall, $x_0$, is

$$\Psi(x_1) = \Psi(x_0) exp\left(\frac{i}{\hbar}\int_{x_0}^{x_1} i\{2m^*[V(x) - E]\}^{1/2}\,dx\right) \equiv \Psi(x_0)e^{-\gamma/2}$$

where $\hbar$ is the reduced Planck constant, $m^*$ is the effective mass of the electron, $E$ is the energy of the electron, $V$ is the height of the potential barrier, $x_1 - x_0$ means the thickness of the barrier, and $V(x)$ is the potential energy at position $x$. The second part represents the probability of the particle tunnelling through the potential barrier. The first part is the decay constant, which is a measure of how quickly the wave function decays inside the barrier. It comes in a semi-classical way that the electron inside the well has a kinetic energy $T = E_{above-well} + V_0$, so its velocity is $v = [2m^*T]^{1/2}$ and it bangs against the wall at a frequency $f = {}^{v}/_{2L}$, where $L$ is the total superlattice length. Upon collision, there is an escape probability by quantum tunnelling $e^{-\gamma}$. As a result, the probability of escape rate is

$$R = \frac{[2m^*T]^{1/2}}{2mL}e^{-\gamma}$$

Substituting the given values from our experimental measures, $L$=200 nm, $x_1 = 2 \times 0.37\ nm$, along with $m^* = 0.21m_0$, we find a match between the recorded lifetime and calculated results:

for peak at 706 nm, $T = 1.756\ eV$, the mean lifetime is $\tau = {}^{1}/_{R} = 2.0\ ps$;

for peak at 756 nm, $T = 1.64\ eV$, the mean lifetime is $\tau = {}^{1}/_{R} = 2.8\ ps$.

This is in accordance with the slight increase in the ultrafast component of the decays measured at a peak with a longer wavelength, as shown in Fig. 1f and Supplementary Fig. 5d.

**Reference**

1. Chiang, Y.-H., Anaya, M. & Stranks, S. D. Multisource Vacuum Deposition of Methylammonium-Free Perovskite Solar Cells. *ACS Energy Lett.* **5**, 2498–2504 (2020).
2. Dai, L. *et al.* Slow carrier relaxation in tin-based perovskite nanocrystals. *Nat. Photonics* **15**, 696–702 (2021).
3. Weller, M. T., Weber, O. J., Frost, J. M. & Walsh, A. Cubic Perovskite Structure of Black Formamidinium Lead Iodide, α-[HC(NH $_2$ ) $_2$ ]PbI $_3$ , at 298 K. *J. Phys. Chem. Lett.* **6**, 3209–3212 (2015).
4. Li, Z., Park, J.-S. & Walsh, A. Evolutionary exploration of polytypism in lead halide perovskites. *Chem. Sci.* **12**, 12165–12173 (2021).
5. Li, Z., Park, J.-S., Ganose, A. M. & Walsh, A. From Cubic to Hexagonal: Electronic Trends across Metal Halide Perovskite Polytypes. *J. Phys. Chem. C* **127**, 12695–12701 (2023).
6. Ferrer Orri, J. *et al.* Unveiling the Interaction Mechanisms of Electron and X-ray Radiation with Halide Perovskite Semiconductors using Scanning Nanoprobe Diffraction. *Adv. Mater.* **34**, 2200383 (2022).
7. Rothmann, M. U. *et al.* Structural and Chemical Changes to CH $_3$ NH $_3$ PbI $_3$ Induced by Electron and Gallium Ion Beams. *Adv. Mater.* **30**, 1800629 (2018).

8. Li, Y. *et al.* Unravelling Degradation Mechanisms and Atomic Structure of Organic-Inorganic Halide Perovskites by Cryo-EM. *Joule* **3**, 2854–2866 (2019).
9. Francis, C. & Voyles, P. M. pyxem: A Scalable Mature Python Package for Analyzing 4-D STEM Data. *Microsc. Microanal.* **29**, 685–686 (2023).
10. Cautaerts, N. *et al.* Free, flexible and fast: Orientation mapping using the multi-core and GPU-accelerated template matching capabilities in the Python-based open source 4D-STEM analysis toolbox Pyxem. *Ultramicroscopy* **237**, 113517 (2022).
11. Shi, Y., Wang, W., Gong, Q. & Li, D. Superpixel segmentation and machine learning classification algorithm for cloud detection in remote-sensing images. *J. Eng.* **2019**, 6675–6679 (2019).
12. Achanta, R. *et al.* SLIC Superpixels Compared to State-of-the-Art Superpixel Methods. *IEEE Trans. Pattern Anal. Mach. Intell.* **34**, 2274–2282 (2012).
13. Duran, E. C. *et al.* Correlated electron diffraction and energy-dispersive X-ray for automated microstructure analysis. *Comput. Mater. Sci.* **228**, 112336 (2023).
14. Doherty, T. A. S. *et al.* Stabilized tilted-octahedra halide perovskites inhibit local formation of performance-limiting phases. *Science* **374**, 1598–1605 (2021).
15. Woodward, D. I. & Reaney, I. M. Electron diffraction of tilted perovskites. *Acta Crystallogr. B* **61**, 387–399 (2005).
16. Howard, C. J. & Stokes, H. T. Group-Theoretical Analysis of Octahedral Tilting in Perovskites. *Acta Crystallogr. B* **54**, 782–789 (1998).
17. Glazer, A. M. The classification of tilted octahedra in perovskites. *Acta Crystallogr. B* **28**, 3384–3392 (1972).

18. Weber, O. J. *et al.* Phase Behavior and Polymorphism of Formamidinium Lead Iodide. *Chem. Mater.* **30**, 3768–3778 (2018).

19. Fabini, D. H. *et al.* Reentrant Structural and Optical Properties and Large Positive Thermal Expansion in Perovskite Formamidinium Lead Iodide. *Angew. Chem. Int. Ed.* **55**, 15392–15396 (2016).

20. Chen, T. *et al.* Entropy-driven structural transition and kinetic trapping in formamidinium lead iodide perovskite. *Sci. Adv.* **2**, e1601650 (2016).

21. Frohna, K. *et al.* Nanoscale Chemical Heterogeneity Dominates the Optoelectronic Response over Local Electronic Disorder and Strain in Alloyed Perovskite Solar Cells. *ArXiv210604942 Cond-Mat Physicsphysics* (2021).

22. Weadock, N. J. *et al.* The nature of dynamic local order in CH3NH3PbI3 and CH3NH3PbBr3. *Joule* **7**, 1051–1066 (2023).

23. Dubajic, M. *et al.* Dynamic Nanodomains Dictate Macroscopic Properties in Lead Halide Perovskites. (2024) doi:10.48550/ARXIV.2404.14598.

24. Leutenegger, S., Chli, M. & Siegwart, R. Y. BRISK: Binary Robust invariant scalable keypoints. in *2011 International Conference on Computer Vision* 2548–2555 (IEEE, Barcelona, Spain, 2011). doi:10.1109/ICCV.2011.6126542.

25. Shabunin, M. CiteOpenCV. (2017).

26. Lowe, D. G. Distinctive Image Features from Scale-Invariant Keypoints. *Int. J. Comput. Vis.* **60**, 91–110 (2004).

27. Riba, E., Mishkin, D., Ponsa, D., Rublee, E. & Bradski, G. Kornia: an Open Source Differentiable Computer Vision Library for PyTorch. (2019) doi:10.48550/ARXIV.1910.02190.

28. Liu, J. & Bu, F. Improved RANSAC features image-matching method based on SURF. *J. Eng.* **2019**, 9118–9122 (2019).

29. Tustison, N. J. *et al.* The ANTsX ecosystem for quantitative biological and medical imaging. *Sci. Rep.* **11**, 9068 (2021).

30. Frohna, K. *et al.* Multimodal operando microscopy reveals that interfacial chemistry and nanoscale performance disorder dictate perovskite solar cell stability. (2024) doi:10.48550/ARXIV.2403.16988.

31. Razeghi, M. *Fundamentals of Solid State Engineering*. (Springer International Publishing, Cham, 2019). doi:10.1007/978-3-319-75708-7.

32. Shankar, R. *Principles of Quantum Mechanics*. (Plenum Press, 1994).